# Continuous-variables quantum cryptography: asymptotic and finite-size security analysis

Panagiotis Papanastasiou

Doctor of Philosophy

University of York
Computer Science

October, 2018

*Dedication*

To my beloved Mother

# Abstract

In this thesis we study the finite-size analysis of two continuous-variables quantum key distribution schemes. The first one is the one-way protocol using Gaussian modulation of thermal states and the other is the measurement-device-independent protocol. To do so, we adopt an efficient channel parameter estimation method based on the assumption of the Gaussian variables and the central limit theorem introduced by Ruppert *et al.* [Phys. Rev. A **90**, 062310 (2014)]. Furthermore, we present a composable security analysis of the measurement device independent protocol for coherent attacks with a channel parameter estimation that is not based on the central limit theorem.

We also investigated, in the asymptotic regime, an asymmetric situation for the authenticated parties against the eavesdropper caused by fast-fading channels. Here we assume that the eavesdropper has the full control of the communication channel and can instantaneously change its transmissivity in every use of it. We assumed the simple model of a uniform fading and addressed the cases of one-way protocols, continuous-measurement-device-independent protocol in symmetric configuration and its star network extension for three users. Finally, we extended the asymptotic study of the one-way protocols using an arbitrary number of phase-encoded coherent states assuming a thermal loss channel without using a Gaussian approximation.

# Contents

# List of Figures

# Acknowledgements

I would like thank all the people in my research group, here in York university, for their support. First of all, I would like to thank my Supervisor Prof. Stefano Pirandola for trusting me with this work. His guidance and experience were provided in every step on this long way called PhD research and for this, I have no words to express my gratitude. Then I would like to thank Dr. Carlo Ottaviani, who I consider as my mentor. His support was more than vital to me for achieving my goals in the PhD. I would also like to thank Dr. Cosmo Lupo for the wonderful collaboration and all his advises on scientific and life matters. I am also grateful for all the help I received from my co-supervisor Prof. Samuel Braunstein, and the TAP members Dr. Roger Colbeck and Prof. John Clark all these years.

A lot of thanks go to my colleagues and comrades in this adventure, Riccardo Laurenza and Thomas Cope. I will miss working with them in the same environment sharing emotions and experiences. I hope that we will meet again in the near future and collaborate. I would also like to express my gratitude for the nice environment of the Computer Science Department and all the efforts made by my colleges here to keep the PhD process as easy as possible.

Many thanks also to my friends here in York and specially Dr. Athanasios Zolotas for his amazing hospitality during the last days of writing this thesis. Knowing that they were there for me any time that I need them made the difference in managing to complete all this work. Most importantly, I would like to thank my family for all the support they gave me all these years being abroad chasing my goals. Special thanks to my mother, Alexiou Panagiota, my big supporter to every step I took in my life till now, being vigilant at every moment. I dedicate this work to her as the least thing I can do to express my gratitude.

# Declaration

I declare that the research described in this thesis is original work and I am the sole author. This work has not previously been presented for an award at this, or any other, University. All sources are acknowledged as References. Except where stated, all of the work contained within this thesis represents the original contribution of the author.

Some parts of this thesis have been published in conference proceedings and journals; where items were written and published jointly with collaborators, the author of this thesis is responsible for the material presented here. For each published item the primary author is the first listed author.

- P. Papanastasiou, C. Ottaviani, and S. Pirandola, "Gaussian one-way thermal quantum cryptography with finite-size effects", Phys. Rev. A **98**, 032314 (2018) [26].

- P. Papanastasiou, C. Ottaviani, and S. Pirandola, "Finite-size analysis of measurement-device-independent quantum cryptography with continuous variables", Phys. Rev. A **98**, 032314 (2018) [34].

- C. Lupo, C. Ottaviani, P. Papanastasiou, and S. Pirandola, "Continuous-variable measurement-device-independent quantum key distribution: Composable security against coherent attacks", Phys. Rev. A **97**, 052327 (2018) [35]
  My contribution here was mostly on the part of channel parameter estimation

- P. Papanastasiou, C. Weedbrook, and S. Pirandola, "Continuous-variable quantum key distribution in uniform fast-fading channels", Phys. Rev. A **97**, 032311 (2018) [47].

- P. Papanastasiou, C. Lupo, C. Weedbrook, and S. Pirandola, "Quantum key distribution with phase-encoded coherent states: Asymptotic security analysis in thermal-loss channels", Phys. Rev. A **98**, 012340 (2018) [64].

# Chapter 1

# Introduction

Quantum Information is the study of Information Theory based on the Principles of Quantum Mechanics [1]. As such, it combines disciplines as Computer Science, Physics and Mathematics. Quantum properties and interactions between systems are now taken into consideration that were firstly noticed in the quantum world of microscopic particles. In fact, these properties and interactions can provide with advancements to technological fields such as quantum computation, communication and metrology. More specifically, one of their contribution in terms of communication is the Quantum Key Distribution (QKD).

Here, protocols describing the public exchange of (quantum) signals between traditionally two or more authenticated parties and their post-processing can guaranty in principle, based on the lows of Quantum Mechanics, the creation of a secret shared random data string, called the key [2, 3]. This can be used later for message encryption based for example on unconditionally secure processes such as the one-time pad primitive [4]. In order such a primitive to work, the parties must have the same key, which is as large as the message to be sent, exchanged in advance after a clandestine meeting. Every time they should exchange a different key for a different message. Otherwise accumulation of messages encrypted with the same key can reveal information about the key putting in danger the security of the communication. This means that there is a need for a regular large-secret-key distribution mechanism using public channels to meet the demand of the amount of telecommunication nowadays. One candidate for this mechanism is QKD. Nowadays a popular way to treat the previous gap is the asymmetric encrypting schemes, e.g. RSA protocol [5], that their security is based on the computational power of the given eavesdropper, the potential adversary of the authenticated parties trying to reveal their secret key and overhear their communication. However, progress in computational capability provided by, e.g., a Quantum computer, can threaten this kind of security.

Fortunately, apart from the first protocol introduced by Bennet and Brassard [6], further progress has been made the last years towards not only theoretical studies but also experimental implementations of QKD protocols. Here, we will focus on research mostly done in a continuous-variables (CV) framework [7] in the sense of using the continuous electromagnetic field degrees of freedom in contrast to the originally used discrete photon degrees of freedom. This CV-QKD version [8] allows the use of common components regarding current technology for being implemented with reasonable performances for metropolitan network areas.

The central protocols encapsulating the basic ideas of CV-QKD are these using a particular kind of quantum states of light, called coherent states (see Sec. 2.4.1.2), as signals modulated by a Gaussian distribution [9–11]. For these protocols, we have also security analyses incorporating finite-size effects [12,14,15]. In this case, their performance is evaluated with respect to a given number of signal exchange in contrast to the initial step of an asymptotic security analysis under the assumption of a very large, asymptotically infinite, number of exchanged signals. This assumption implies a more realistic description and evaluation of a given protocol closer to its performance in a practical implementation. Moreover, it comprises an intermediate step for a security analysis under a composable framework [16], where, for example, the overall security of a cryptographic task including a QKD protocol can be assessed with respect to the latter's security. Here, a given protocol breaks down to smaller tasks that all contribute to its security. The result of each task is evaluated with respect to a small parameter connected to the distance from the ideal situation. In the end, these parameters sum up to a larger parameter associated with the protocol's security. Recently, in Ref. [17,18], proofs of composable security were presented for the protocols with coherent states and Gaussian distribution taking into consideration the most general (powerful) attacks.

A variation of the protocols described above is this of using a particular kind of quantum states, called thermal states (see 2.4.1.1), as the noisy counterpart of the coherent states mentioned above. This can be seen as an error in the preparation procedure of the signal states. However, this trusted thermal noise [23] can be connected to the frequency of the electromagnetic field and provide with schemes working in different regimes than the optical one, for instance, schemes operating in microwaves. The asymptotic security of schemes using thermal states with Gaussian modulation was studied in Ref. [19–23] and an experiment was done in Ref. [25]. In this thesis, and to the best of my knowledge, we present for the first time a non-asymptotic security analysis of such a protocol using thermal states (see Ref. [26]) by incorporating finite size effects based on the method provided by Ref. [15].

In such QKD schemes, the two authenticated parties exchange signals publicly usually through a telecommunication channel. Then we assume mainly that the eavesdropper has the opportunity to interact with the signals as they travel from one user to another. But in general, an attack is feasible to be made during the preparation or the detection of the signals, namely a side-channel attack [3, 27, 28]. In this case, the detection attacks are considered a more common practice for the eavesdropper. Nevertheless, protocols were proposed in Ref. [29–31] that take into consideration countermeasures against side-channel attacks on the detection process. These protocols are considered to be measurement device independent (MDI) in the sense that the detection of the signals are made by a third party (authenticated or not), i.e., an intermediate relay, and the outcomes are broadcast. Due to the existence of the relay, such an end-to-end setting can be extended to a network configuration as in Ref. [32,33]. In Ref. [34], we investigated the performance of this protocol assuming finite-size effects based on an non-asymptotic security analysis as previously. Furthermore, we participated in the work for providing a composable security analysis for the CV-MDI-QKD (see Ref. [35,36]).

Let us assume now, that the communication channel is not stable in the sense that its ability to transmit (transmissivity) signals changes in time. This inconsistence follows usually from some pattern which is described by a probability distribution. This phenomenon is called the fading of the channel [37] and can have several implications on the performance of the protocol. Usually, this comes into effect and has implications for a given protocol when we assume free-space links [38] susceptible to environmental conditions such as atmospheric turbulence [39–45]. In case that the changes are slow enough so that the parties can estimate the fading distribution, the performance of the protocol can be averaged out over it, as for example in Ref. [46]. There are cases though that the channel is under the full control of the eavesdropper, who might have the ability to change instantaneously the transmissivity of the link for every signal. This results in an asymmetric situation for the users against the eavesdropper: the parties know the particular distribution only by the end of the signal exchange. We investigated this worst case scenario in Ref. [47] for the protocol with coherent states and Gaussian modulation, the CV-MDI-QKD protocol in the symmetric configuration and its three-user star network extension adopting a simple assumption of uniform fast fading channels. The simplicity of the uniform distribution can be justified by the assumption of an adversary tampering with the transmissivity of the channel in a mechanical way.

An alternative way for CV-QKD that keeps some properties of its discrete counterpart is the use of a given number of coherent states, each one encoding a given letter in an alphabet, instead of using a Gaussian distribution to modulate them. Such protocols have an advantage on their performance compared with the fully continuous ones regarding the existence of more efficient error correction codes for discrete variables [8]. Asymptotic analysis has been provided for such protocols in Ref. [48]. We extended this analysis for the case of a thermal loss channel without using a Gaussian approximation assumption.

This thesis is split into two parts. The first one is introducing the main concept of CV-QKD and discusses the schemes and studies that I have been based on in order to be able to contribute in the recent advances of CV-QKD. My contribution is presented in the second part. So we start with the structure of the first part: In Chap. 2, we introduce the quantum states of the electromagnetic field and their representation in phase space, which justifies the term of continuous-variables. We also discuss matters such as their evolution through channels and their detection as signals. Finally, we introduce quantities that quantify the amount of information carried by such states. In Chap. 3, we introduce the QKD with CV using as an example the the one-way protocols with coherent states modulated by a Gaussian distribution. We also introduce their discrete counterpart, i.e., a protocol with an arbitrary number of phase-encoded coherent states. Afterwards, in Chap. 4 and Chap. 5, we address the asymptotic analysis of the thermal state protocol and the CV-MDI protocol respectively. And in the last chapter of this part (Chap. 6), we present the efficient channel parameter estimation method incorporating finite-size effects for the protocols using coherent states with Gaussian modulation.

The structure of the second part continues as follows: Based on this previous method and by adapting it appropriately, we present an non-asymptotic security analysis for the thermal state protocol and for the CV-MDI protocol in Chap. 7 and Chap. 8 respectively. In Chap. 9, we discuss about a channel parameter estimation without assuming the central limit theorem and explain how the non-assymptotic security analysis in a composable framework functions for the case of the CV-MDI protocol. In Chap. 10, we investigate the fast fading channel assumption for protocols with coherent states, measurement-device independent protocol and a network star configuration for three-users. Finally, in Chap. 11, we address the case of thermal loss channel for the arbitrary number phase-encoded coherent state protocol.

# Part I

# Preliminaries

# Chapter 2

# Continuous-variables states

## 2.1 Introduction

In current telecommunication, the electromagnetic field plays the central role of the signal carrier. In this chapter, we will introduce its quantized counterpart summarized by the bosonic systems, which can be associated with CV systems. In particular, we will present the formalism of bosonic systems and the corresponding quantum states. Furthermore, we are going to address their representation in the phase space, as it gives us insight in our later discussion. In fact, the phase-space representation is directly connected with the term CV. This representation is analogous to a classical system with its "position" corresponding to the in-phase component of the field and its "momentum" corresponding to the out-of phase component of the field. All the following concepts are described more thoroughly in Ref. [1, 7, 8, 49].

## 2.2 Quantized electromagnetic field

In classical physics, the two fundamental quantities describing the electromagnetic field are the electric filed $\mathbf{E}(\mathbf{r}, t)$ and the magnetic field $\mathbf{B}(\mathbf{r}, t)$, both functions of time $t$ and space $\mathbf{r}$. The two fields are satisfying the Maxwell equations given by

$$\nabla \times \mathbf{E} = -\frac{\partial \mathbf{B}}{\partial t}, \tag{2.1}$$

$$\nabla \times \mathbf{B} = \frac{1}{c^2}\frac{\partial \mathbf{E}}{\partial t}, \tag{2.2}$$

$$\nabla \cdot \mathbf{E} = 0, \tag{2.3}$$

$$\nabla \cdot \mathbf{B} = 0, \tag{2.4}$$

for the empty space, i.e., vacuum, where the densities of currents and charges are zero and $c$ denotes the speed of light. The Maxwell equations interconnect the evolution of these two fields resulting in a propagation of an electromagnetic wave. In fact, solving these equations within an infinite square box, the field is decomposed into planar waves or bosonic modes. A single mode of the field is a propagating wave with a given frequency $f = 2\pi\omega$, propagation direction $k$ and polarization $z$ and its electric field can be expressed as

$$\mathbf{E}(\mathbf{r}, t) = \mathrm{i}Ez\ [a\mathbf{u}(\mathbf{r})e^{\mathrm{i}\omega t} - \text{c.c.}], \tag{2.5}$$

where $E$ includes all the physical constants. The propagation in space is described by $\mathbf{u}(\mathbf{r})$ whereas the evolution in time by $e^{\mathrm{i}\omega t}$. The complex conjugate term describes the reverse propagation wave component in contrast to the propagation of the first term component in the brackets. In classical electromagnetism the amplitude $a$ is a complex number and $a^*$ is its complex conjugate. A rigorous way to quantize the field is by replacing the amplitudes in Eq. (2.5) with the mutually adjoint quantum operators $\hat{a}$ and $\hat{a}^\dagger$ respectively, which follow the bosonic commutation relations

$$[\hat{a}, \hat{a}^\dagger] = 1, \quad [\hat{a}, \hat{a}] = [\hat{a}^\dagger, \hat{a}^\dagger] = 0. \tag{2.6}$$

As a result, the electric and magnetic field are now quantum operators denoted by $\hat{\mathbf{E}}$ and $\hat{\mathbf{B}}$. On that account, the energy of the mode is expressed by

$$\hat{H} = \frac{1}{2}\int \epsilon_0 \hat{\mathbf{E}}^2 + \frac{1}{\mu_0}\hat{\mathbf{B}}^2\ d\mathbf{r}, \tag{2.7}$$

where $\mu_0\epsilon_0 = c^{-2}$, $\epsilon_0$ is the electric permittivity and $\mu_0$ is the magnetic permeability of free space. Using Eq. (2.5) and natural units ($h = 2$), the energy simplifies to

$$\hat{H} = 2\hat{n} + 1, \tag{2.8}$$

where $\hat{n} = \hat{a}\hat{a}^\dagger$.

We can associate each isolated bosonic mode to a vector space $\mathcal{H}$ called Hilbert space, where its unit vectors $|\psi\rangle \in \mathcal{H}$ describe a given state $\hat{\rho} = |\psi\rangle\langle\psi|$ of the mode, where $\langle\psi|$ stands for the Hermitian conjugate of $|\psi\rangle$.

**Remark**: For unit vectors $|\psi_i\rangle \in \mathcal{H}$ for $i = 1, 2, \ldots, N$ we have that the unit vector $|\psi\rangle = \sum_{i=1}^{N} c_i|\psi_i\rangle \in \mathcal{H}$, where $\sum_{i=1}^{N} |c_i|^2 = 1$ and $c_i \in \mathbb{C}$ .

A basis of $\mathcal{H}$ can be constructed by the number operator. To do so, one solves the eigensystem

$$\hat{n}|n\rangle = n|n\rangle, \tag{2.9}$$

where $n = 0, 1, \cdots + \infty$, and $|n\rangle$ are the eigenkets of $\hat{n}$. The eigenset $\{|n\rangle\}_{n=0}^{\infty}$ forms an orthonormal basis and spans a countable Hilbert space called the Fock space. It is called the number state basis because each eigenket describes a state with exactly $n$ energy excitations of the system, i.e., photons. Nevertheless, as one can see, even if the mode has no excitations, its energy is not zero. As a result, the state without excitations is called the vacuum and its energy is called the vacuum shot noise which we assume to be equal to 1. Moreover, the operators $\hat{a}$ and $\hat{a}^\dagger$ are called the ladder operators and their action in the number state basis is given by

$$\hat{a}|0\rangle = 0, \quad \hat{a}|n\rangle = \sqrt{n}|n-1\rangle \quad (\text{for } n \geq 1), \tag{2.10}$$

$$\hat{a}^\dagger|n\rangle = \sqrt{n+1}|n+1\rangle \quad (\text{for } n \geq 0). \tag{2.11}$$

A bosonic system is a system that consist of $N$ quantized radiation field modes, i.e., bosonic modes, corresponding to $N$ quantum harmonic oscillators. A bosonic state then is associated with a Hilbert space $\mathcal{H}^{\otimes N} = \bigotimes_{i=1}^{N} \mathcal{H}_i$ equal to the tensor product of $N$ bosonic mode Hilbert spaces $\mathcal{H}_i$. Accordingly, one can express the ladder operator of a given bosonic system with respect to the ladder operators of its modes in a vectorial form as $\hat{\mathbf{c}} := (\hat{a}_1, \hat{a}_1^\dagger \ldots \hat{a}_N, \hat{a}_N^\dagger)$. As a result, the bosonic commutation relations for $\hat{\mathbf{c}}$, by virtue of the symplectic form

$$\boldsymbol{\omega} := \begin{pmatrix} 0 & 1 \\ -1 & 0 \end{pmatrix} \tag{2.12}$$

are expressed as

$$[\hat{\mathbf{c}}_i, \hat{\mathbf{c}}_j^\dagger] = \boldsymbol{\Omega}_{ij} \quad (i, j = 1, \ldots, 2N), \qquad \boldsymbol{\Omega} := \bigoplus_{k=1}^{N} \boldsymbol{\omega}. \tag{2.13}$$

A state of $M$ isolated systems each one consisting of $N_i$ modes $i = 1, 2, \ldots, M$ is described by a tensor product of states

$$|\phi\rangle\langle\phi| = |\phi_1\rangle\langle\phi_1| \otimes |\phi_2\rangle\langle\phi_2| \otimes \cdots \otimes |\phi_M\rangle\langle\phi_M|, \tag{2.14}$$

where $|\phi_i\rangle\langle\phi_i|$ are the states of each isolated system. Let us assume that there is an ensemble of systems each one found in a state $|\phi_i\rangle\langle\phi_i|$ for $i = 1, 2, \ldots, L$ with a probability $p_i$. Then the average state of this ensemble will be given by

$$\hat{\rho} = \sum_{i=1}^{L} p_i |\phi_i\rangle\langle\phi_i|, \tag{2.15}$$

where $\sum_{i=1}^{L} p_i = 1$ and are called mixed states in contrast with the pure states defined as the states that can be associated with vectors $|\psi\rangle \in \mathcal{H}$. The mixed states encapsulate the notion of classical randomness associated with not knowing the initial conditions of a process. On the contrary, quantum randomness or uncertainty that is an intrinsic property of a quantum system is described by the principle of superposition (see remark in Sec. 2.2).

Now consider a system consisting of two subsystems $A$ and $B$. The state of the composite system is $\hat{\rho}_{AB}$ while the state $\hat{\rho}_A$ of subsystem $A$ is given by the partial trace of $\hat{\rho}_{AB}$ over the the subsystem $B$ expressed as

$$\hat{\rho}_A = \mathrm{Tr}_B \hat{\rho}_{AB}. \tag{2.16}$$

and its action is given in the following definition.

**Definition 2.2.1** The partial trace operation is the linear operation that for any two vectors $|a_1\rangle$, $|a_2\rangle$ in the states space of $A$ and any two vectors $|b_1\rangle$, $|b_2\rangle$ in the state space of $B$ is defined as follows

$$\mathrm{Tr}_B(|a_1\rangle\langle a_2| \otimes |b_1\rangle\langle b_2|) := |a_1\rangle\langle a_2| Tr(|b_1\rangle\langle b_2|) = |a_1\rangle\langle a_2|\langle b_2|b_1\rangle. \tag{2.17}$$

We can see that for a composite system in a pure state described by a superposition of states

$$\hat{\rho}_{AB} = \sum_{i=1}^{L} c_i|a_i\rangle|b_i\rangle \sum_{j=1}^{L} c_j^*\langle a_j|\langle b_j| = \sum_{i,j=1}^{L} c_i c_j^* |a_i\rangle\langle a_j| \otimes |b_i\rangle\langle b_j|, \tag{2.18}$$

where $\langle b_i|b_j\rangle = \delta_{ij}$ and $\sum_{i=1}^{L} |c_i|^2 = 1$, the partial trace operation over $B$ will be given by

$$\hat{\rho}_B = \mathrm{Tr}_B \hat{\rho}_{AB} = \sum_{i=1}^{L} |c_i|^2 |a_i\rangle\langle a_i| \tag{2.19}$$

which is a mixed state. In fact quantum mechanical systems have the property according to which the exact knowledge of a system's state (pure state) does not reflect the exact knowledge about its subsystems state (mixed states).

## 2.3 Phase-space representation

At this point, one associates a 2-dimensional vector space, namely the phase-space $\mathcal{K}$, spanned by the quadratures $q$ and $p$ of a given bosonic mode with a Hilbert space $\mathcal{H}$. These are the continuous eigenvalues $q, p \in \mathbb{R}$ of the quadrature operators $\hat{q}$ and $\hat{p}$ and are defined as

$$\hat{q}|q\rangle = q|q\rangle \quad \text{and} \quad \hat{p}|p\rangle = p|p\rangle, \tag{2.20}$$

where the eigenkets are connected by

$$|q\rangle = \frac{1}{2\sqrt{\pi}} \int dp\; e^{-\mathrm{i}qp/2} \quad \text{and} \quad |p\rangle = \frac{1}{2\sqrt{\pi}} \int dq\; e^{\mathrm{i}qp/2}. \tag{2.21}$$

The states $|q\rangle$ and $|p\rangle$ lie outside the Hilbert space of the system since they are not normalizable. Nevertheless, they form an orthogonal set. In particular, one has

$$\langle q|q'\rangle = \delta(q-q') \quad \text{and} \quad \langle p|p'\rangle = \delta(p-p'), \tag{2.22}$$

where $\delta$ stands for the Dirac-delta function. The quadrature eigenstates form a basis for the Hilbert space $\mathcal{H}$ in the sense that for any given $|\phi\rangle \in \mathcal{H}$ exists its decomposition in the position basis $\{|q\rangle\}$ given by $|\phi\rangle = \int dq\phi(q)|q\rangle$, where $\phi(q) = \langle q|\phi\rangle$ is called the wave function of the position for the state $|\phi\rangle$. Analogously, one can proceed for $|p\rangle$. The relation between the quadrature operators and the ladder operators are given by

$$\hat{q} = \hat{a} + \hat{a}^\dagger \quad \text{and} \quad \hat{p} = \mathrm{i}(\hat{a}^\dagger - \hat{a}), \tag{2.23}$$

resulting in the canonical commutation relation

$$[\hat{q}, \hat{p}] = 2\mathrm{i}. \tag{2.24}$$

In phase space, a classical state of a mode, where its electric field is described by Eq. (2.5), corresponds to a single point $x = (q,p)^T$. From an intuitive point of view, the quantum description of the field is the extension of the classical one, where due to the introduction of quantum noise, a manifestation of Heisenberg's principle, the single point is exchanged with a continuum set of points taken with different probabilities. Before defining more rigorously this probability formulation, one introduces the vectorial form of the quadrature operators for a bosonic system of $N$ modes. In that event, the quadrature operators are arranged in the vectorial form as

$$\hat{\mathbf{x}} := (\hat{q}_1, \hat{p}_1, \ldots, \hat{q}_N, \hat{p}_N)^T, \tag{2.25}$$

where the canonical commutation relations in natural units ($h = 2$) are given by

$$[\hat{x}_i, \hat{x}_j] = 2\mathrm{i}\boldsymbol{\Omega}_{ij} \tag{2.26}$$

derived by Eq. (2.13). Subsequently, one has that

$$\hat{\mathbf{x}}^T|\mathbf{x}\rangle = \mathbf{x}^T|\mathbf{x}\rangle \tag{2.27}$$

with $\mathbf{x} \in \mathbb{R}^{2N}$ and $|\mathbf{x}\rangle := (|x_1\rangle, \ldots, |x_{2N}\rangle)$.

For any bosonic state $\hat{\rho}$, there is an equivalent representation in terms of a quasi-probability distribution called Wigner function over the CV $\mathbf{x} \in \mathbb{R}^{2N}$, eigenvalues of the quadrature operators $\hat{\mathbf{x}}$, which span a real symplectic space $\mathcal{K} := (\mathbb{R}^{2N}, \mathbf{\Omega})$, i.e., the phase space. In order to introduce this representation, one needs first to introduce the Weyl operator

$$D(\boldsymbol{\xi}) := \exp\left(\mathrm{i}\hat{\mathbf{x}}^T \mathbf{\Omega} \boldsymbol{\xi}\right), \quad \boldsymbol{\xi} \in \mathbb{R}^{2N}. \tag{2.28}$$

Then the Wigner characteristic function will have the following form

$$\chi(\boldsymbol{\xi}) = \mathrm{Tr}[\hat{\rho} D(\boldsymbol{\xi})], \tag{2.29}$$

where by using its Fourier transform one obtains

$$W(\mathbf{x}) = \frac{1}{(2\pi)^{2N}} \int_{\mathbb{R}^{2N}} d^{2N}\boldsymbol{\xi} \, \exp\left(-\mathrm{i}\mathbf{x}^T \mathbf{\Omega} \boldsymbol{\xi}\right) \chi(\boldsymbol{\xi}). \tag{2.30}$$

This is the Wigner function which is normalized to 1 but it can be negative thus characterized as a quasi-probability distribution.

As a quasi-probability distribution, a Wigner function is fully characterized by its statistical moments. The first statistical moment corresponding to the mean $\bar{\mathbf{x}}$ is the displacement vector

$$\bar{\mathbf{x}} := \langle \hat{\mathbf{x}} \rangle = \mathrm{Tr}[\hat{\mathbf{x}}\hat{\rho}]. \tag{2.31}$$

The second moment is called the covariance matrix (CM) $\mathbf{V}$ with elements

$$V_{ij} := \frac{1}{2}\langle \{\Delta\hat{x}_i, \Delta\hat{x}_j\} \rangle, \quad \Delta\hat{x}_i = \hat{x}_i - \langle \hat{x}_i \rangle, \tag{2.32}$$

where $\{,\}$ is the anti-commutator. Using the CM of Eq. (2.32) and the commutation relations in Eq. (2.26), one can express the uncertainty principle as

$$\mathbf{V} + \mathrm{i}\mathbf{\Omega} \geq 0, \tag{2.33}$$

which implies that $\mathbf{V} > 0$, i.e., is a positive definite matrix. The Wigner function may be used to represent the quantum state of a mode. In particular, its contour depicts the set of the most probable points in the phase space and provides us with an intuitive picture especially suitable for quantum states with a Gaussian Wigner function.

With the phase-space representation, we exploit the quantum mechanical nature of light keeping the same approach as in the classical regime, i.e., describing it as a field or wave and not as particles. This allows us to use it as a carrier for (quantum) signals encoded on its modulated intensity (energy) exploiting the current technology designed for the classical electromagnetic field. In addition to this, the mathematical tools based on this description provide us with a concise and intuitive way to describe the quantum states of light via the Wigner function, where their description with CV becomes evident. As we will see later, another important tool studied in this chapter is the CM. In fact, the states with Gaussian Wigner functions are described only by their first two statistical moments, in the sense that all the information for them are included in their mean and CM.

## 2.4 Gaussian states

Gaussian states are bosonic states having a Wigner representation which is a Gaussian. Their characteristic function has the following form

$$\chi(\mathbf{x}) = \exp\left(-\frac{1}{2}\boldsymbol{\xi}^T(\boldsymbol{\Omega}\mathbf{V}\boldsymbol{\Omega}^T)\boldsymbol{\xi} - \mathrm{i}(\boldsymbol{\Omega}\bar{\mathbf{x}})^T\boldsymbol{\xi}\right) \tag{2.34}$$

and their Wigner function is given by

$$W(\mathbf{x}) = \frac{1}{(2\pi)^N\sqrt{\mathrm{Det}\mathbf{V}}} \exp\left(-(1/2)(\mathbf{x}-\bar{\mathbf{x}})^T\mathbf{V}^{-1}(\mathbf{x}-\bar{\mathbf{x}})\right). \tag{2.35}$$

One of their significant properties is that both the states and their dynamics are completely characterized by the two first moments $\bar{\mathbf{x}}$ and $\mathbf{V}$. In fact, one is able to express a given Gaussian state as

$$\hat{\rho}^G = \hat{\rho}(\bar{\mathbf{x}}, \mathbf{V}). \tag{2.36}$$

**Remarks**:

a) For a global Gaussian state

$$\hat{\rho}^G_{\mathrm{global}}(0, \mathbf{V}_{\mathrm{global}}) = \bigotimes_{i=1}^{N} \hat{\rho}_i(0, \mathbf{V}_i)$$

holds that $\mathbf{V}_{\mathrm{global}} = \bigoplus_{i=1}^N \mathbf{V}_i$.

b) The mean value $\bar{\mathbf{x}}$ can always be configured by local operations on each mode (e.g., displacements defined given in Eq. (2.44), which are not affecting the inter-mode correlations of the CM.

In order to be able to study the Gaussian states and their properties, we study here the Gaussian transformations as a way to describe mathematically how a given Gaussian state can be derived from another one. This is necessary to our later discussion so as to identify connections between Gaussian states which form specific classes.

#### 2.4.0.1 Gaussian unitary operators and symplectic transformations

For the reversible transformation of a state $\hat{\rho}$, e.g., time evolution, there are trace preserving unitary quantum operations called unitary quantum channels. A general rule for such a transformation is $\hat{\rho} \rightarrow \hat{U}\hat{\rho}\hat{U}^\dagger$, where the transformation is represented by the unitary matrix $\hat{U}$. When $\hat{U}$ transforms a Gaussian state to another Gaussian state, then is called a Gaussian channel. These unitary operators can be obtained from a second order Hamiltonian $\hat{H}$ on the field operators, where a general form can be written as follows

$$\hat{H} = \mathrm{i}(\hat{a}^\dagger \alpha + \hat{a}^\dagger \mathbf{F} \hat{a} + \hat{a}^\dagger \mathbf{G} \hat{a}^{\dagger T}) + \mathrm{H.c.}, \tag{2.37}$$

where $\alpha \in \mathbb{C}^N$, $\mathbf{F}$ and $\mathbf{G}$ are $N \times N$ complex matrices, and H.c. stands for Hermitian conjugate. Accordingly, the unitary operator will have the following form

$$\hat{U} = \exp\left(-\mathrm{i}\hat{H}/2\right). \tag{2.38}$$

On the other hand, in the Heisenberg picture, this kind of unitary corresponds to a linear unitary Bogoliubov transformation

$$\hat{\mathbf{a}} \rightarrow \hat{U}^\dagger \hat{\mathbf{a}} \hat{U} = \mathbf{A}\hat{\mathbf{a}} + \mathbf{B}\hat{\mathbf{a}}^\dagger + \alpha \tag{2.39}$$

where the $N \times N$ complex matrices $\mathbf{A}$ and $\mathbf{B}$ satisfy $\mathbf{A}\mathbf{B}^T = \mathbf{B}\mathbf{A}^T$ and $\mathbf{A}\mathbf{A}^\dagger = \mathbf{B}\mathbf{B}^\dagger + \mathbf{I}$. In terms of the quadrature operators, they are described by affine maps such

$$(\mathbf{S}, \mathbf{d}) : \hat{\mathbf{x}} \rightarrow \mathbf{S}\hat{\mathbf{x}} + d \tag{2.40}$$

where $d \in \mathcal{R}^{2N}$ and $\mathbf{S}$ is a $2N \times 2N$ real matrix. For such a transformation to preserve the commutation relations in Eq. (2.26), it is required that $\mathbf{S}\boldsymbol{\Omega}\mathbf{S}^T = \boldsymbol{\Omega}$, namely $\mathbf{S}$ to be symplectic. Finally, the action of such operations to the statistical moments are summarized by the following relations

$$\bar{\mathbf{x}} \rightarrow \mathbf{S}\bar{\mathbf{x}} + d \quad \text{and} \quad \mathbf{V} \rightarrow \mathbf{S}\mathbf{V}\mathbf{S}^T. \tag{2.41}$$

### 2.4.1 Examples of Gaussian states

In this section, we present the most common classes of Gaussian states with respect to our research and they are usually connected to specific Gaussian unitary transformations. These states are not only useful in the sense of providing theoretical intuition but also can be produced efficiently in an experimental setting using the current technology.

#### 2.4.1.1 Vacuum and thermal states

The vacuum state is the eigenstate of the annihilation operator with zero eigenvalue, i.e., $\hat{a}|0\rangle = 0$, while its CM is the identity matrix. This state is characterized by the fact that both the quadratures reach symmetrically the minimum variance according to Heisenberg's principle. This variance, intrinsic in the theory of quantum mechanics and manifested in Heisenberg's principle, can be seen from an information theoretic point of view as noise induced by nature, thus is called quantum shot noise.

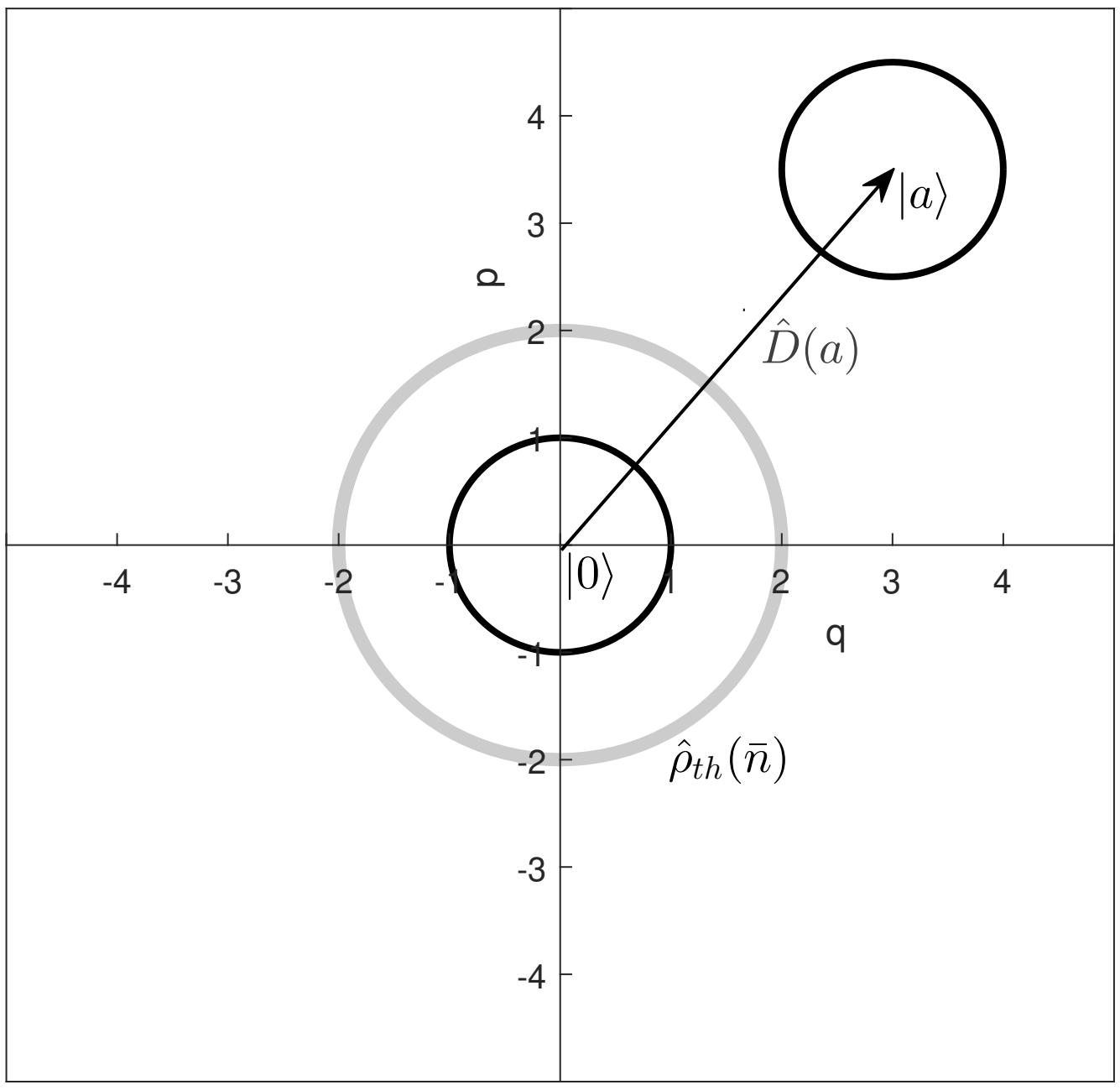

*Figure 2.1: A vacuum state $|0\rangle$ is associated with a circle of unit radius (quantum shot noise) centred in the origin (zero mean). We can obtain a coherent states $|a\rangle$ after applying a displacement operator $\hat{D}(a)$ which changes only the first moment of the vacuum state to $\bar{x} = (Re[a], Im[a])$. A thermal state is associated with a circle with variance $2\bar{n}+1$ larger than the shot noise (gray circle) centred in the origin.*

The vacuum state is a special case of Gaussian states, called thermal states and their phase space contour is depicted in Fig. 2.1. These states have the same variance for both quadratures and maximize the von Neumann entropy (see Eq.(2.77)) for a fixed amount of energy $\bar{n} = \mathrm{Tr}[\hat{\rho}\hat{a}^{\dagger}\hat{a}]$ which corresponds to the average number of photons, i.e., bosonic excitations. In particular, its representation with number-states is given by

$$\hat{\rho}(\bar{n}) = \sum_{n=0}^{+\infty} \frac{\bar{n}^n}{(\bar{n}+1)^{n+1}} |n\rangle\langle n|. \tag{2.42}$$

while its CM is

$$\mathbf{V} = (2\bar{n}+1)\mathbf{I}. \tag{2.43}$$

#### 2.4.1.2 Coherent states

In order to study the second class of states, one needs first to define the displacement operator a complex version of the Weyl operator introduced in Eq. (2.28).

**Definition 2.4.1** The displacement operator is given by

$$D(a) := \exp\left(a\hat{a}^\dagger - a^*\hat{a}\right), \tag{2.44}$$

where $a = (q + \mathrm{i}p)/2$ is the complex amplitude and transforms the quadrature operators as $\hat{\mathbf{x}} \to \hat{\mathbf{x}} + \mathbf{d}_a$, where $\mathbf{d}_a = (q, p)^T$.

The displacement operator takes its name by the fact that its action on the vacuum, illustrated in Fig. 2.1, creates a state $|a\rangle = D(a)|0\rangle$ that is an eigenstate of the annihilator operator $\hat{a}|a\rangle = a|a\rangle$. Such states are called coherent states, they have the same CM with the vacuum, while their mean value is $\bar{\mathbf{x}} = \mathbf{d}_a$. Moreover, the square of the absolute value of the amplitude a is equal with the mean number of photons of the state, i.e., $|a|^2 = \bar{n}$, and for $\bar{n} \to \infty$ the given mode is describing a classical electromagnetic field. The number state representation of a coherent state is given by

$$|a\rangle = \exp\left(-\frac{1}{2}|a|^2\right) \sum_{n=0}^{\infty} \frac{a^2}{\sqrt{(n!)}} |n\rangle. \tag{2.45}$$

On that account, for a given pair of coherent states $|a\rangle$,$|b\rangle$ there is an overlap given by

$$|\langle a|b\rangle|^2 = \exp\left(-|b-a|^2\right). \tag{2.46}$$

Thus, the class of the coherent states form an over-complete basis being non-orthogonal.

#### 2.4.1.3 Squeezed states

It is first helpful to describe the squeezed states through another unitary transformation represented by the squeezing operator.

**Definition 2.4.2** The one mode squeezing operator is defined as

$$S(r) := \exp\left(r(\hat{a}^2 - (\hat{a}^\dagger)^2)/2\right), \tag{2.47}$$

for squeezing parameter $r \in \mathbb{R}$ and transforms the quadrature operators as

$$\hat{\mathbf{x}} \to \mathbf{S}(r)\hat{\mathbf{x}}, \quad \mathbf{S}(r) := \begin{pmatrix} e^{-r} & 0 \\ 0 & e^{r} \end{pmatrix}. \tag{2.48}$$

A squeezed state is generated by applying this operator to the vacuum and its number state representation is

$$|0, r\rangle = \frac{1}{\sqrt{\cosh\, r}} \sum_{n=0}^{\infty} \frac{\sqrt{(2n)!}}{2^n n!} \tanh\, r^n |2n\rangle. \tag{2.49}$$

As depicted in Fig. 2.2, the one mode squeezing operator shrinks the quantum noise in quadrature $\hat{q}$ for $r < 0$, whereas for $r > 0$ shrinks the noise in quadrature $\hat{p}$. Furthermore, in the limit of infinite squeezing, namely $r \to \infty$, the realized states are the asymptotic quadrature eigenstates, i.e., the states in Eq. (2.20) for $q = 0$ and $p = 0$ respectively. What is more, its CM is given by $V = \mathbf{S}(r)\mathbf{S}(r)^T = \mathbf{S}(2r)$, for which one quadrature noise variance is below the quantum shot noise while the other is anti-squeezed above it. A general squeezed state is realized by applying the displacement operator to a squeezed vacuum state depicted in Fig. 2.2. The final state has the same CM while its mean value will be $\bar{x} = (q, p)$ given by the amplitude of the displacement $a = (q + \mathrm{i}p)/2$.

#### 2.4.1.4 Rotated states

One more unitary operation is needed before one is able to study a general formula for a one-mode Gaussian state. This element is called the phase rotation operator.

**Definition 2.4.3** The rotation operator, for a proper rotation with angle $\theta$, is defined as $R(\theta) = \exp\left(-\mathrm{i}\hat{a}^\dagger\hat{a}\right)$ and transforms the quadratures as

$$\hat{\mathbf{x}} \to \mathbf{R}(\theta)\hat{\mathbf{x}}, \quad \mathbf{R}(\theta) = \begin{pmatrix} \cos\theta & \sin\theta \\ -\sin\theta & \cos\theta \end{pmatrix}. \tag{2.50}$$

By using the singular value decomposition, any $2 \times 2$ matrix can be decomposed as $\mathbf{S} = \mathbf{R}(\theta)\mathbf{S}(r)\mathbf{R}(\phi)$. Therefore, any Gaussian state can be described by a mean value $\bar{\mathbf{x}} = d$ and a CM $\mathbf{V} = (2\bar{n} + 1)\mathbf{R}(\theta)\mathbf{S}(2r)\mathbf{R}(\phi)$. For $\bar{n} = 0$, this CM describes the most general one-mode pure Gaussian state corresponding to a rotated, squeezed and displaced vacuum state $|a, \theta, r\rangle = D(a)R(\theta)S(r)|0\rangle$.

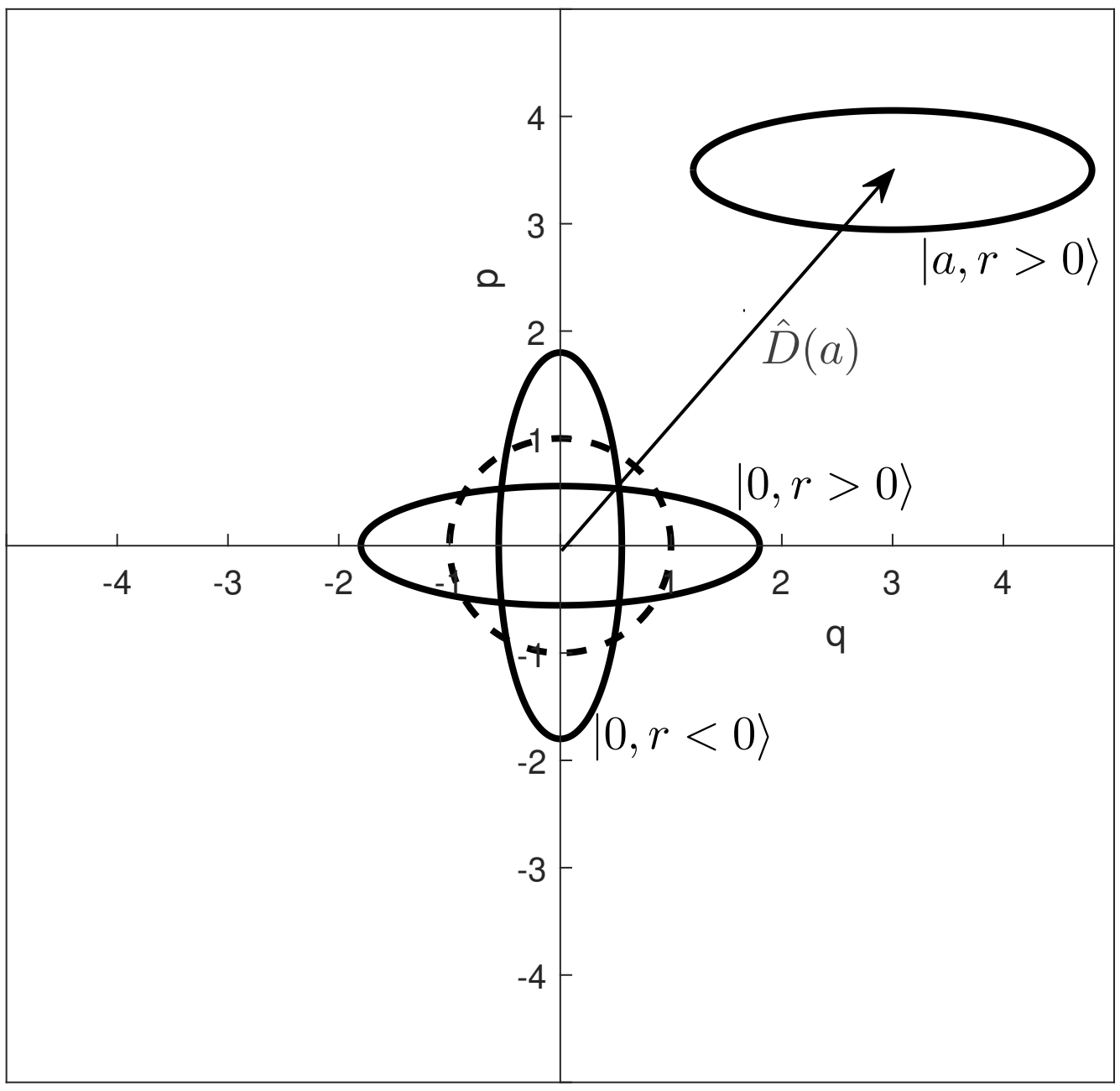

*Figure 2.2: A vertical ellipse is associated with a p-quadrature squeezed state $|0, r < 0\rangle$ and a horizontal with a q-quadrature squeezed state $|0, r > 0\rangle$ respectively. The displacement operator $D(a)$ is acting on a squeezed state affecting only its first moment witch finally goes to $\bar{\mathbf{x}} = (Re[a], Im[a])$. For comparison, the vacuum state is presented here with a circle of unit radius (dashed circle).*

#### 2.4.1.5 Two-mode squeezed vacuum states

A two-mode squeezed vacuum (TMSV) state describes a two-mode entangled sate. In order to describe the creation of a TMSV state, one has to study the following two-mode operation.

**Definition 2.4.4** For the annihilation operators $\hat{a}$ and $\hat{b}$ of two modes the two-mode squeezing operator is defined as

$$S_2(r) = \exp\left(r(\hat{a}\hat{b} - \hat{a}^\dagger\hat{b}^\dagger)/2\right) \tag{2.51}$$

for the two-mode squeezing parameter $r$ while the quadrature operator $\hat{\mathbf{x}} := (\hat{q}_a, \hat{p}_a, \hat{q}_b, \hat{p}_b)$ is transformed as

$$\hat{\mathbf{x}} \to \mathbf{S}_2(r)\hat{\mathbf{x}}, \quad \mathbf{S}_2(r) := \begin{pmatrix} \text{cosh}r\mathbf{I} & \text{sinh}r\mathbf{Z} \\ \text{sinh}r\mathbf{Z} & \text{cosh}r\mathbf{I} \end{pmatrix} \tag{2.52}$$

and $\mathbf{Z} = diag(1, -1)$.

By applying $\mathbf{S}_2(r)$ to a pair of modes $A$ and $B$ in vacuum states,

$$|\nu\rangle_{\text{TMSV}} = \hat{\mathbf{S}}_2(r)\left(|0\rangle_A \otimes |0\rangle_B\right)$$

one obtains

$$\hat{\rho}_{\text{TMSV}} = |\nu\rangle\langle\nu|_{\text{TMSV}}, \quad |\nu\rangle_{\text{TMSV}} = \sqrt{1-\lambda^2}\sum_{n=0}^{\infty}(-\lambda)^n|n\rangle_A|n\rangle_B, \tag{2.53}$$

with $\lambda = \tanh r \in [0,1]$ and $\nu = \cosh 2r$. The CM of this state is

$$\mathbf{V}_{\text{TMSV}}(\nu) := \begin{pmatrix} \nu\mathbf{I} & \sqrt{\nu^2-1}\mathbf{Z} \\ \sqrt{\nu^2-1}\mathbf{Z} & \nu\mathbf{I} \end{pmatrix}. \tag{2.54}$$

An TMSV state is characterized by its maximal correlations between the two modes. These correlations, which are increasing with $\nu$ represent a typical form of entanglement. For $\nu = 1$, one obtains $\mathbf{V}_{\text{TMSV}}(1) = \mathbf{I} \oplus \mathbf{I}$, which corresponds to the tensor product of two vacuum states $\hat{\rho}_{\text{TMSV}} = |0\rangle\langle 0|_A \otimes |0\rangle\langle 0|_A$. Therefore, the TMSV state is separable only in this case, whereas for $\nu \to \infty$ the state becomes an ideal Einstein-Podolsky-Rosen (EPR) state with maximal correlations, namely $\hat{q}_A \to \hat{q}_B$ and $\hat{p}_A \to -\hat{p}_B$. In other words, the positions of the two modes become correlated and the momenta become anti-correlated. An important observation is that the reduced states of TMSV states, i.e.,

$$\hat{\rho}_A = \text{Tr}_B\left(\hat{\rho}_{\text{TMSV}}\right) \quad \text{and} \quad \hat{\rho}_B = \text{Tr}_A\left(\hat{\rho}_{\text{TMSV}}\right)$$

are two identical thermal states $\hat{\rho}_A = \hat{\rho}_B = \hat{\rho}_{\text{th}}(\nu)$ with CM equal to $\nu\mathbf{I}$, where $\nu = 2\bar{n}+1$ quantifies the mean thermal photon number in each mode.

### 2.4.2 Symplectic decomposition

An important property of the CM of a Gaussian state is their connection with special diagonal matrices. According to Williamson's theorem (see Ref. [8]), there is a symplectic transformation for every positive-definite real matrix of even dimension that can transform the given matrix to its diagonal form. By applying this to an $N$-mode CM $\mathbf{V}$, one obtains that

$$\mathbf{V} = \mathbf{S}\mathbf{V}^{\oplus}\mathbf{S}^T, \quad \mathbf{V}^{\oplus} := \bigoplus_{k=1}^{N}\nu_k\mathbf{I} \tag{2.55}$$

The diagonal matrix $\mathbf{V}^{\oplus}$ is called the Williamson form of $\mathbf{V}$ and the quantities $\nu_k$ are its symplectic eigenvalues.

**Remarks**:

a) The above property provides an equivalent way to express the uncertainty principle of Eq. (2.33) as

$$\mathbf{V} \geq 0, \quad \mathbf{V}^{\oplus} \geq \mathbf{I}. \tag{2.56}$$

This can be stated differently by saying that a CM must be positive definite and its symplectic eigenvalues must satisfy $\nu_k \geq 1$.

b) According to Eq. (2.43), $\mathbf{V}^{\oplus}$ corresponds to the CM of a product state of $N$ thermal modes where their variances are given by the symplectic eigen-spectrum of $\mathbf{V}$. For example, the Williamson's form

$$\mathbf{V}^{\oplus} = \bigoplus_{i=1}^{N} \nu_i \mathbf{I} \tag{2.57}$$

of an $N$-mode CM is associated with the Gaussian state

$$\hat{\rho}(0, \mathbf{V}^{\oplus}) = \bigotimes_{i=1}^{N} \hat{\rho}_{\text{th}}(\nu_i). \tag{2.58}$$

In that event, one notices that an arbitrary Gaussian state can be decomposed, up to a unitary $\hat{U}_{\mathbf{S},\mathbf{d}} = \hat{D}(d)\hat{U}_S$, into a tensor product of thermal states as

$$\hat{\rho}(\mathbf{d}, \mathbf{V}) = \hat{U}_{\mathbf{S},\mathbf{d}} \hat{\rho}(0, \mathbf{V}^{\oplus}) \hat{U}^{\dagger}_{\mathbf{S},\mathbf{d}}. \tag{2.59}$$

## 2.5 Measurement and evolution of Gaussian states

In this section, we are concerned with basically two concepts which are the detection and the evolution of bosonic states. Firstly, we present the concept of evolution of a Gaussian state, e.g., through a communication channel, described by the action of Gaussian channels. Afterwards we present the homodyne and heterodyne measurement of a bosonic mode, since they are the main detections applied in our study dictated by the fact that are commonly used in the current technological implementations.

### 2.5.1 Gaussian channels

At this point and in order to facilitate the following discussion we present the notion of purification, where a given an arbitrary state $\hat{\rho}_A$ can be mapped to a pure state of a larger system $\hat{\rho}_{RA}$.

**Definition 2.5.1** A purification of a density operator with spectral decomposition $\hat{\rho}_A = \sum_x p_X(x)\ |x\rangle_A\langle x| \in \mathcal{D}(\mathcal{H}_A)$ is a pure bipartite state $|\psi\rangle_{RA}$ on a reference system $A$, with the property that the reduced state on system $A$ is equal to $\hat{\rho}_A = \text{Tr}_R |\psi\rangle_{RA}\langle\psi|$.

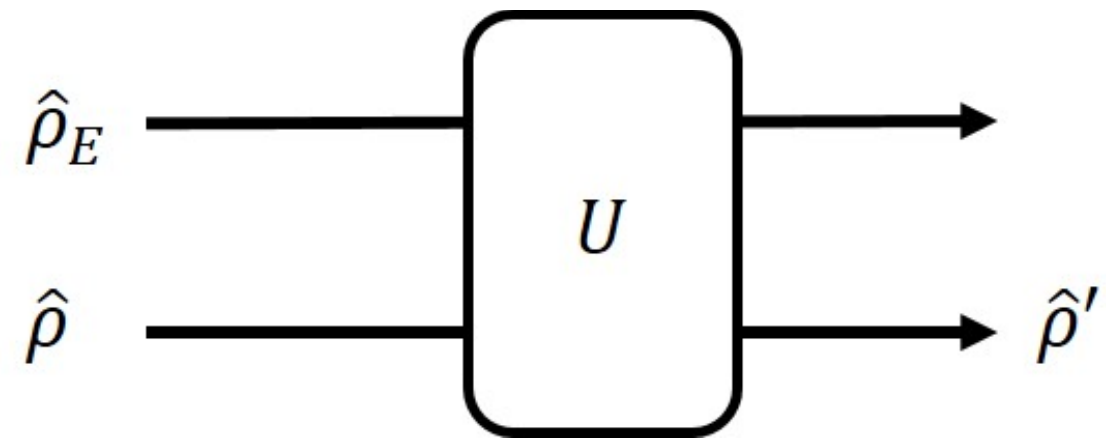

*Figure 2.3: This is the dilation of the channel $\mathcal{E}$ for an input state $\hat{\rho}$ and environment state $\hat{\rho}_E$ under the global unitary evolution $U$ with output $\hat{\rho}' = \mathcal{E}(\hat{\rho})$. If $\hat{\rho}_E = |\Phi\rangle_E\langle\Phi|$ is a pure state then the dilation is called Stinespring dilation.*

In that regard, the evolution denoted by a channel $\mathcal{E}$ of an arbitrary state can be described as a unitary evolution of a larger pure system as illustrated in Fig. 2.3. This is called the dilation of the channel, named after the expansion of the Hilbert space used to describe the system of interest as part of a larger pure system. More specifically, a $N$-mode Gaussian channel for an arbitrary state $\hat{\rho} \in \mathcal{D}(\mathcal{H}^{\otimes})$ is a completely positive and trace preserving linear map $\mathcal{E}_G : \hat{\rho} \to \mathcal{E}_G(\hat{\rho}) \in \mathcal{D}(\mathcal{H}^{\otimes})$. A way to represent such a map is by a Gaussian unitary operation acting on a larger system part of which is the input state $\hat{\rho}$. The rest of the system which is called the environment $E$ is found in a multi-mode Gaussian pure state $|\Phi\rangle_E$ and without loss of generality in a multi-mode vacuum $|\Phi\rangle_E$=$|0\rangle_E$. Then the output of the channel is given after tracing out the environment, namely

$$\mathcal{E}(\hat{\rho}) = \mathrm{Tr}_E \left( U(\hat{\rho} \times |\Phi\rangle_E\langle\Phi|)U^\dagger \right). \tag{2.60}$$

With regard to the first and second statistical moments, the action of the channel can be expressed as

$$\bar{x} \to \mathbf{T}\bar{x} + d, \quad \mathbf{V} \to \mathbf{TVT} + \mathbf{N}, \tag{2.61}$$

where $d \in \mathbb{R}^{2N}$ is a displacement vector, while $\mathbf{T}$ and $\mathbf{N} = \mathbf{N}^T$ are $2N \times 2N$ real matrices satisfying the complete positivity condition

$$\mathbf{N} + \mathrm{i}\boldsymbol{\Omega} - \mathrm{i}\mathbf{T}\boldsymbol{\Omega}\mathbf{T}^T \geq 0. \tag{2.62}$$

The correspondence to the Gaussian unitary operators is obtained by setting $\mathbf{T} := \mathbf{S}$ to be symplectic and $\mathbf{N} = 0$. For a one-mode channel, setting $N = 1$, one derive the following relations

$$\mathbf{N} = \mathbf{N}^T \geq 0, \quad \det\mathbf{N} \geq (\det\mathbf{T} - 1)^2. \tag{2.63}$$

According to the following definition, a Gaussian channel can be decomposed to a simpler form called the canonical form.

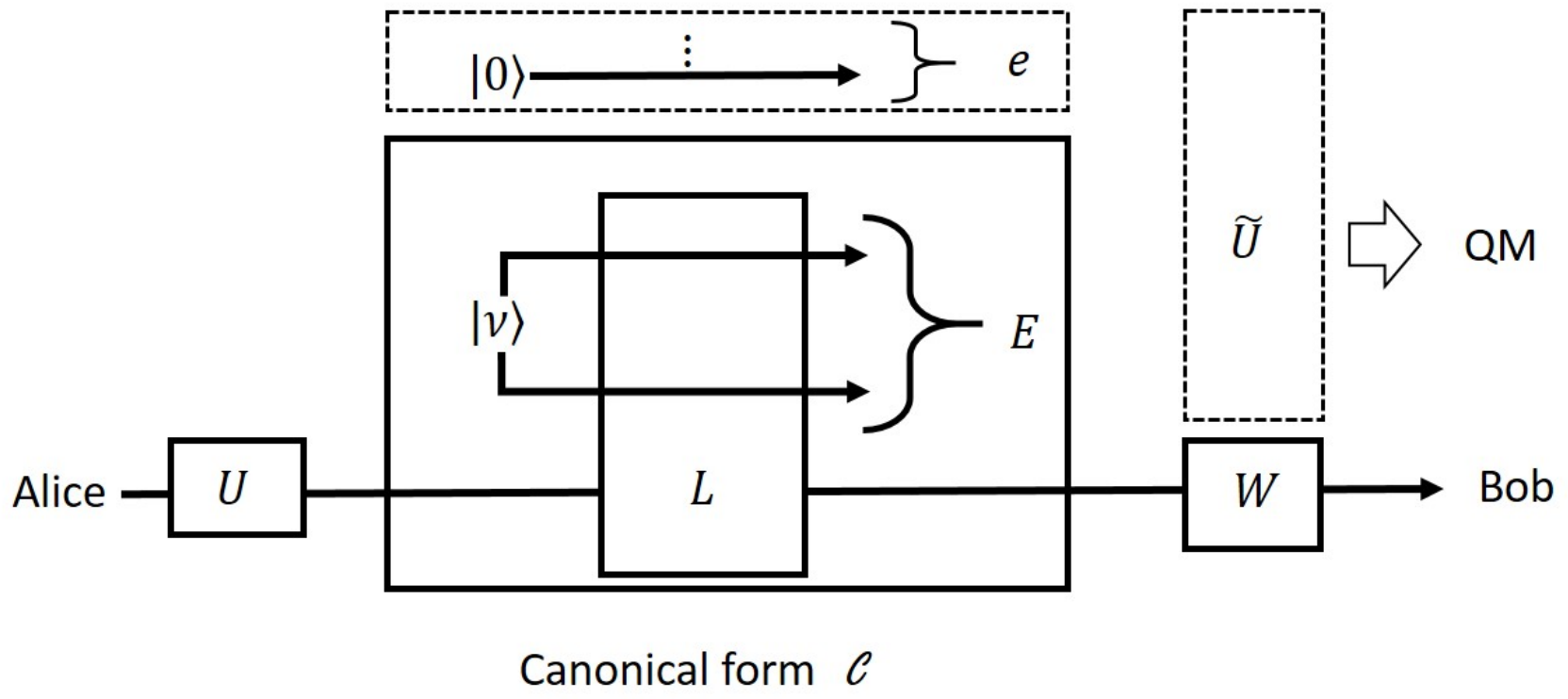

*Figure 2.4: Here is the Stinespring dilation of a channel unique up to partial isometries $\tilde{U}$. Due to this fact, the environment can be expressed as an ensemble of modes $\{E, e\}$, where the modes in $e$ are found in vacuum states. Furthermore, this channel is expressed through its canonical form $\mathcal{C}$ with $U$ and $W$ the input and output modes respectively. The canonical form is further reduced to a three-mode Gaussian unitary operator $L$ acting on the input mode and to a two-mode squeezed vacuum state $|\nu\rangle\langle\nu|$.*

**Definition 2.5.2** A Gaussian channel $\mathcal{G}(\mathbf{d}, \mathbf{T}, \mathbf{N})$ can always be decomposed as

$$\mathcal{G}(\hat{\rho}) = W[\mathcal{C}(U\hat{\rho}U^{\dagger})]W^{\dagger} \tag{2.64}$$

where $W$, $U$ are the output-input Gaussian unitary operators. Then $\mathcal{C} = \mathcal{C}(\mathbf{d}_c; \mathbf{T}_c; \mathbf{N}_c)$ is called the canonical form and is a Gaussian channel with $\mathbf{d}_c = 0$ and $\mathbf{T}_c$, $\mathbf{N}_c$ diagonal.

The explicit expression of the $\mathbf{T}_c$ and $\mathbf{N}_c$ depend on three quantities that are preserved by the action of the Gaussian unitary operators. These are the generalized transmittance $\tau := \det\mathbf{T}(-\infty < \tau < \infty)$, the rank $r := [\mathrm{rank}(\mathbf{T})\mathrm{rank}(\mathbf{N})]/2(\text{for } r = 0, 1, 2)$ and the temperature $\bar{n}$ a non-negative number defined by

$$\bar{n} := \begin{cases} (\det N)^{1/2} & \text{for } \tau = 1 \\ \frac{(\det N)^{1/2}}{|1-\tau|} - \frac{1}{2} & \text{for } \tau \neq 1. \end{cases} \tag{2.65}$$

#### 2.5.1.1 Thermal-loss channel

The canonical form of a given one-mode channel can be further simplified through its Stinespring dilation. In particular, a generic $\mathcal{C}$ is transformed to a three-mode Gaussian unitary operator $U_L$ corresponding to the symplectic transformation $\mathbf{L} = \mathbf{L}(\tau, r)$, which mixes the input mode with an TMSV state $|\nu\rangle\langle\nu|_{\mathrm{TMSV}}$ with thermal variance $\nu = 2\bar{n} + 1$. So for any canonical form of an one-mode Gaussian channel, one can assign the pair

$\{L(\tau, r), |\nu\rangle\langle\nu|_{\text{TMSV}}\}$ as illustrated in Fig. 2.4. By setting $r = 2$ and $0 < \tau < 1$, one can study a special case of one-mode Gaussian channel describing the evolution of a Gaussian state across a communication channel, such as an optical fiber, between two parties. This channel is called the thermal-loss channel and is defined as $\mathcal{L}(\tau, r) := \mathcal{C}(0 < \tau < 1, 2, \bar{n})$. It simulates the attenuation of the input signals and their combination with thermal noise according to $\hat{x} \to \sqrt{\tau}\hat{x} + \sqrt{1-\tau}\hat{x}_{\text{th}}$, where $\hat{x}_{\text{th}}$ is in a thermal state with $\bar{n}$ mean number of photons, which is the reduced state of $\hat{\sigma}_E = |\nu|\rangle\langle\nu|$. For expressing this channel, one can define a unitary Gaussian operation describing the interaction between the two modes.

**Definition 2.5.3** Let us assume two modes with ladder operators $\hat{a}$ and $\hat{b}$ and $\tau = \cos^2\theta$ is the transitivity of the beam splitter for $\theta \in [0, \pi/2]$. For $\tau = 1/2$, the beam splitter is called balanced. Then the transformation is defined as

$$B(\theta) = \exp\left(\theta(\hat{a}^\dagger\hat{b} - \hat{a}\hat{b}^\dagger)\right) \tag{2.66}$$

and the quadrature operators $\hat{\mathbf{x}} := (\hat{q}_a, \hat{p}_a, \hat{q}_b, \hat{p}_b)$ are transformed as

$$\hat{\mathbf{x}} \to \mathbf{B}(\tau)\hat{\mathbf{x}}, \quad \mathbf{B}(\tau) := \begin{pmatrix} \sqrt{\tau}\mathbf{I} & \sqrt{1-\tau}\mathbf{I} \\ -\sqrt{1-\tau}\mathbf{I} & \sqrt{\tau}\mathbf{I} \end{pmatrix}. \tag{2.67}$$

By virtue of the beam splitter operation, one obtains the following relation

$$\mathcal{L}(\hat{\rho}) = \text{Tr}_E\left(U_L(\hat{\rho} \otimes \hat{\sigma}_E)U_L^\dagger\right), \tag{2.68}$$

where $U_L = B(\theta) \otimes \mathbf{I}$.

A communications scenario can be described by two parties wanting to communicate secretly against a third party who tampers with the channel, namely the eavesdropper. In that event, an entangling cloner can simulate the third party's strategy which consist of two main stages. Firstly, the eavesdropper creates two copies of the signal from the sender. Then resends one copy to the receiver using a channel with the same transmittance $\tau$, while keeps the other output of the beam splitter for a measurement. This can be modeled using the previous description (see also Ref. [61]). Eve has two modes in a TMSV state, which is used as a source of thermal noise attributed only to Eve and known by her. One of these modes (a thermal state) interacts with Alice's input mode via a beam splitter operation with transmissivity $g = 1 - \tau$ then it travels to Bob. Alice's mode is kept to be measured along with the remaining mode which is used as reference for the injected noise. Eve can now estimate Bob's arriving mode. In other words, by this description Eve purifies the channel giving her the advantage to acquire all the information lost by Alice and Bob's system during the channel propagation.

### 2.5.2 Measuring Gaussian states

#### 2.5.2.1 Homodyne detection

This detection is the most common Gaussian detection measuring the quadrature $\hat{q}$ or $\hat{p}$ of a given mode. The measurement operators describing this measurement are the projectors of the quadrature basis, namely the infinite squeezed states $|q\rangle\langle q|$ or $|p\rangle\langle p|$ respectively. The outcome of such a measurement is given by a probability distribution

$$P(q) = \int W(q,p)dp \quad or \quad P(p) = \int W(q,p)dq, \tag{2.69}$$

which is the marginal probability of the Wigner function of the state over the conjugate quadrature.

For a multi-mode state, the partial homodyne measurement, e.g., measuring only one quadrature of one mode, can be realized by integrating over both quadratures of the non-measured modes. In fact, the CM

$$\mathbf{V} = \begin{pmatrix} \mathbf{A} & \mathbf{C} \\ \mathbf{C}^T & \mathbf{B} \end{pmatrix}. \tag{2.70}$$

of a $N+1$-mode state comprised of two subsystems, $\mathbf{A}$ with $N$ modes and $\mathbf{B}$ with 1 mode, is transformed as

$$\tilde{\mathbf{V}} = \mathbf{A} - \mathbf{C}(\mathbf{\Pi}\mathbf{B}\mathbf{\Pi})^{-1}\mathbf{C}^T \tag{2.71}$$

where $\mathbf{A}$ is the CM of subsystem $A$ and $\mathbf{B}$ is the CM of subsystem $B$. Matrix $\mathbf{C}$ describes the correlations between system $A$ and $B$. Moreover, $\mathbf{\Pi}$ is equal to $\mathbf{\Pi}_q = \mathrm{diag}(1,0)$ or $\mathbf{\Pi}_p = \mathrm{diag}(0,1)$ for measuring $\hat{q}$ or $\hat{p}$ respectively. The power in $-1$ denotes here the pseudo-inverse of the matrix included in the parenthesis.

#### 2.5.2.2 Heterodyne detection

Another basic measurement is the heterodyne detection, which is the measurement of both quadratures of a mode. In theory, this measurements corresponds to a projection on coherent states, namely $E(a) := \pi^{-1/2}|a\rangle a$. This is accomplished by combining the measured mode with an ancillary vacuum mode through a balanced beam splitter and then homodyne each of the outputs with respect to different quadratures. Accordingly, the output CM after a partial heterodyne detection is

$$\tilde{\mathbf{V}} = \mathbf{A} - \mathbf{C}(\mathbf{B} + \mathbf{I})^{-1}\mathbf{C}^T \tag{2.72}$$

**Remark**:A generalized technique for modelling any partial Gaussian measurement is implemented by appending ancillary modes to the original system, applying a Gaussian unitary, part of the modes are being homodyned, another part is discarded. The remaining part is the output system.

#### 2.5.2.3 Bell detection

A Bell detection is a detection applied to two modes. Here two modes are mixed in a balanced beam splitter and then each of its outputs is detected with respect to a conjugate homodyne detection. This measurement is applied by the relay of Chap. 9.

## 2.6 Entropy and symplectic decomposition

In our later discussion, we are going to use quantum continuous-variable systems in order to encode information. The notion of information describes the knowledge that the receiver of a message acquires once he decodes it or the ignorance before doing so. Messages which have high probability to occur, contain almost no information. On the other hand, messages that are rare to be received maximize the amount of information.

In order to express more rigorously the previous discussion, let us consider a source emitting signals over an alphabet with a certain probability. A stochastic variable $X = \{p(x), x\}$ is associated to that source and takes its values from a set $\mathcal{X}$, i.e., the alphabet, with a given probability distribution $p(x)$. Accordingly, the the Shannon entropy of the source is expressed as

$$H(X) = -\sum_{x\in\mathcal{X}} p(x)\log_2 p(x). \tag{2.73}$$

If the variable $x$ is continuous, the sum corresponds to an integral [1] over the probability density derived from the distribution $p(x)$. Subsequently, one can define the conditional entropy [50] of two stochastic variables $X$ and $Y$ between two stochastic variables as follows.

**Definition 2.6.1** For two variables $(X, Y) \sim p(x, y)$, the conditional entropy $H(Y|X)$ is defined as

$$H(X|Y) = \sum_{x\in\mathcal{X}} p(x)H(Y|X = x) \tag{2.74}$$

[1]The corresponding quantity is called differential entropy. Although it has differences from the entropy defined in Eq. (2.73), they are not affecting the definition of mutual information presented later.

where $H(Y|X=x) = -\sum_{y\in\mathcal{Y}} p(y|x)\log_2 p(y|x)$ and $p(y|x) = \frac{p(x,y)}{p(x)}$.

This allows us to define another important quantity that accounts for the information known for a variable $Y$ given that we know variable $X$.

**Definition 2.6.2** The mutual information between two stochastic variables $Y$ and $X$ is defined as

$$I(Y:X) = H(Y) - H(Y|X). \tag{2.75}$$

**Remark**: For Gaussian stochastic variables (see Theorem 4.3.1 in Ref [51]), this turns to be equal to

$$I(Y:X) = \frac{1}{2}\log_2\frac{V_Y}{V_{Y|X}}, \tag{2.76}$$

where $V_Y$ and $V_{Y|X}$ are the variances of the probability distribution $Y$ and conditional probability distribution of $Y|X$.

On the other hand, due to Heisenberg's principle, quantum mechanics describe states that have intrinsic uncertainty that can be quantified by the von Neumann entropy [1].

**Definition 2.6.3** The von Neumann entropy of a quantum state is defined as

$$S(\hat{\rho}) = -\mathrm{Tr}\left(\hat{\rho}\log_2\hat{\rho}\right) = -\sum_x \lambda_x \log\lambda_x \tag{2.77}$$

for $\lambda_x$ being the eigenvalues of $\hat{\rho}$.

**Remarks**:

a) A thermal state $\hat{\rho}_{\mathrm{th}}(\bar{n})$ is defined as the state that maximizes the von Neumann entropy for a given mean number of energy excitations, namely photons.

b) According to the Schmidt decomposition [1], the local spectra of the two reduced density matrices $\hat{\rho}_A$ and $\hat{\rho}_B$ of a global pure state $\rho_{AB}$ are equal, namely $\{\lambda_x^A\} = \{\lambda_x^B\}$ resulting in $S(\hat{\rho}_A) = S(\hat{\rho}_B)$.

c) A pure state has zero von Neumann entropy

At this point, we will focus on the entropic properties of Gaussian states, which will be the main carrier of signals in our following discussion about secret key distribution. In particular, their entropic analysis can be calculated effectively due to the fact that their description is complete with only their first two moments. In fact, it turns out that the von

Neumann entropy of a Gaussian state can be expressed by the corresponding symplectic eigen-spectrum of its CM

$$S(\hat{\rho}) = \sum_{k=1}^{N} h(\nu_k), \tag{2.78}$$

where the $\nu_k$ have been defined in Eq. (2.55) and

$$h(x) := \left(\frac{x+1}{2}\right)\log_2\left(\frac{x+1}{2}\right) - \left(\frac{x-1}{2}\right)\log_2\left(\frac{x-1}{2}\right). \tag{2.79}$$

The von Neumann entropy is invariant under unitary evolution because is dependent only on the eigenvalues of a given state (see Eq. (2.77)). Thus the entropy of a state $\hat{\rho}(\mathbf{d}, \mathbf{V})$ is the same with the entropy of the corresponding state $\hat{\rho}(0, \mathbf{V}^{\oplus})$. According to Eq. (2.57), the entropy of the latter is given as the sum of each of the thermal states in the tensor product [1].

**Remark**:According to the Stinespring dilation of a Gaussian channel, e.g. the thermal loss channel, the output state of its environment is given by the the modes $E$ that interacted with a Gaussian travelling state plus the ancillary modes $e$, which without loss of generality can be assumed as a tensor product of pure vacuum states, altogether evolved by the unitary operator $\tilde{U}$ (see Fig. 2.4). From the discussion above, we notice that the von Neumann entropy is not dependent on $\tilde{U}$ and also that the systems in $e$ are not giving any contribution to it. Thus the only relevant modes for the calculation of the von Neuman entropy of the environment are the interacting ones.

From equation Eq. (2.42) and the definition of the von Neuman entropy we can calculate the von Neumann entropy $S_{\text{th}}$ of a thermal state with $\bar{n}$ mean photon number

$$S_{\text{th}} = -\sum_{n=0}^{\infty} \frac{\bar{n}^n}{(\bar{n}+1)^{n+1}} \log_2 \frac{\bar{n}^n}{(\bar{n}+1)^{n+1}} \tag{2.80}$$

where by using the series $\sum_{n=0}^{\infty} r^n = \frac{1}{1-r}$ and $\sum_{n=0}^{\infty} n r^n = \frac{r}{(r-1)^2}$ for $|r| < 1$ we obtain Eq. (2.79) for $x = 2\bar{n} + 1$. The later function, for large $x \gg 1$ is reduced to

$$h_{\infty}(x) = \log_2(\frac{\mathrm{e}}{2}x) + O(\frac{1}{x}). \tag{2.81}$$

Based on the von Neumann entropy, we define another quantity that describes the maximum (classical) accessible information from a source that emits quantum states as signals parametrized by a stochastic variable $a$, i.e., a quantum ensemble $\mathcal{A} = \{\hat{\rho}_a, p_a\}$ with a mean state $\hat{\rho}_A = \sum_a p_a \hat{\rho}_a$.

**Definition 2.6.4** For a quantum source $A$ the Holevo information is defined as

$$\chi(A:a) = S(\hat{\rho}_A) - \sum_a p_a S(\hat{\rho}_a). \tag{2.82}$$

**Remarks**: a) This quantity can be interpreted as the amount of information that someone has over the variable $a$ given that he has access to system $A$. b) Let us assume now that the the state $\hat{\rho}_a$ is a Gaussian state and that its CM is independent on $a$, for example $a$ characterizes the displacement of the state in the phase space. Since its von Neumann entropy is dependent only on the CM the sum in the above equation will reduce to $S(\hat{\rho}_a)$ the von Neumann entropy of only one of the states of the ensemble. c) The discussion in the remark after Eq. (2.78) holds also for the Holevo information inherited by the von Neumann entropy.

## 2.7 Conclusion

In this chapter, we focused on the electromagnetic field and on its quantum states. We also discussed about their evolution. These states accept a CV description when we represent them in the phase space. In particular, the Gaussian members of the bosonic quantum states are good candidates for signal sources in quantum technological implementation, such as QKD, as we will see in the next chapter. This advantage stems from the fact that they are easy to produce in a lab, demand the same telecommunications infrastructure as the classical electromagnetic field and at the same time acquire the key quantum properties for the aforementioned applications. In the next chapter, we will explore this key quantum properties and present quantum key distribution protocols using continuous-variables states. These protocols are basic in the sense that they belong to the first CV-QKD protocols showing that quantum key distribution can be implemented in this framework. Moreover, other protocols discussed in this thesis can be seen as variations of them.

# Chapter 3

# Quantum key distribution with continuous-variables

## 3.1 Introduction

Quantum cryptography utilizes the properties of quantum states, i.e., generally non-orthogonal ones, in order to secure communication between two parties, named conventionally Alice and Bob. In particular, Alice and Bob create a shared random secret key that can use along with a symmetric cryptographic protocol such as the one-time pad. To do so, Alice encodes her random variable in quantum states and sends them to Bob through a communication (quantum) channel (see for example Eq. (2.60)). By measuring these signals (decoding), Bob receives a random variable that is correlated to Alice. Then they apply classical post processing procedures to their variables such as parameter estimation, error correction and privacy amplification procedures [2, 3].

The security of such a quantum information protocol for key distribution is based on the fact that quantum states in general cannot be perfectly copied, i.e., additional noise will be added during the procedure of cloning. This is due to a manifestation of the Heisenberg principle called the no-cloning theorem [52]. Therefore, an attempt for interception of the signals (quantum states) from someone that has access to the communication channel, called traditionally Eve, is leading at the same time to the interception's detection by the parties due to the injected noise. In other words, a copy and resend strategy by the eavesdropper can be tracked and quantified by the honest parties.

In fact, by channel parameter estimation, the parties evaluate the correlations between their data strings and estimate the amount of information Eve has on their signals. Then by applying error correction, they end up with common data strings. Both of these steps may need sacrifice of part of the data string (see also Chap. 6). Finally, by privacy amplification the parties reduce their data string by an amount indicated by the parameter estimation step so as Eve's knowledge on the remaining data string has been eliminated to an insignificant amount.

The steps of classical post processing require a classical communication channel between the parties apart from the quantum channel used for transferring the quantum signals mentioned before. Although this channel is not private it should be authenticated meaning that the two parties should acquire a pre-shared key in order to be able to recognize each other. This fact makes actually the QKD a key expansion task rather than a key creation one.

In a CV setting the quantum states that are used are bosonic states, e.g., coherent states, which form an non-orthogonal basis (see Eq. (2.46)) and can be efficiently produced in an experimental configuration and sent through optical fibre communication channels using the current technology. More specifically, if these protocols use a Gaussian random variable to encode on the Gaussian states then are called fully Gaussian protocols. They were firstly introduced in Ref. [9–11]. In the following, we present such one-way protocols. In this thesis we assume that these protocols are the central elements of our study in the sense that protocols we present later can be considered as their variations. In the following sections, we use these fully Gaussian protocols as examples for introducing CV-QKD.

## 3.2 Gaussian modulation of coherent states

Here we describe the one-way protocols using Gaussian modulation of coherent states in the prepare and measure representation. These protocols can be described as one protocol with different versions as explained in the following discussion.

Alice prepares coherent states $|a\rangle\langle a|$ with amplitude $a = Q_M + \mathrm{i}P_M$ by displacing vacuum states according to a random variable $\mathbf{X}_M = (Q_M, P_M)$ following the normal distribution with variance $\mathbf{V}_M = (\mathrm{Var}(Q_M), \mathrm{Var}(P_M))$. We assume here that both quadratures are modulated symmetrically, i.e, $\mathrm{Var}(Q_M) = \mathrm{Var}(P_M) =: V_M$. As a result, Alice's input mode $A$ is described by quadrature operators

$$\hat{Q}_A = Q_M + \hat{Q}_0 \quad \text{and} \quad \hat{P}_A = P_M + \hat{P}_0, \tag{3.1}$$

where $\hat{Q}_0 = \hat{a} + \hat{a}^\dagger$ and $\hat{P}_0 = \mathrm{i}(\hat{a}^\dagger - \hat{a})$ with $\hat{a}$ and $\hat{a}^\dagger$ being annihilation and creation operators of mode $A$, which initially was in the vacuum state and had variance $V_0 = 1$ corresponding to the quantum fluctuations. This mode propagates through the channel controlled by Eve and arrives to Bob as the output mode $B$. Bob chooses randomly between the quadrature $p$ or $q$ and applies a homodyne or heterodyne detection on mode $B$. If Bob uses the homodyne detection, he ends up with a string of values of $Q_B$ and $P_B$. By informing Alice for his choice of quadrature, Alice keeps only the relevant quadrature of the pairs $(Q_M, P_M)$ resulting with a string of values $Q_M$ and $P_M$ with the same quadrature sequence as Bob's string. This defines the version of the protocol with switching [9] and without loss of generality, from now on, we will assume that Bob always chooses to make a measurement on only one quadrature, e.g., quadrature $q$. Another option for Bob is to apply a heterodyne detection which defines the no-switching version of the protocol [11]. Here Bob keeps both the pairs $(Q_B, P_B)$, which along with Alice's pairs $(Q_M, P_M)$ constitute of the two data stings from which they are going to extract the secret key.

## 3.3 Security analysis

The security of a protocol is usually described as a state discrimination problem. In this case Alice, Bob and Eve share a state that is compared with an ideal secure state. This is quantified by the distance of the two states which should be less than a very small parameter $\epsilon$. The ideal state that is shared between the parties and Eve should be

$$\hat{\rho}_{\text{id}} = \hat{\rho}_{AB} \otimes \hat{\rho}_E \tag{3.2}$$

implying that there are no correlations between the system of Alice and Bob $AB$ and the eavesdropper $E$. Then we have that

$$D(\hat{\rho}_{ABE}, \hat{\rho}_{\text{id}}) \leq \epsilon, \tag{3.3}$$

where $D$ is the trace distance [53]. In other words, the previous relation states that the state $\hat{\rho}_{ABE}$ cannot be distinguished from $\hat{\rho}_{\text{id}}$ with a probability $1 - \epsilon$, in the sense, that if $\epsilon \to 0$ then the two states are identical.

**Remark**: A security analysis under a composable framework takes into account different tasks of a protocol. For the $i$-th task there is a target state that is indistinguishable from a corresponding ideal state with a probability $\epsilon_i$. Then the security parameter

$$\epsilon = \sum_i \epsilon_i$$

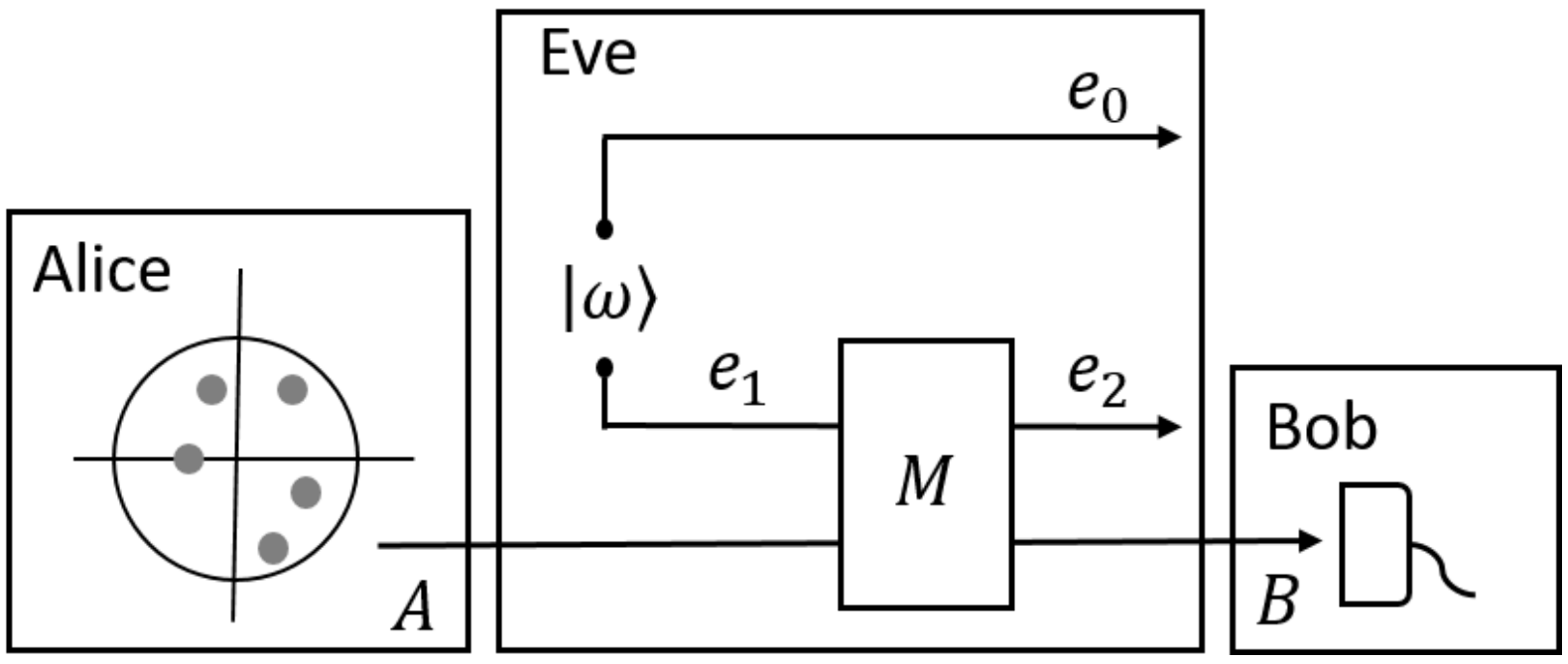

*Figure 3.1: Schematic configuration of the protocol: Alice prepares coherent states (gray disks) in mode A with modulated amplitude by a Gaussian distribution with variance $V_M = \mu - 1$ and $\mu$ equal with the radius of the black circle (phase space representqation of Alice's average state). Then mode A is send through the channel found under the control of Eve to Bob. For most of the canonical forms the channel can be represented by Eve possesing two modes $e_0$ and $e_1$ found in a TMSV state with variance $\omega$ and a symplectic transformation M acting only on mode A and $e_1$. Then the output mode B is measured by Bob either with a homodyne measurement applyed randomly on its q or p quadrature or else with a heterodyne measurement.*

is given as the sum of the parameters $\epsilon_i$. A detailed discussion about composable security is found in Ref. [16]. In Chap. 9, we will see how this concept is applied to the MDI protocol of Chap. 5.

In order to assess the performance of a given protocol in terms of security, one can introduce the secret key rate in the asymptotic limit, i.e., for infinite number of exchanged signals $N \gg 1$. The following formula is based on the privacy amplification procedure (leftover hash lemma [3]). According to this, secret key bits can be still extracted from a random variable with large probability although it is verified that a given amount has already been disclosed to an eavesdropper. Later it was generalized in the quantum regime by Ref. [13] (see also Ref. [54] for the case of CV) giving

$$R_\infty = \max\{0, I_{AB} - I_E\}, \tag{3.4}$$

where $I_{AB}$ stands for the mutual information (see Eq.(2.75)) between Alice's and Bob's classical variables, while $I_E$ stands for the accessible information of Eve on Alice's or Bob's classical variable depending on the direction of the reconciliation process. The secret key rate is manifested as a lower bound because usually Eve's accessible information is assumed to be as large as the laws of nature allow realizing the worst case scenario for Alice and Bob. In order to calculate the secret key rate, we need some assumptions for the eavesdropping

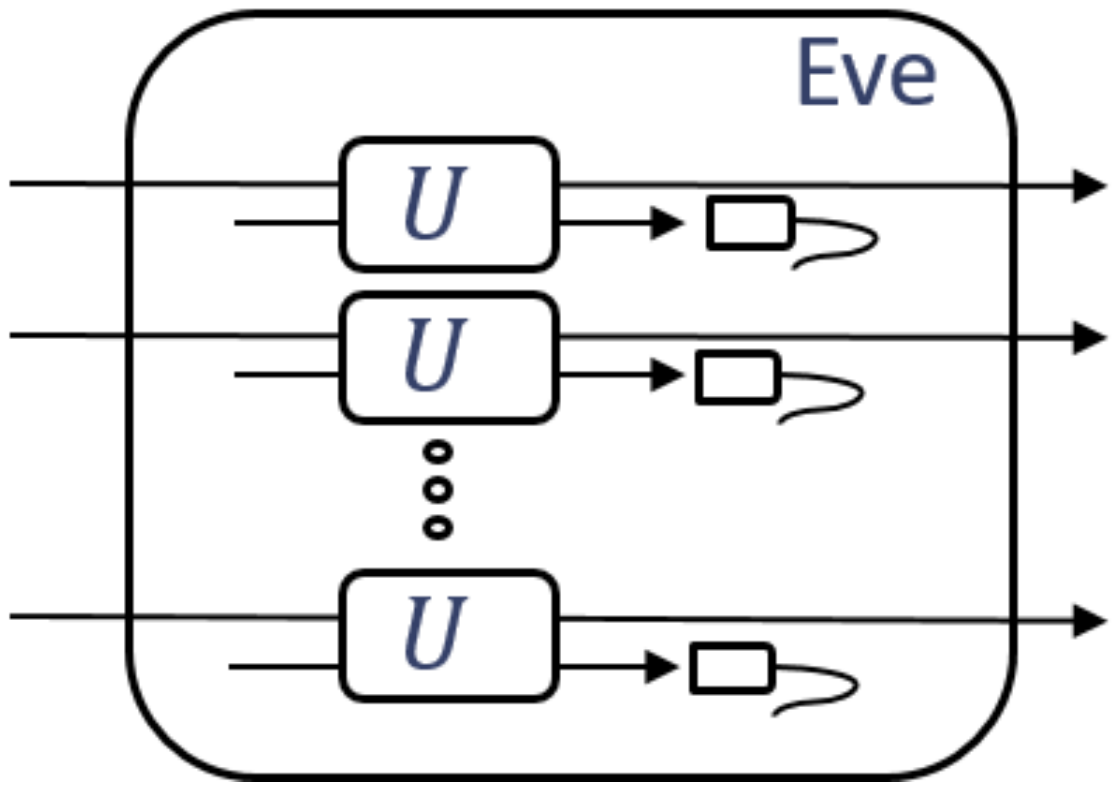

*Figure 3.2: Individual attacks are described by the interaction of each Eve's ancillary mode with a signal by a unitary operator $U$. The output mode is measured directly after the interaction.*

attack.

## 3.4 Eavesdropping

The classification of the three type of the attacks that characterize Eve's strategy in order to intercept the communication between Alice and Bob are the individual, collective and coherent attacks in increasing order of power. In order to describe an attack, one can adopt the dilation of the channel (see Fig. 2.3), which is under Eve's full control. Therefore, the signal states travelling through the channel interact with Eve's ancillary modes which are processed finally by Eve. Different configurations for the ancillary initial state and the interaction described by the unitary operation result in attacks with different potential [2, 3].

### 3.4.1 Individual attacks

During individual attacks depicted in Fig. 3.2, Eve prepares ancillary modes for every signal state between Alice and Bob. Each signal mode and ancillary mode interact via a unitary operator $U$ and Eve's output mode is measured before the classical communication through the public channel takes place. Therefore, both the quantities $I_{AB}$ and $I_E$ in the secret key rate of Eq. (3.44) are calculated by the mutual information of the classical variables of the parties involved. These variables are dependent on the unitary operation as well as the applied measurements [55].

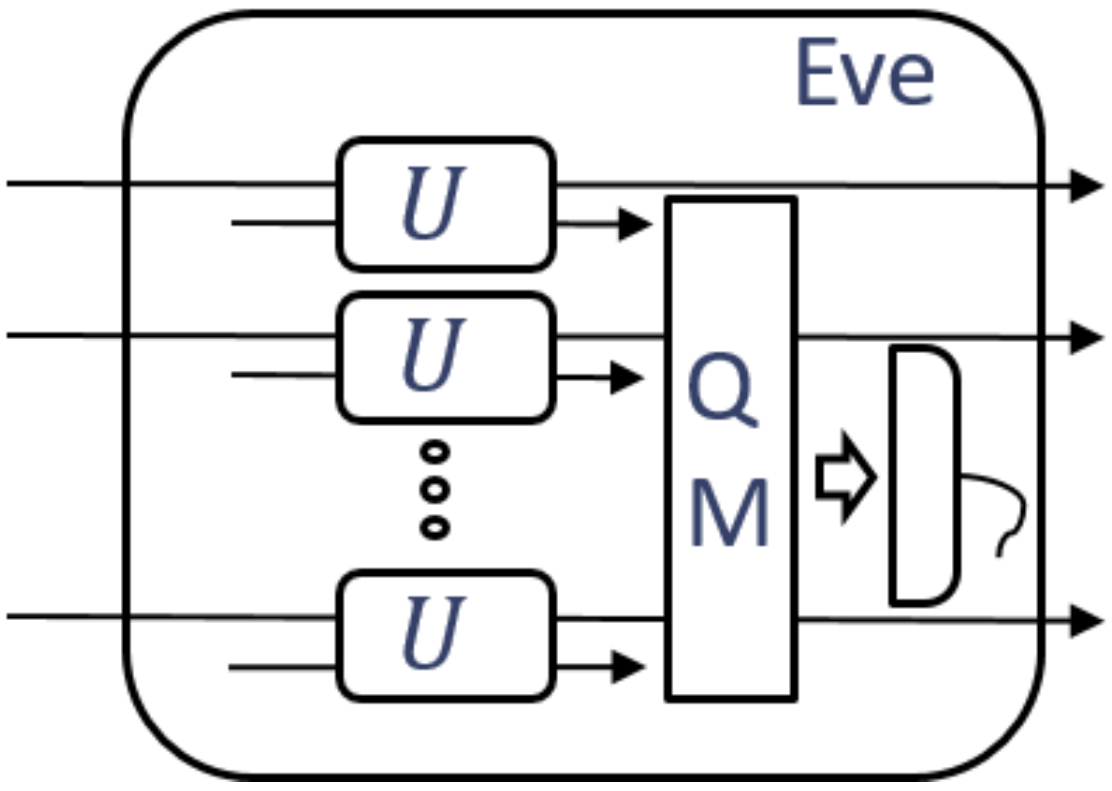

*Figure 3.3: In collective attacks Eve is interacting with each signal separately during the interception by correlating an ancillary mode which is stored in a quantum memory. After the end of the protocol (classical communication) Eve applies an optimal measurement simultaneously to all the ancillary modes for every use of the channel. Thus collective attacks are considered more powerful than the individual ones.*

### 3.4.2 Collective attacks

In the case of a collective attack, Eve prepares her ancillary modes, where each one is interacting through a unitary operator $U$ with a signal sent from Alice to Bob. After the interaction, Eve's output modes are stored in a quantum memory. This gives the advantage to Eve to apply an optimal coherent measurement to the quantum memory after the reconciliation process between Alice and Bob through the public channel. Due to this setting, Eve's information about Alice's or Bob's variable is restricted theoretically only by the laws of nature. Thus Eve's accessible information $I_E$ in Eq. (3.44) is quantified by the Holevo bound $\chi(E:x)$ (see Eq. (2.82)) for $x$ being Alice's or Bob's variable respectively.

The most effective strategy for Eve is to use a Gaussian unitary operator (see Fig. 3.3) in order to interact with the signals [56–59]. More specifically, one associates to a pure global state $\hat{\rho}_{ABE}$ with CM $\mathbf{V}_{ABE}$, that describes the systems of the parties and Eve, its Gaussian counterpart $\hat{\rho}^{\mathrm{G}}_{ABE}$ with the same CM. Then, it turns out that we obtain the following bound for the secret key rate

$$K(\hat{\rho}_{ABE}) \geq K(\hat{\rho}^{\mathrm{G}}_{ABE}). \tag{3.5}$$

As we have seen previously, the most realistic Gaussian channel simulating an optical-fibre communication channel is the entangling cloner (see Sec. 2.5.1.1). From now on, we will use this kind of attack in order to calculate the secret key rate.

### 3.4.3 Coherent attacks

The most powerful attack for Eve is the coherent attack illustrated in Fig. 3.4. In this case, Eve prepares a global ancillary system which interacts by a global unitary operator with all the signals sent through the channel. Eve stores the outputs of the ancillary systems in a quantum memory, waits for the reconciliation process between Alice and Bob over the public channel to be finished, and applies an optimal joint measurement on the quantum memory. The study of the security against these attacks is extremely complex and in the most general case cannot be solved analytically. Nevertheless, there have been results that allow to tackle with the complexity of the problem and reduce it to a simplified scenario. In particular, in Ref. [60], it has been proved also for infinite dimensional systems (CV systems) the quantum de Finetti theorem. According to this theorem, the study of these attacks can be reduced to collective attacks whenever the shared state between Alice and Bob after many uses of the channel $\rho_{AB}^{\otimes n}$ is invariant under any permutation of the systems. This can be satisfied in general by randomization of the classical data during the classical post-processing.

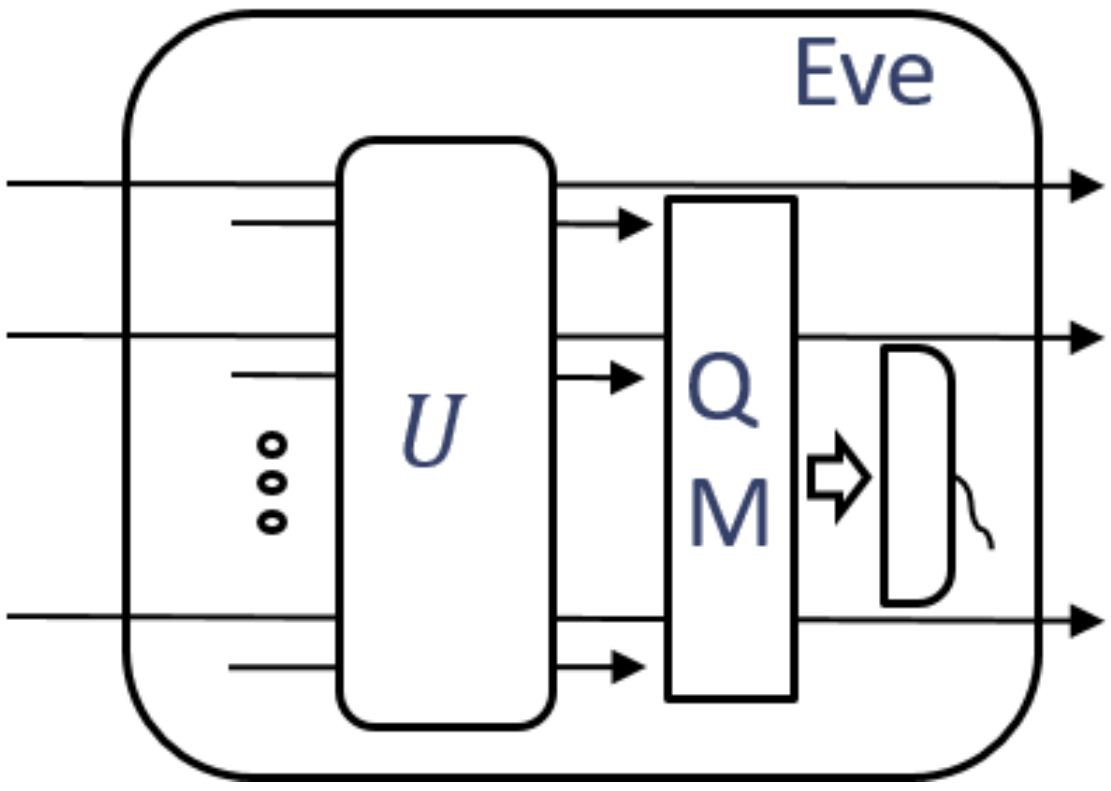

*Figure 3.4: During a coherent attack Eve prepares a global ancillary system which interacts with every signal with a global unitary operator. The outputs of the interaction corresponding to the ancillary system are stored in a quantum memory, to which an optimal joint measurement is applied after the end of the protocol and the classical communication between Alice and Bob.*

## 3.5 Asymptotic secret key rate

### 3.5.1 Mode propagation

In the worst case scenario, Eve has the full control of the channel. For assessing the security of the protocol, we use the assumption of collective attacks. More specifically, since our protocol is fully Gaussian, in the sense that Alice modulates coherent states with a Gaussian modulation, the worst attack would be describe by a Gaussian channel. The most realistic situation of a Gaussian channel is that of a thermal-loss channel (see Sec. 2.5.1.1), e.g., optical fibre telecommunication channels, with transmissivity $\tau$ and thermal noise $\omega = 2\bar{n} + 1$, as in Fig. 3.1. The last two modes $e_0$ and $e_1$ belong to Eve and their CM is given by

$$\mathbf{V}_{e_1 e_0} = \begin{pmatrix} \omega\mathbf{I} & \sqrt{\omega^2 - 1}\mathbf{I} \\ \sqrt{\omega^2 - 1}\mathbf{I} & \omega\mathbf{I} \end{pmatrix}, \quad \mathbf{Z} = \begin{pmatrix} 1 & 0 \\ 0 & -1 \end{pmatrix}. \tag{3.6}$$

The corresponding beam-splitter symplectic transformation is expressed as $\mathbf{L}(\tau) = \mathbf{M}(\tau) \oplus \mathbf{I}$ where $\mathbf{M}(\tau) = \mathbf{B}(\tau)$ as in Eq. (2.67).

According to this propagation, Bob's mode is described by the quadrature operators

$$\hat{Q}_B = \sqrt{\tau}(Q_M + \hat{Q}_0) + \sqrt{1-\tau}\hat{Q}_{e_1}, \tag{3.7}$$

$$\hat{P}_B = \sqrt{\tau}(P_M + \hat{P}_0) - \sqrt{1-\tau}\hat{P}_{e_1} \tag{3.8}$$

for the switching protocol and

$$\hat{Q}_{B'} = \sqrt{\frac{\tau}{2}}(Q_M + \hat{Q}_0) + \sqrt{\frac{\tau - 1}{2}}\hat{Q}_{e_1} + \sqrt{\frac{1}{2}}\hat{Q}_{\text{vac}}, \tag{3.9}$$

$$\hat{P}_{B'} = -\sqrt{\frac{\tau}{2}}(P_M + \hat{P}_0) + \sqrt{\frac{\tau - 1}{2}}\hat{P}_{e_1} + \sqrt{\frac{1}{2}}\hat{P}_{\text{vac}} \tag{3.10}$$

for the no-switching protocol, where $\hat{Q}_0, \hat{P}_0, \hat{Q}_{\text{vac}}, \hat{P}_{\text{vac}}$ have variance $V_0 = V_{\text{vac}} = 1$. The variables $Q_{B'}$ and $P_{B'}$ for the protocol with heterodyne detection (see Sec. 2.5.2) comes from the fact that Bob applies a heterodyne measurement by mixing the incoming mode $B$ with a vacuum mode associated with the operators $\hat{Q}_{\text{vac}}$ and $\hat{P}_{\text{vac}}$ by a balanced beam splitter. To every quadrature operator $\hat{X}$ with variance $V$ we can assign a random variable $X$ that describes a potential measurement outcome (its eigenvalues as in Eq. (2.20)). Therefore, we can describe the variables $Q_B$ ($Q_{B'}$) and $P_B$ ($P_{B'}$) in terms of signal and

noise as follows:

$$Q_B = \sqrt{\tau} Q_M + Q_N, \quad Q_N = \sqrt{\tau} Q_0 + \sqrt{1-\tau} Q_{e_1}, \tag{3.11}$$

$$P_B = \sqrt{\tau} P_M + P_N, \quad Q_N = \sqrt{\tau} P_0 - \sqrt{1-\tau} P_{e_1} \tag{3.12}$$

$$Q_{B'} = \sqrt{\frac{\tau}{2}} Q_M + Q_{N'}, \quad Q_{N'} = \sqrt{\frac{\tau}{2}} Q_0 + \sqrt{\frac{\tau-1}{2}} Q_{e_1} + \sqrt{\frac{1}{2}} Q_{\text{vac}}, \tag{3.13}$$

$$P_{B'} = -\sqrt{\frac{\tau}{2}} P_M + P_{N'}, \quad P_{N'} = -\sqrt{\frac{\tau}{2}} P_0 + \sqrt{\frac{\tau-1}{2}} P_{e_1} + \sqrt{\frac{1}{2}} P_{\text{vac}} \tag{3.14}$$

and their corresponding variances are

$$V_B = \tau V_M + V_N, \tag{3.15}$$

$$V_N = \tau + (1-\tau)(\omega) = 1 + V_\epsilon = 1 + (1-\tau)(\omega - 1) \tag{3.16}$$

$$V_{B'} = \frac{\tau}{2} V_M + V_{N'}, \tag{3.17}$$

$$V_{N'} = \frac{\tau + (1-\tau)\omega + 1}{2} = \frac{2 + (1-\tau)(\omega-1)}{2} = \frac{V_N + 1}{2}, \tag{3.18}$$

where $Q_N$ ($Q_{N'}$) and $P_N$ ($P_{N'}$) are the noise variables of the quadratures of mode $B$ respectively with variance $V_N$ ($V_{N'}$ ).

Eve's mode after the propagation will be described by the quadrature operators

$$\hat{Q}_{e_2} = -\sqrt{1-\tau}(Q_M + \hat{Q}_0) + \sqrt{\tau} \hat{Q}_{e_1}, \tag{3.19}$$

$$\hat{P}_{e_2} = -\sqrt{1-\tau}(P_M + \hat{P}_0) + \sqrt{\tau} \hat{P}_{e_1} \tag{3.20}$$

with variance

$$V_{e_2} = (1-\tau)(V_M + 1) + \tau + \tau(\omega - 1) = (1-\tau)\mu + \tau\omega \tag{3.21}$$

and the CM of her average state will be given by

$$\mathbf{V}_{e_2 e_0} = \begin{pmatrix} (\tau\omega + (1-\tau)\mu)\,\mathbf{I} & \sqrt{\tau(\omega^2-1)}\mathbf{Z} \\ \sqrt{\tau(\omega^2-1)}\mathbf{Z} & \omega\mathbf{I}, \end{pmatrix}, \tag{3.22}$$

with $\mu = V_M + 1$. The CM of Eve's conditional state in direct reconciliation is obtained by setting $\mu = 1$ in the variance of the relevant quadrature with respect the applied measurement. For instance, in the case of a homodyne measurement we have

$$\mathbf{V}_{e_2 e_0 | Q_M} = \begin{pmatrix} \mathbf{V} & \sqrt{\tau(\omega^2-1)}\mathbf{Z} \\ \sqrt{\tau(\omega^2-1)}\mathbf{Z} & \omega\mathbf{I} \end{pmatrix}, \quad \mathbf{V} = \text{diag}\{(1-\tau) + \tau\omega, V_{e_2}\} \tag{3.23}$$

while for a heterodyne measurement

$$\mathbf{V}_{e_2 e_0 | Q_M P_M} = \begin{pmatrix} (\tau\omega + (1-\tau))\,\mathbf{I} & \sqrt{\tau(\omega^2-1)}\mathbf{Z} \\ \sqrt{\tau(\omega^2-1)}\mathbf{Z} & \omega\mathbf{I}. \end{pmatrix}, \tag{3.24}$$

In the case of reverse reconciliation, we apply the homodyne and heterodyne formulas of Eq. (2.71) and Eq. (2.72) to mode $B$ of the CM of $\hat{\rho}_{e_2e_1B}$ in order to calculate the conditional one. These covariance matrices are given by

$$\mathbf{V}_{e_2e_0B} = \begin{pmatrix} V_{e_2}\mathbf{I} & \sqrt{\tau(\omega^2-1)}\mathbf{Z} & \mathbf{D} \\ \sqrt{\tau(\omega^2-1)}\mathbf{Z} & \omega\mathbf{I} & \mathbf{d} \\ \mathbf{D} & \mathbf{d} & \mathrm{diag}\{V_{\hat{Q}_B}, V_{\hat{P}_B}\}\mathbf{I} \end{pmatrix}, \tag{3.25}$$

where the matrices $\mathbf{D}$ and $\mathbf{d}$ are describing the correlations of mode $B$ with mode $e_2$ and $e_0$ respectively. By using Eq. (3.7), Eq. (3.9) respectively and Eq. (3.19) along with Eq. (3.22), we have

$$\mathbf{D} = \sqrt{\tau(1-\tau)}(V_M + 1 - \omega)\mathbf{I}, \tag{3.26}$$

$$\mathbf{d} = \sqrt{1-\tau}(\omega^2-1)\mathbf{Z}. \tag{3.27}$$

The resulting conditional covariance matrices are

$$\mathbf{V}_{e_2e_0|Q_B} = \begin{pmatrix} \mathbf{A} & \mathbf{C} \\ \mathbf{C} & \mathbf{B} \end{pmatrix}, \tag{3.28}$$

with

$$\mathbf{A} = \begin{pmatrix} \frac{(V_M+1)\omega}{\tau(V_M+1-\omega)+\omega} & 0 \\ 0 & \tau(V_M+1)+(1-\tau)\omega \end{pmatrix}, \tag{3.29}$$

$$\mathbf{B} = \begin{pmatrix} \frac{1-\tau+\tau(V_M+1)\omega}{\tau(V_M+1)+(1-\tau)\omega} & 0 \\ 0 & \omega \end{pmatrix}, \tag{3.30}$$

$$\mathbf{C} = \begin{pmatrix} \sqrt{\tau(\omega^2-1}\left(\frac{(V_M+1)}{\tau(V_M+1)+(1-\tau)\omega}\right) & 0 \\ 0 & -\sqrt{\tau(\omega^2-1} \end{pmatrix}, \tag{3.31}$$

and

$$\mathbf{V}_{e_2e_0|Q_{B'}P_{B'}} = \begin{pmatrix} a\mathbf{I} & c\mathbf{Z} \\ c\mathbf{Z} & b\mathbf{I} \end{pmatrix}, \tag{3.32}$$

with

$$a = \frac{(1-\tau)(V_M+1) + [\tau + (V_M+1)]\omega}{1+\tau(V_M+1)+(1-\tau)\omega}, \tag{3.33}$$

$$b = \frac{(1-\tau) + [1+\tau(V_M+1)]\omega}{1+\tau(V_M+1)+(1-\tau)\omega}, \tag{3.34}$$

$$c = \sqrt{\tau(\omega^2-1}\left(\frac{(V_M+2)}{\tau(V_M+1)+(1-\tau)\omega+1}\right). \tag{3.35}$$

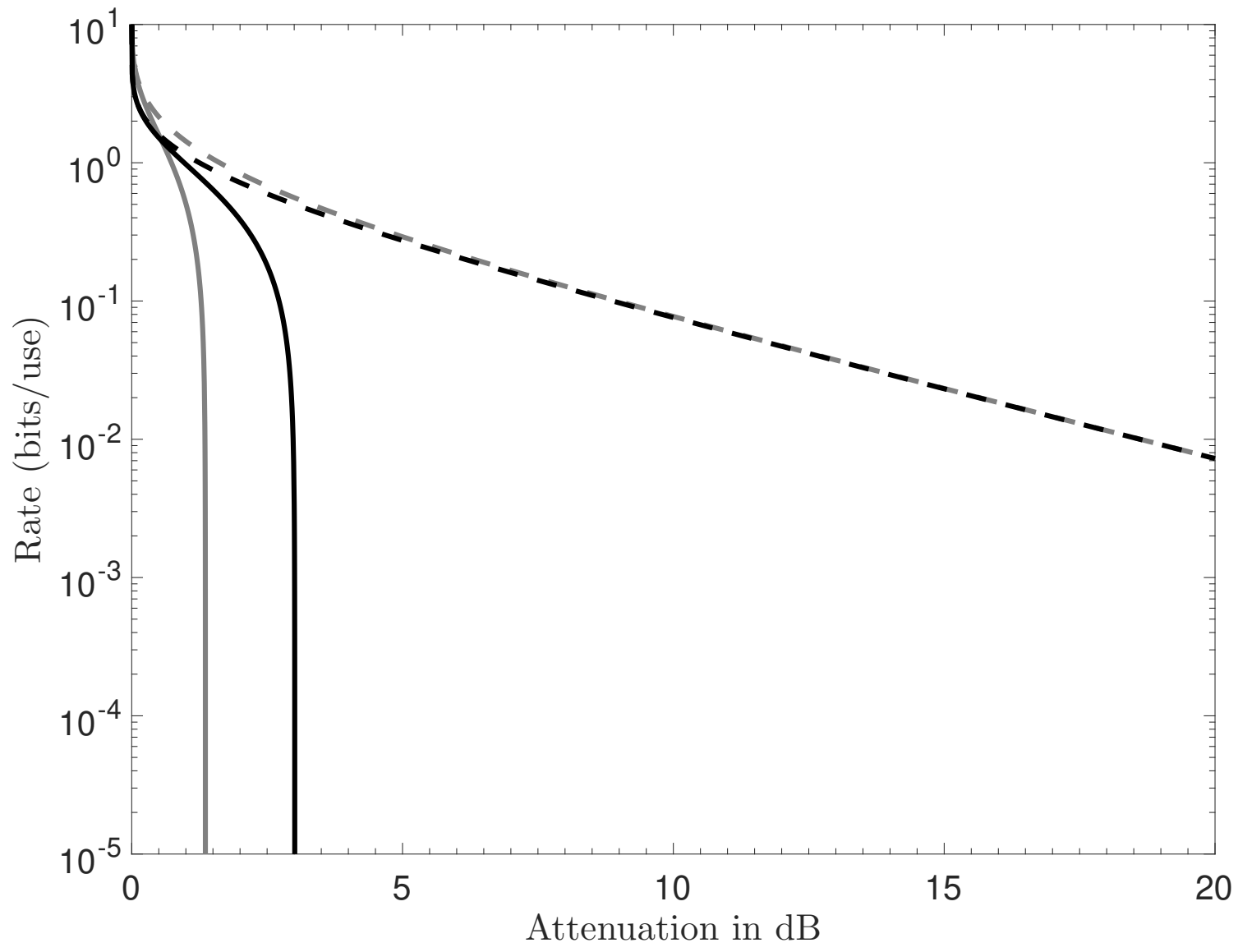

*Figure 3.5: The secret key rate for direct reconciliation (solid lines) and reverse reconciliation (dashed lines) versus the attenuation in dB. The black lines correspond to the protocol with the homodyne detection while the grey lines to the protocol with the heterodyne detection. We have plotted the ideal case for $\xi = 1$ and $V_M \to \infty$ given by the corresponding formulas in Eq. (3.40–3.43). We can see that the rates with reverse reconciliation can beat the 3dB limit in contrast to the rates with direct reconciliation.*

### 3.5.2 Mutual information and Holevo information

The mutual information is not dependent on the reconciliation direction but is dependent on Bob's measurement. In particular, $I_{AB}$ is given by

$$I_{\text{hom}} = \frac{1}{2}\log_2 \frac{V_{\hat{Q}_B}}{V_{\hat{Q}_B|Q_M}} \tag{3.36}$$

$$= \frac{1}{2}\log_2 \frac{\tau(V_M+1)+(1-\tau)\omega}{\tau+(1-\tau)\omega} \tag{3.37}$$

for a protocol with homodyne measurement. Both the quadratures are treated in the same way by this protocol. Thus, even though Bob in reality is switching between the quadratures for applying the measurement, for the mutual information calculation, we can assume that Bob is always choosing to measure with respect to one of them (e.g. $\hat{Q}_B$).

On the other hand, for a protocol with heterodyne detection the contribution of both quadratures have to be taken into account. Therefore, the formula of the mutual informa-

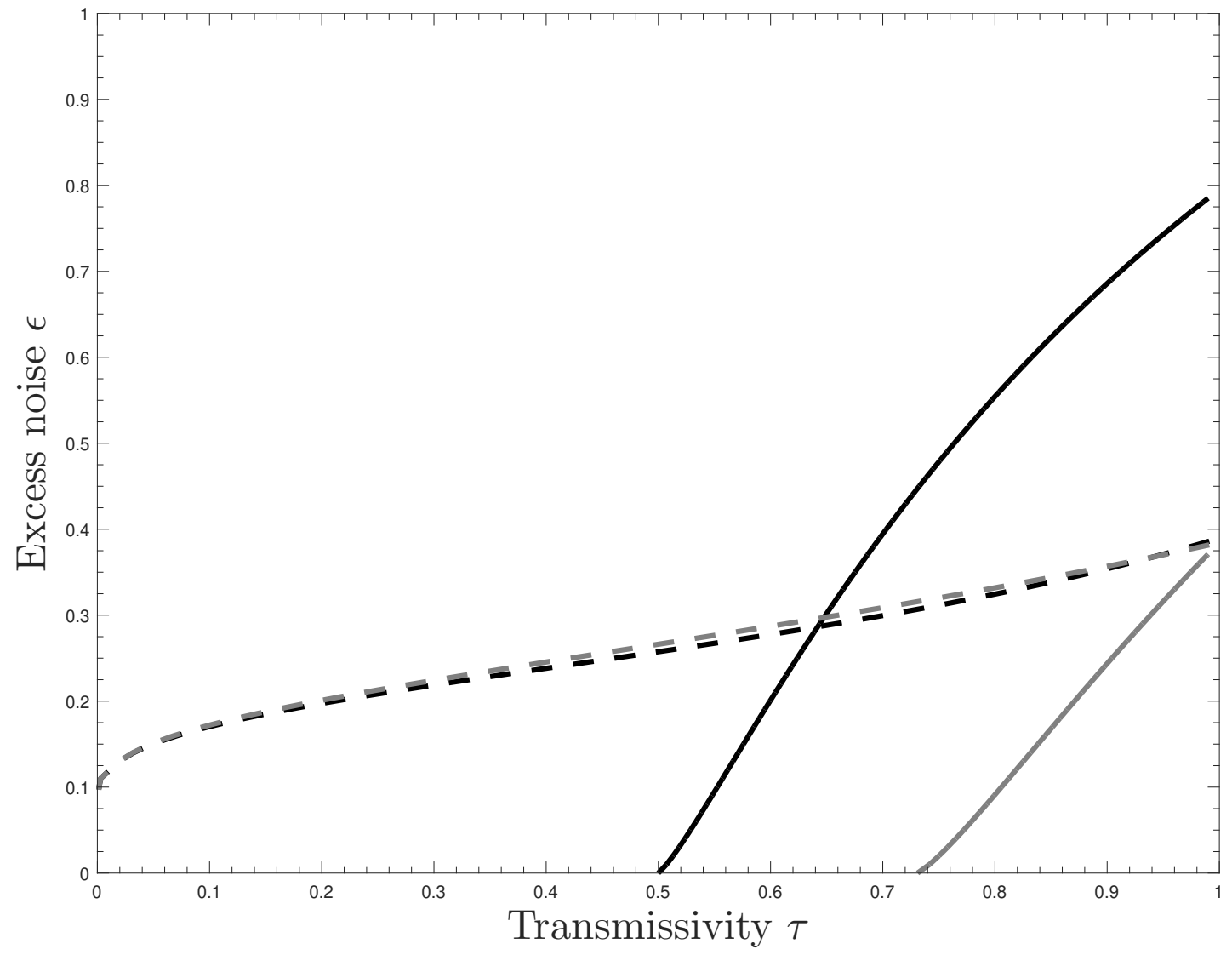

*Figure 3.6: The security thresholds for of the direct reconciliation (solid lines) and reverse reconciliation (dashed lines) with respect the transmissivity $\tau$ and excess noise $\epsilon$. The black lines correspond to the switching protocol (homodyne detection) and the gray ones to the protocol without switching (heterodyne detection).*

tion $I_{AB}$ in that case is given by

$$I_{\text{het}} = \frac{1}{2}\log_2 \frac{V_{\hat{Q}_{B'}}}{V_{\hat{Q}_{B'}|Q_M}} + \frac{1}{2}\log_2 \frac{V_{\hat{P}_{B'}}}{V_{\hat{P}_{B'}|Q_M}} \tag{3.38}$$

$$= \log_2 \frac{\tau(V_M+1) + (1-\tau)\omega + 1}{\tau + (1-\tau)\omega + 1}, \tag{3.39}$$

where we have replaced from Eq. (4.21–4.24). From the CMs of Sec. 3.5.1, we obtain the symplectic eigenvalues of Eve's average and conditional state depending on Bob's measurement and the reconciliation direction. By using Eq. (2.82), the following remarks and Eq. (2.78), we can calculate the Holevo information $I_E = \chi(E:x)$ where $x$ is $Q_A$,$P_A$ or $Q_B, P_B$.

### 3.5.3 Secret key rate for infinite modulation

We can have analytic expressions for the secret key rates of each of the four variations of the protocol by assuming infinite modulation $\mu \to \infty$ for $\mu = V_M + 1$. Then we have,

$$R^{\blacktriangleright}_{\text{hom}}(\tau,\omega) = \frac{1}{2}\log_2 \frac{\tau}{1-\tau}\frac{\tau\omega+1-\tau}{\tau+(1-\tau)\omega} + h\left(\sqrt{\frac{(\tau+(1-\tau)\omega)\omega}{\tau\omega+1-\tau}}\right) - h(\omega), \tag{3.40}$$

$$R^{\blacktriangleright}_{\text{het}}(\tau,\omega) = \log_2 \frac{2\tau}{\mathrm{e}(1-\tau)[\tau+(1-\tau)\omega+1]} + h\left(\tau+(1-\tau)\omega\right) - h(\omega), \tag{3.41}$$

$$R^{\blacktriangleleft}_{\text{hom}}(\tau,\omega) = \frac{1}{2}\log_2 \frac{1}{1-\tau}\frac{\omega}{\tau+(1-\tau)\omega} - h(\omega), \tag{3.42}$$

$$R^{\blacktriangleleft}_{\text{het}}(\tau,\omega) = \frac{1}{2}\log_2 \frac{\tau}{1-\tau}\frac{2}{\tau+(1-\tau)\omega+1} + h\left(\frac{(1-\tau\omega)+1}{\tau}\right) - h(\omega), \tag{3.43}$$

where $\blacktriangleright$ ($\blacktriangleleft$) stands for the direct (reverse) reconciliation. We have plot the previous rate in Fig. 3.5 in terms of attenuation in $dB$ by substituting $\tau = 10^{-\frac{dB}{10}}$. In the same manner, we can have expressions for the secret key rate dependent on the excess noise $\epsilon = \frac{(1-\tau)(\omega-1)}{\tau}$. In Fig. 3.6, we plotted the security thresholds in terms of $\epsilon$ and $\tau$.

### 3.5.4 Numerical calculation of the secret key rate

In the case the reconciliation code is not efficient enough in order to provide with the maximum potential correlations between the honest party variables, the secret key rate of Eq. (3.44) is modified by adding the reconciliation efficiency parameter $\xi$ [54], namely

$$R_\infty = \xi I_{AB} - I_E. \tag{3.44}$$

This in fact can be seen as a result of finite-size effects (see Chap. 6, Chap. 9). More specifically, in the ideal case of infinite number of signal exchange, we expect that the classical post-processing can result in a flawless error correction procedure so that the parties achieve the maximum correlations between their variables. This allows the whole mutual information to take part on the secret key rate calculation meaning that $\xi = 1$. In a realistic scenario, when the signal exchange is finite, the parties share only a fraction of the mutual information so that $0 < \xi < 1$.

In this case, the calculation of secret key rate function is not straight forward since then there is a dependence not only on the channel parameters, e.g., transmissivity $\tau$ and thermal noise $\omega$ but also on the modulation $\mu = V_M + 1$. In fact, we can see that for every transmissivity $\tau$ there is a different value for $\mu$ that gives the highest rate (see Fig. 3.7). That is why we usually consider the asymptotic secret key rate as a function

$$R(\xi,\mu,\omega,\tau) = \xi I_{AB} - I_E \tag{3.45}$$

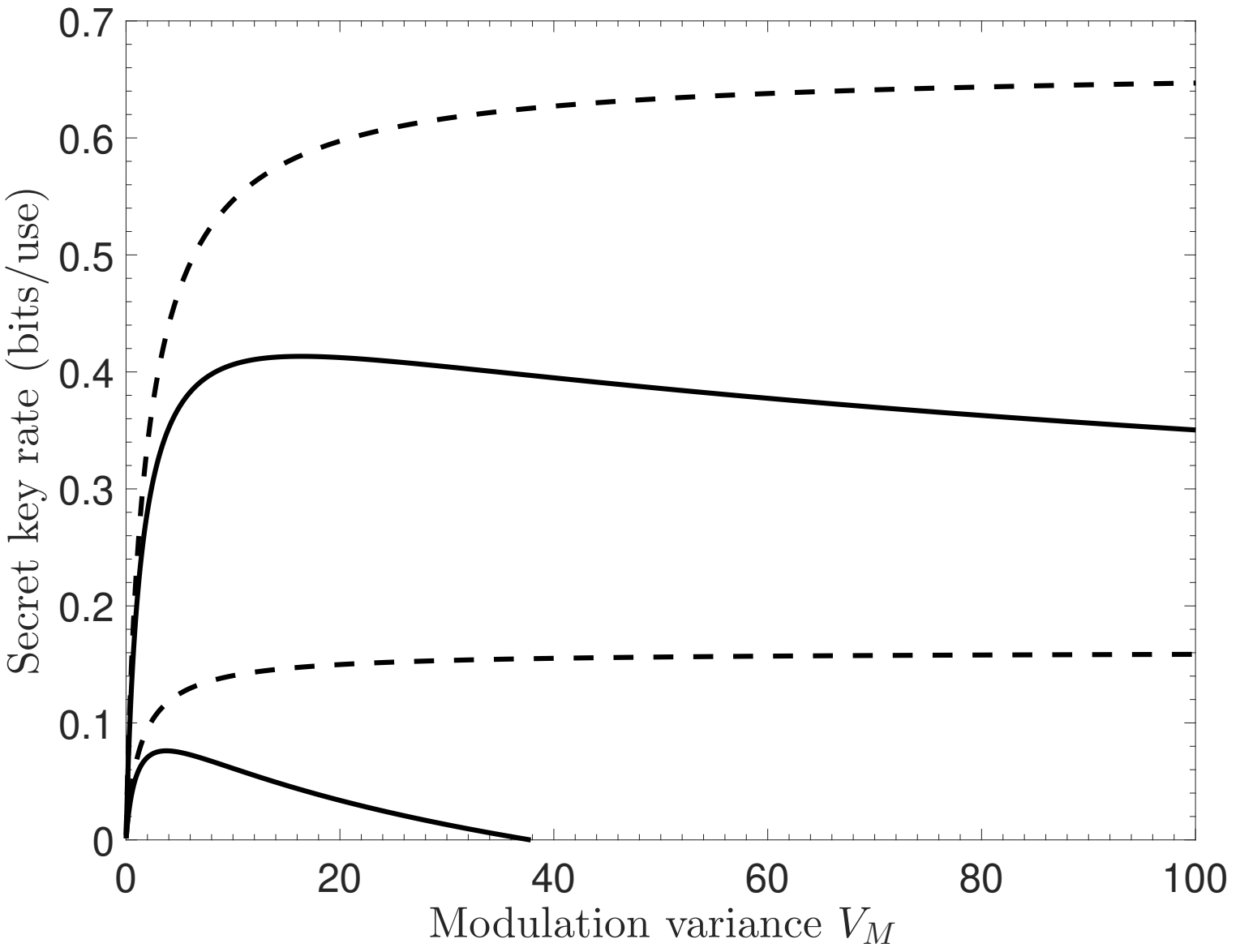

*Figure 3.7: Secret key rate versus modulation variance $V_M$ of the reverse reconciliation protocol using homodyne detection. The solid lines correspond to reconciliation efficiency $\xi = 0.9$, the higher for transmissivity $\tau = 0.6$ and the lower for $\tau = 0.2$. The dashed lines have been plotted fro $\xi = 1$ for the corresponding transmissivities. We can see that the value of the secret key rate saturates after a certain value of modulation for the case $\xi = 1$. This is not the same for cases $\xi < 1$ that there is a maximum for the secret key rate different for different values of transmissivities when for instance $\xi = 0.9$.*

and we present results with numerical calculations after optimizing over the modulation $\mu = V_M + 1$.

## 3.6 Phase-encoded coherent states

A potential treatment to the previous deficiency of the reconciliation codes with respect to a Gaussian (continuous) distribution of coherent states can be given by using discrete modulation schemes for which very good reconciliation codes exist [8] (see also Fig. 3.12. These schemes are usually described by a given number $N$ of coherent states centred on discrete points on phase space. If these points share the same distance from the center of the phase space and as a consequence the same energy and only vary with respect their discrete phase differences, then we have a scheme as in Ref. [48] using phase-encoded coherent states.

Consider a discrete variable $X_A = \{a_k, p_k\}$ with $k = 1, 2, \ldots, N$ following a uniform distribution $p_k = 1/N$. Alice prepares coherent states with amplitude $a_k = ze^{i\phi_k}$, where $\phi_k = \frac{2\pi}{N}k$. The coefficient $z$ describes a fixed radius in the phase space and accounts for the square root of the mean number of photons $\bar{n}$ as shown in Fig. 3.8 for a four-state protocol $N = 4$. For each use of the channel, the coherent state is prepared on mode $A$ with a phase chosen from the $\phi_k$ and amplitude $z$, and is sent through a thermal-loss channel (see Sec. 2.5.1.1). Its output $B$ is detected by Bob by the application of a heterodyne measurement. This channel description is not providing an optimal attack assumption for such a protocol but still remains the most realistic simulation of the current technology communication channels such as optical fibres.

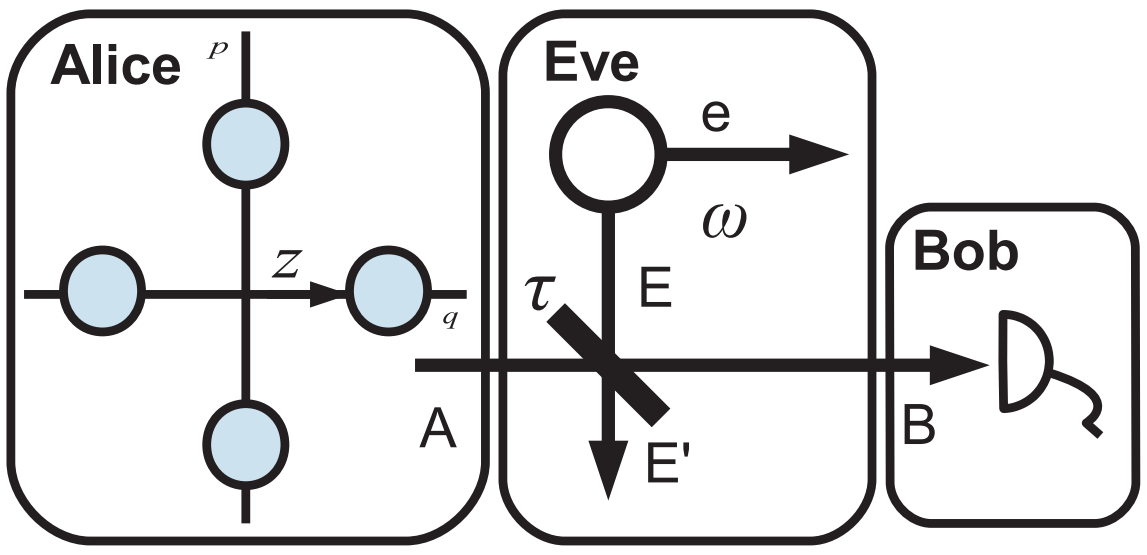

*Figure 3.8: Alice prepares mode A in one of the four coherent states with radius z and sends it to Bob through a thermal-loss channel dilated into an entangling-cloner attack. In particular, the beam splitter has transmissivity τ, characterizing the channel loss, and the variance ω ≥ 1 of Eve's TMSV state provides additional thermal noise to the channel. Eve's output modes are stored in a quantum memory measured at the end of the protocol, i.e., after the entire quantum communication and Alice and Bob's classical communication. At the output of the channel, Bob applies an heterodyne detection to mode B. An upper bound on the performance of the parties can be computed by assuming that also Bob has a quantum memory that he measures at the end of the entire communication process.*

### 3.6.1 Upper bound for the secret key rate

We calculate here an upper bound for the secret key rate for the case of a pure loss channel by setting $\omega = 1$. This upper bound is based on the fact that Bob has a quantum memory so that he may apply an optimal joint detection. This is achieved by calculating Bob's accessible information on Alice's variable by the use of the Holevo function (see Eq. (2.82)). By virtue of this analysis, we can investigate the the effect of changing the values of parameters $z$ and $N$. For example, we can search for the conditions under which the protocol has similar performance for $N = 4$ coherent states compared with the case of $N \to \infty$ coherent states.

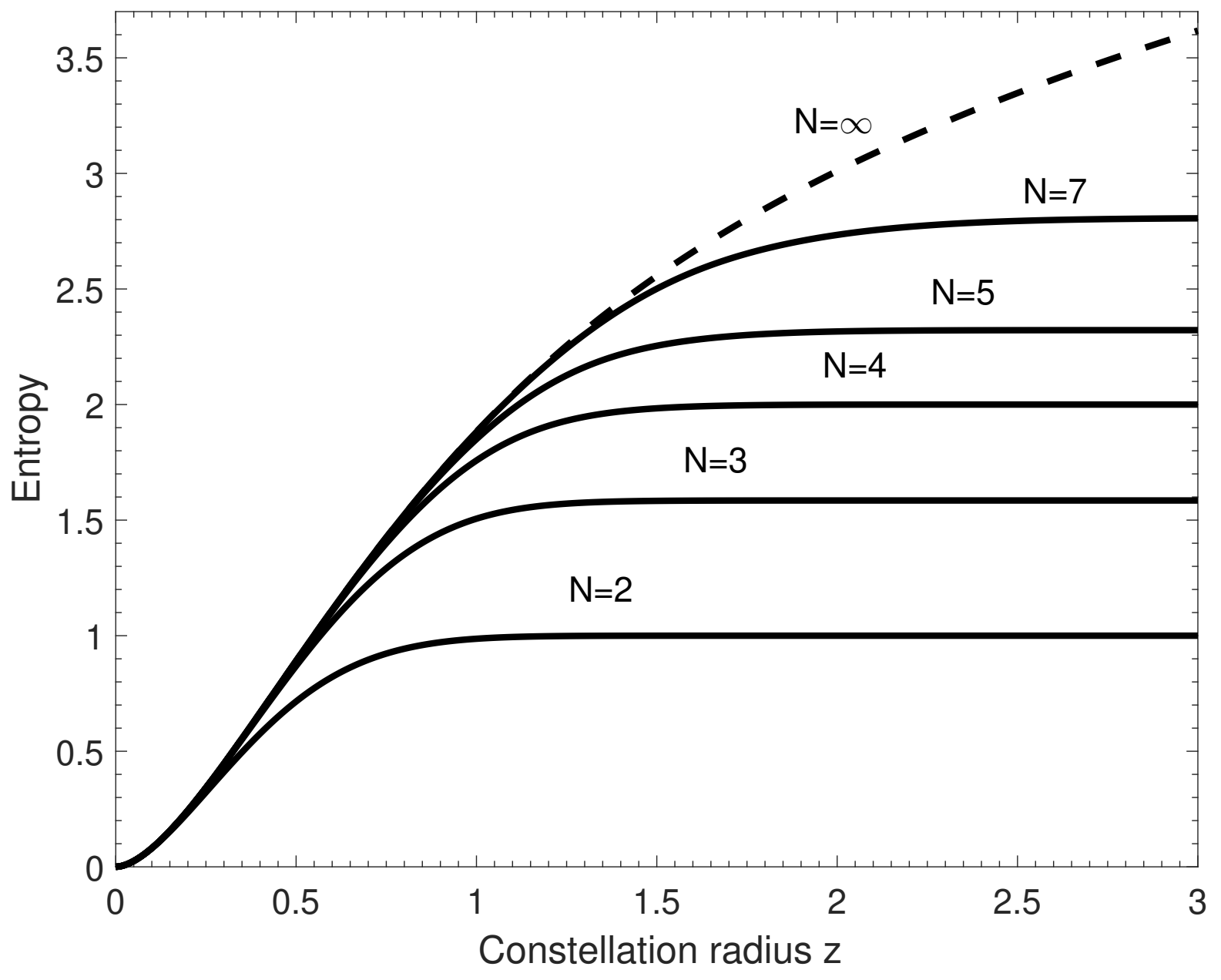

*Figure 3.9: The von Neumann entropy $S(\rho_A)$ of the Alice's average state $\rho_A$ for different number $N$ over the radius $z$ of the constellation circle (solid lines). We plotted also the entropy of the continuous uniform distribution ($N \to \infty$) of the constellation states (dashed line).*

Alice's average state before the channel propagation is given by an ensemble of $N$ coherent states $|a_k\rangle$ distributed with the same probability $p_k = 1/N$

$$\rho_A = \frac{1}{N} \sum_{k=0}^{N-1} |a_k\rangle\langle a_k|, \tag{3.46}$$

which is parametrized by $N$ and $z$. The von Neumann entropy $S(\rho_A)$ (see Eq. 2.77) has been plotted in Fig. 3.9 with respect to the radius of the encoding scheme $z$ for different $N$. In Appendix B.1, we use a preliminary Gram-Schmidt procedure in order to compute this entropy. We notice that the entropy $S(\rho_A)$ is increasing as the number of states in the circle becomes larger. For any given $N$, the entropy saturates to a constant value after a certain value of the radius $z$. We also compare with the case of $N \to \infty$, where we have calculated the corresponding average state in Appendix B.2.

By assuming a pure loss channel with transmissivity $\tau \in (0, 1)$ and $\omega = 1$, we can obtain Bob's average state

$$\rho_B = \frac{1}{N} \sum_{k=0}^{N-1} |\sqrt{\tau} a_k\rangle\langle \sqrt{\tau} a_k|. \tag{3.47}$$

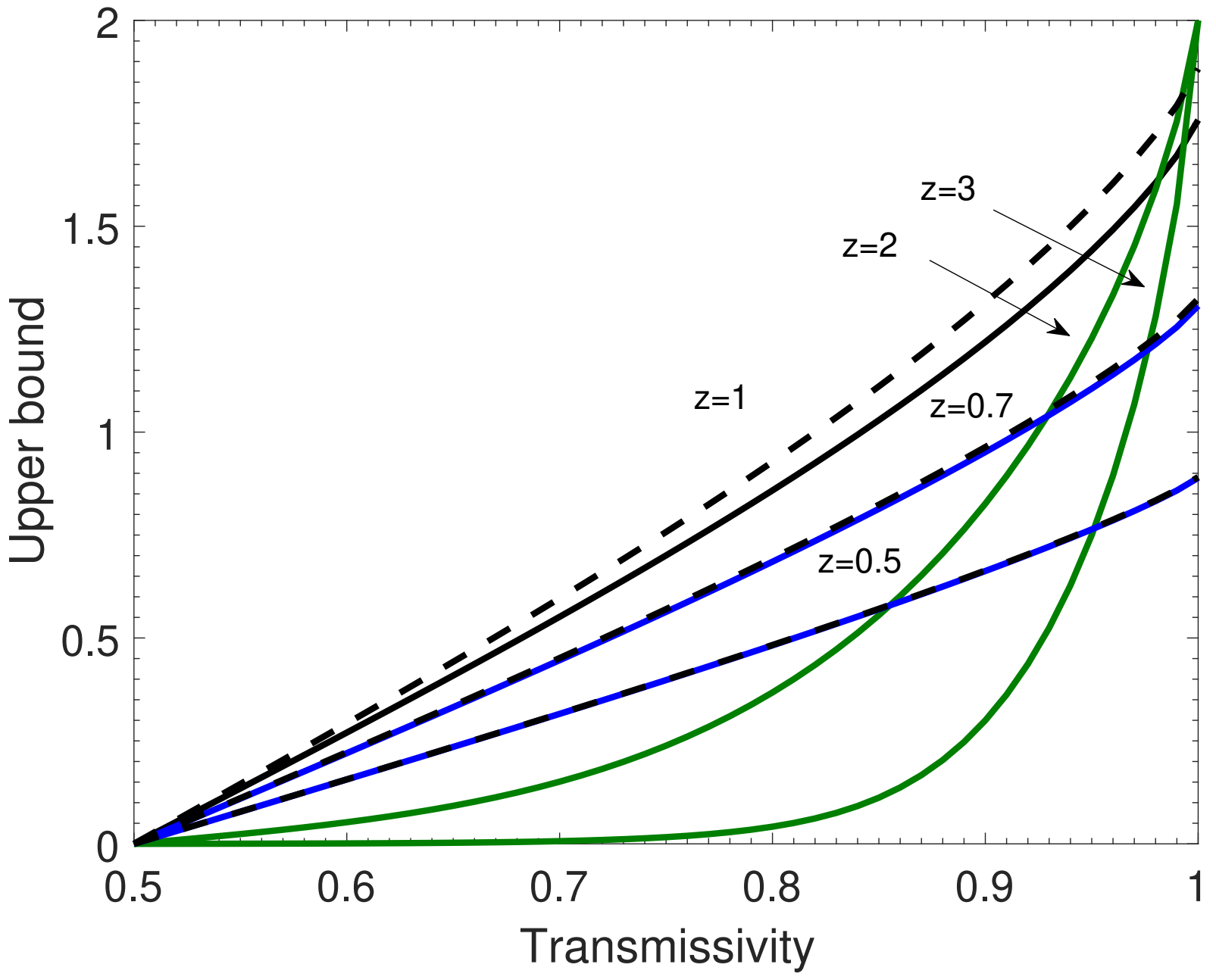

*Figure 3.10: The optimal secret key rate of Eq. (7) for N=4 is plotted over the transmissivity $\tau$ for different values of the radius $z$ of the constellation. We can see that for values $z < 1$ the rate decreases as $z$ is decreasing (blue lines) while for the $z > 1$ the rate decreases as $z$ increases till it gets to zero for $z = 10^6$ (green lines). We also plotted the optimal secret key rate for the continuous uniform distribution of states (black dashed lines). We see that for $z < 0.6$ the two rates become almost identical. This corresponds to a saturation point for the 4-state protocol , so that it makes no difference to use four coherent states or an infinite number.*

Then the corresponding Holevo information in this case will be given by

$$\chi(B : a_k) = S(\rho_B) - \frac{1}{N}\sum_{k=0}^{N-1} S(|a_k\rangle\langle a_k|). \tag{3.48}$$

This relation is simplified by the fact that a coherent state is a pure state and thus its von Neumann entropy is zero leading to $\chi(B : a_k) = S(\rho_B)$ (see remarks after Def. 2.6.3). We follow the same steps as above for this Holevo information calculation. Accordingly, the eavesdropper keeps the other output of the beam splitter, mode $E'$, and her average state will be given by

$$\rho_{E'} = \frac{1}{N}\sum_{k=0}^{N-1} |\sqrt{1-\tau}a_k\rangle\langle\sqrt{1-\tau}a_k| \tag{3.49}$$

and her accessible information by

$$\chi(E' : a_k) = S(\rho_{E'}). \tag{3.50}$$

Therefore, the optimal secret key rate will be given by Eq. (3.44), where we have set

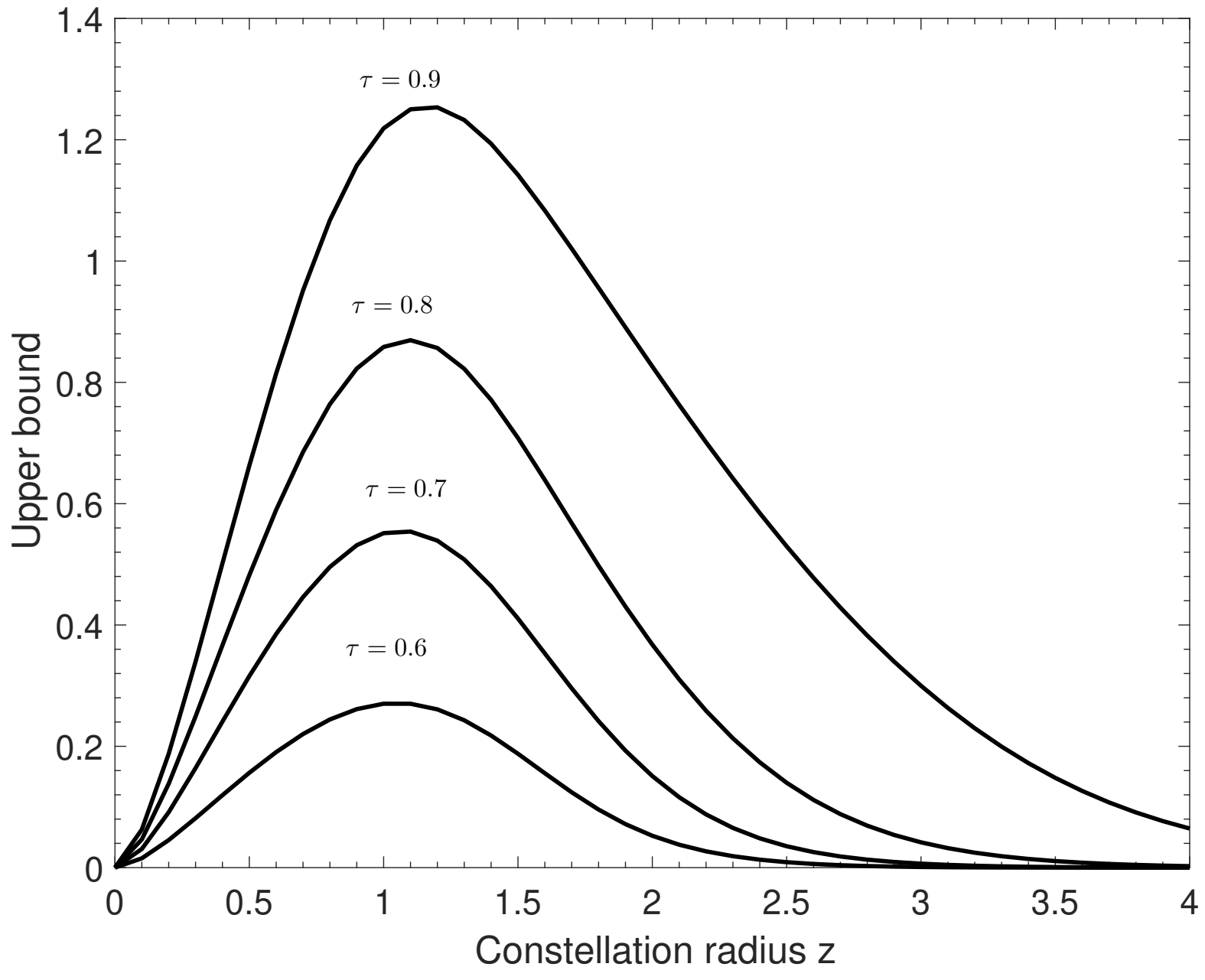

*Figure 3.11: We plotted the optimal rate (upper bound) as a function of the radius z with fixed transmissivities* $\tau = 0.6, 0.7, 0.8, 0.9$. *We can observe that there are certain values that maximize the rate for a given transmissivity.*

$I_{AB} = \chi(B : a_k)$ and $I_E = \chi(E' : a_k)$, which finally is turning into

$$R^{\text{opt}} = S(\rho_B) - S(\rho_{E'}). \tag{3.51}$$

In Fig. 3.10, we plotted this optimal rate for $N = 4$ as a function of the transmissivity $\tau$ and for different values of the radius $z$. We notice that there is an optimal intermediate value for $z$. This means that $z$ cannot be too small resulting in states very close to the vacuum states neither too large leading to almost perfectly distinguishable states. Then, we also show that the optimal performance for the $N = 4$ protocol is very near to that of the protocol with continuous distribution of coherent states for the same values of the radius $z$. In Fig. 3.11, we see that the rate is maximized by a specific value of $z$ for any given value of the transmissivity.

### 3.6.2 Secret key rate

Due to the current technological restrictions, a realistic assumption for the secret key rate is that Bob cannot have a quantum memory and thus an optimal collective measurement. In particular, we assume here that Bob applies to mode $B$ a heterodyne detection in each channel use. Therefore, the corresponding term $I_{AB}$ of Eq. (3.44) is calculated now by using the classical mutual information $I(X_A : X_B)$ (see Eq. (2.75)), where $X_A = \{a_k, p_k\}$ with $p_k = 1/N$ and $X_B = \{b, p(b)\}$ and $b$ is the continuous outcome of Bob's heterodyne detection.

It is obvious that $H(X_A) = \log_2 N$. Then in order to quantify $I(X_A : X_B)$, we need to calculate $H(X_A|X_B)$ through $p(a_k|b)$, i.e., the probability that the state $|a_k\rangle$ was sent through the channel given that Bob measured the amplitude $b$.

To do so, we first calculate the probability that Bob measures $b$ given that the coherent state $|\alpha_k\rangle$ was sent through the channel, $p(b|a_k) = \frac{1}{\pi} e^{-|b-\sqrt{\tau}a_k|^2}$, and by the use of Bayes' rule we obtain

$$p(a_k|b) = \frac{1}{N\pi p(b)} e^{-|b-\sqrt{\tau}a_k|^2}, \tag{3.52}$$

where $p(b) = \frac{1}{N}\sum_{k=0}^{N} p(b|a_k)$.

Then the asymptotic key rate for the direct reconciliation is given by setting $I_{AB} = I(X_A : X_B)$ and $I_E = S(\rho_{E'})$ in Eq. (3.44) obtaining

$$R = I(X_A : X_B) - S(\rho_{E'}), \tag{3.53}$$

which is plotted in Fig. 11.1 for $N = 4$. Afterwards, we consider the case of a pure-loss channel in reverse reconciliation. We just need to re-compute Eve's Holevo bound (now with respect to Bob's outcomes). More specifically, we need to re-compute Eve's conditional entropy. Eve's state conditioned to Bob's outcome $b$ is

$$\rho_{E'|b} = \sum_{k=0}^{N-1} p(a_k|b)|\sqrt{1-\tau}a_k\rangle\langle\sqrt{1-\tau}a_k|, \tag{3.54}$$

where $p(a_k|b)$ is given in Eq. (3.52). We can then compute $S(\rho_{E'|b})$ which is now depending on $b$. Using this quantity, we may write the secret-key rate

$$R = I(X_A : X_B) - S(\rho_{E'}) + \int d^2b\, p(b) S(\rho_{E'|b}). \tag{3.55}$$

This rate is plotted in Fig. 3.12 optimizing over $z$ for $N = 4$. We compare it with the the rate from the fully Gaussian protocol. Also it has been plotted in Fig. 11.4 for the four-state protocol $N = 4$ and radius $z = 0.1$.

## 3.7 Conclusion

Here we presented the one-way protocols which use Gaussian modulation of coherent states under the assumption of an entangling cloner attack and studied their security in the asymptotic regime. We considered the secret key rate as the key quantity to assess the a protocol in terms of security combined with performance in achievable distances. We also presented protocols using discrete encoding. In the following chapters, we will also address the cases of one-way protocols using thermal states instead (see Chap. 4) and the CV-MDI protocol (see Chap. 5) which can expand the CV-QKD to setting allowing for the use of different frequencies and also to network configurations for multiple users. I would like to mention that the plots in this chapter and particularly the study and the plots of Sec. 3.6.1 are my contribution. I chose to put them in the preliminaries context for the sake of the later discussion since they are not belonging in contributions of the recent advancements in CV-QKD.

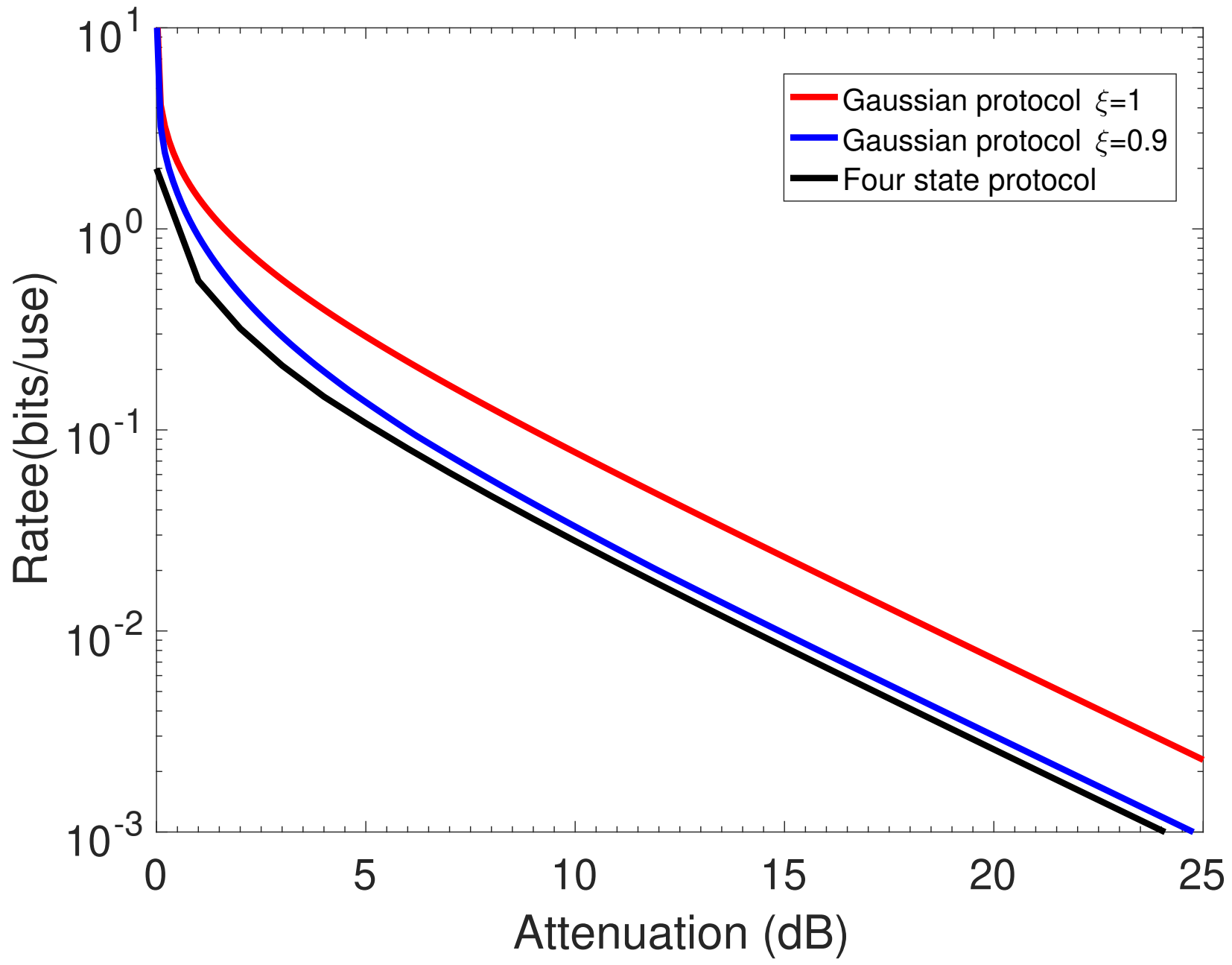

*Figure 3.12: The secret key rate for the protocol with Gaussian modulation (blue line) and using only $N = 4$ states (black line) optimized over $V_M$ and $z$ respectively. For the case of Gaussian modulation we used a reconciliation parameter equal to $\xi = 0.9$. For the 4-state protocol we assumed an ideal reconciliation which is very close to the practical implementations due to the discrete encoding that this protocol supports. We see that the two rates are comparable in the sense that it can be achieved almost the same result by using a simpler and cheaper in resources encoding and decoding process. We have also plotted the ideal case of Gaussian modulation protocol with $\xi = 1$ for the sake of comparison.*

# Chapter 4

# One-way protocols using thermal states

## 4.1 Introduction

In this chapter, we present protocols that use Gaussian displacement of thermal states (instead of coherent states as in Sec. 3.2) investigated in Ref. [19,20]. Our calculations are based on Ref. [22]. Thermal states can be understood as noisy versions of coherent states resulting, for instance, from usage of imperfect devices [23] such as cheap laser sources. Such noise appearing in the preparation can be trusted [23] in the sense that its source has been modeled and calibrated so as to be separated from the untrusted channel noise attributed to Eve. However by increasing the mean photon number we can have laser sources in different frequencies. In particular, this allows for a QKD application in the microwave regime. In order to assess their performance, we assume a quantum thermal loss channel between Alice and Bob that is simulated by an entangling cloner and we follow the way we presented results in Sec. 3.5. Later in Chap. 7 we extend this protocols security in the non-asymptotic regime based on the method presented in Chap. 6.

## 4.2 The protocol and propagation of the modes

In the prepare and measure scheme, Alice prepares thermal states with mean photon number $\bar{n} = \frac{V_{\text{th}}}{2}$. The CM of this state is given by

$$\mathbf{V}_A = \begin{pmatrix} V_{\text{th}} + V_s & 0 \\ 0 & V_{\text{th}} + 1/V_s \end{pmatrix}, \tag{4.1}$$

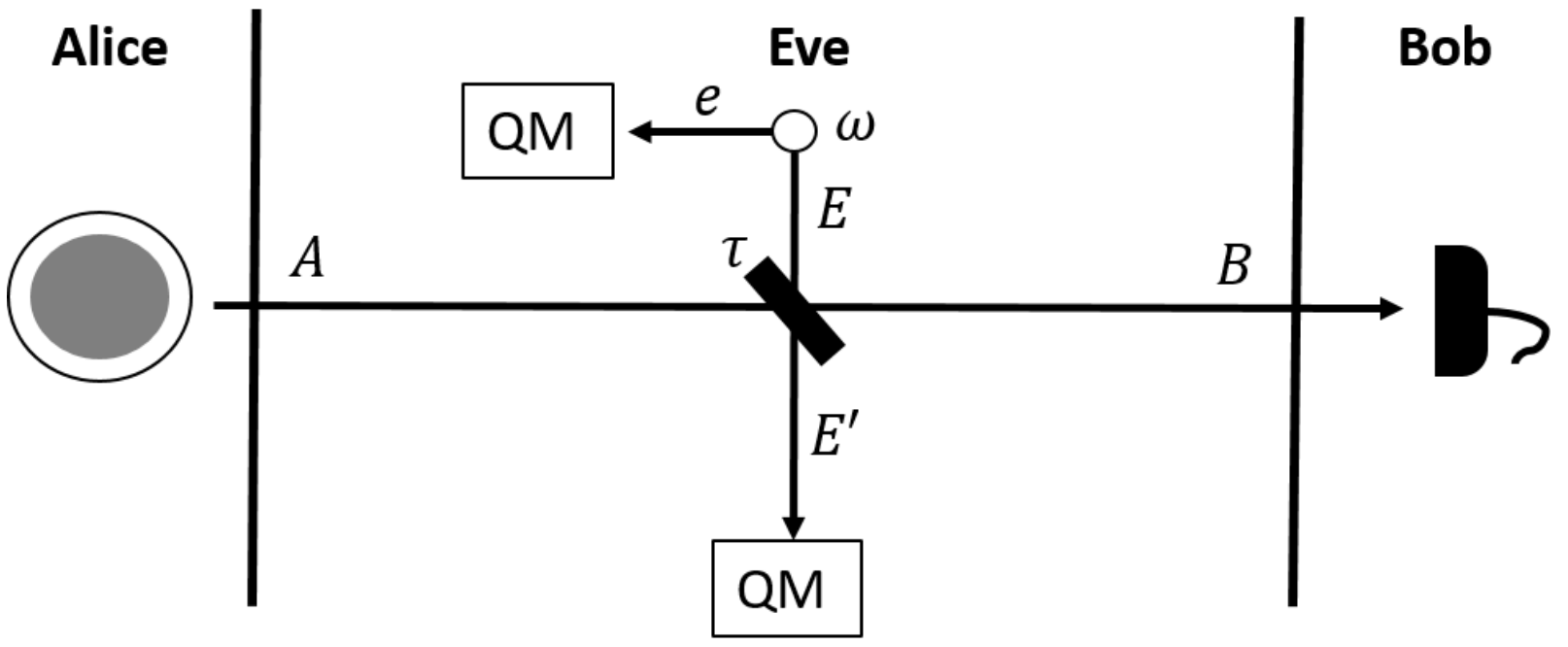

*Figure 4.1: Scheme of the Gaussian protocol using thermal states. Alice prepares a vacuum states with additional thermal trusted noise, i.e., a thermal state (larger white circle). These are travelling through the channel simulated by an entangling cloner. Afterwards, Bob is applying a homodyne measurement.*

where $V_{\text{th}}$ accounts for the variance due to the extra number of thermal photons and $V_s := 1$ is the variance due to quantum fluctuations (quantum shot noise). In case Alice was preparing squeezed states with extra thermal photons then $0 < V_s < 1$. Then she modulates their amplitude by a Gaussian distribution with variance $V_M$ and sends them to Bob through a thermal noise channel as in Fig. 4.1. In the Heisenberg picture, Alice's mode $A$ is described by the quadrature operators $\hat{Q}_A$ and $\hat{P}_A$, where

$$\hat{Q}_A = Q_M + \hat{Q}_0 \quad \text{and} \quad \hat{P}_A = P_M + \hat{P}_0, \tag{4.2}$$

while $\hat{Q}_0 = a + a^\dagger$ and $\hat{P}_0 = \mathrm{i}(a^\dagger - a)$ and $a$, $a^\dagger$ are the creation and annihilation operators of mode $A$. The variables $Q_M$ and $P_M$ are classical stochastic variables following the normal distribution and accounting for the random displacement of the thermal states in the phase space.

Then mode $A$ and Eve's mode $E$ are mixed by a beam splitter (see Fig. 4.1)characterized by transmissivity $\tau$. The outputs of the beam splitter are given by

$$\hat{Q}_B = \sqrt{\tau}\hat{Q}_A + \sqrt{1-\tau}\hat{Q}_E \tag{4.3}$$

$$\hat{P}_B = \sqrt{\tau}\hat{P}_A + \sqrt{1-\tau}\hat{P}_E \tag{4.4}$$

$$\hat{Q}_{E'} = -\sqrt{1-\tau}\hat{Q}_A + \sqrt{\tau}\hat{Q}_E \tag{4.5}$$

$$\hat{P}_{E'} = -\sqrt{1-\tau}\hat{P}_A + \sqrt{\tau}\hat{P}_E. \tag{4.6}$$

We associate to each quadrature operator $\hat{X}$ with variance $V$ a random variable $X$ which describes the its potential measurement outcome following a Gaussian distribution with variance $V$. And then we describe the measurement outcomes in terms of signal and noise

$$Q_B = \sqrt{\tau} Q_M + Q_N, \quad Q_N = \sqrt{\tau} Q_0 + \sqrt{1-\tau} Q_E \tag{4.7}$$

$$P_B = \sqrt{\tau} P_M + P_N, \quad P_N = \sqrt{\tau} P_0 + \sqrt{1-\tau} P_E \tag{4.8}$$

Given that Eve's modes are found in a TMSV states, their CM is given by

$$\mathbf{V}_{Ee} = \begin{pmatrix} \omega & \sqrt{\omega^2-1} \\ \sqrt{\omega^2-1} & \omega \end{pmatrix}, \tag{4.9}$$

so that the variances of the modes $B$ and $E'$ are given by

$$V_{\hat{Q}_B} = \tau(V_M + V_{\text{th}} + V_s) + (1-\tau)\omega \tag{4.10}$$

$$V_{\hat{P}_B} = \tau(V_M + V_{\text{th}} + 1/V_s) + (1-\tau)\omega \tag{4.11}$$

$$V_{\hat{Q}_{E'}} = (1-\tau)(V_M + V_{\text{th}} + V_s) + \tau\omega \tag{4.12}$$

$$V_{\hat{P}_{E'}} = (1-\tau)(V_M + V_{\text{th}} + 1/V_s) + \tau\omega. \tag{4.13}$$

We also group the variances in terms of signal and noise

$$V_{Q_B} = \tau V_M + V_N, \quad V_N = \tau(V_{\text{th}} + V_s) + (1-\tau)\omega = 1 + \tau(V_s - 1 + V_{\text{th}}) + V_\epsilon \tag{4.14}$$

$$V_{P_B} = \tau V_M + V_N, \quad V_N = \tau(V_{\text{th}} + 1/V_s) + (1-\tau)\omega = 1 + \tau(1/V_s - 1 + V_{\text{th}}) + V_\epsilon \tag{4.15}$$

where $V_\epsilon = \tau\epsilon$ is the variance due to the excess noise $\epsilon = \frac{(1-\tau)(\omega-1)}{\tau}$.

As a result, Eve's CM after the propagation of the channel will be given by

$$\mathbf{V}_{E'e} = \begin{pmatrix} V_{E'}\mathbf{I} & \sqrt{\tau\omega^2-1}\mathbf{Z} \\ \sqrt{\tau\omega^2-1}\mathbf{Z} & \omega\mathbf{I} \end{pmatrix}. \tag{4.16}$$

where $V_{E'} = [(1-\tau)(V_M + V_{\text{th}} + 1) + \tau\omega]$ and $V_{\hat{Q}_{E'}} = V_{\hat{P}_{E'}} = V_{E'}$ for $V_s = 1$.

Afterwards Bob Applies a homodyne or heterodyne detection on mode $B$ depending on the version of the protocol. In the case of a homodyne detection, the corresponding conditional variances on Alice's variables $Q_M$ and $P_M$ are given by

$$V_{\hat{Q}_B|Q_M} = \tau(V_{\text{th}} + V_s) + (1-\tau)\omega \tag{4.17}$$

$$V_{\hat{P}_B|P_M} = \tau(V_{\text{th}} + 1/V_s) + (1-\tau)\omega \tag{4.18}$$

$$V_{\hat{Q}_{E'}|Q_M} = (1-\tau)(V_{\text{th}} + V_s) + \tau\omega \tag{4.19}$$

$$V_{\hat{P}_{E'}|P_M} = (1-\tau)(V_{\text{th}} + 1/V_s) + \tau\omega. \tag{4.20}$$

calculated by either using the formula [61]

$$\mathrm{Var}(\hat{Q}|Q) = \mathrm{Var}(\hat{Q}) - \frac{|\langle \hat{Q}Q \rangle|^2}{\mathrm{Var}(Q)}$$

or in our case by simply setting $V_M := 0$ in Eq. (4.10–4.13).

In case of a heterodyne detection, Bob's mode is mixed with the mode $C$ in a balanced beam splitter, which is found in the vacuum state. Then the outputs of the beam splitter $B'$ and $C'$ have the following variances

$$V_{Q_{B'}} = V_{Q_{C'}} = \frac{1}{2}(V_{Q_B} + 1) \tag{4.21}$$

$$V_{P_{B'}} = V_{Q_{C'}} = \frac{1}{2}(V_{P_B} + 1) \tag{4.22}$$

Afterwards conjugate homodyne detections are applied to the two outputs and we obtain the conditional variances

$$V_{Q_{B'}|Q_M} = \frac{1}{2}(\tau(V_{\mathrm{th}} + V_s) + (1-\tau)\omega + 1) \tag{4.23}$$

$$V_{P_{B'}|P_M} = \frac{1}{2}(\tau(V_{\mathrm{th}} + V_s) + (1-\tau)\omega + 1) \tag{4.24}$$

by replacing from Eq. (4.10–4.13) and setting $V_M = 0$.

## 4.3 Secret key rate

The secret key rate in the asymptotic regime is calculated by the mutual information of Alice and Bob minus the accessible information (Holevo function) that Eve has on the Alice's or Bob's variable depending on the reconciliation process. More specifically, the latter is obtained by the difference of the von Neumann entropies corresponding to the total state of Eve after the end of the protocol and to the conditional state of Eve on the given variable. So we are going to present calculations for the four different rates describing the four variations of the protocol depending on Bob's measurement (homodyne vs heterodyne) and on the direction of the reconciliation.

The mutual information is not dependent on the reconciliation direction but is dependent on Bob's measurement. In particular, it is given by

$$I_{\mathrm{hom}} = \frac{1}{2}\log_2 \frac{V_{\hat{Q}_B}}{V_{\hat{Q}_B|Q_M}} = \tag{4.25}$$

$$= \frac{1}{2}\log_2 \frac{\tau(V_M + V_{\mathrm{th}} + V_s) + (1-\tau)\omega}{\tau(V_{\mathrm{th}} + V_s) + (1-\tau)\omega} \tag{4.26}$$

for a protocol with homodyne measurement. Both the quadratures are treated in the same way by this protocol. Thus even though Bob in reality is switching between the quadratures for applying the measurement in order to calculate the mutual information we can assume that Bob is always choosing to measure with respect to one of them (e.g. $\hat{Q}_B$).

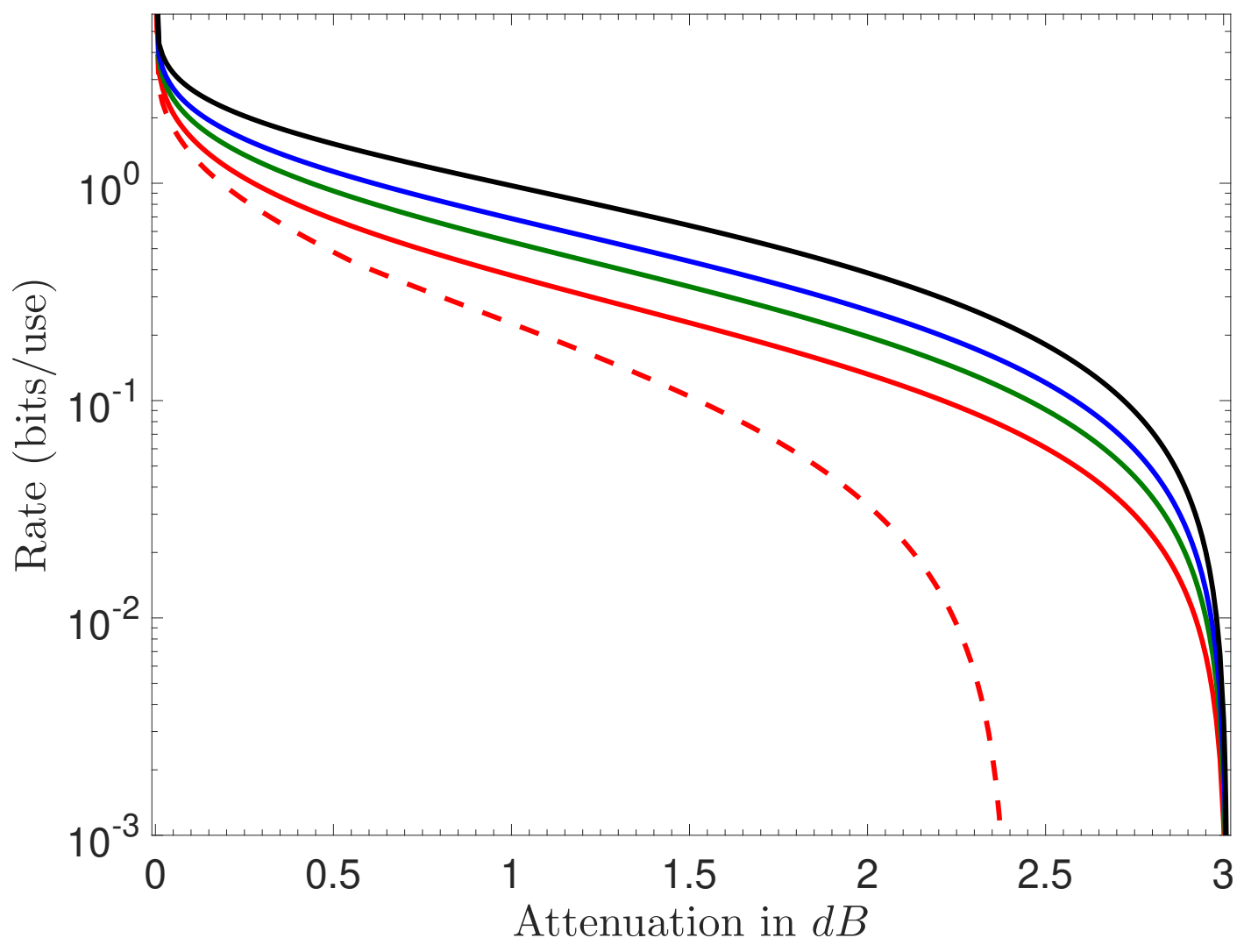

*Figure 4.2: The secret key rate of the direct reconciliation protocol using homodyne detection and thermal states. We assumed a pure loss channel $\omega = 1$, very large modulation $\mu \sim 10^6$ and ideal reconciliation efficiency $\xi = 1$. We have plotted the cases of $V_{th} = 0$ (black line), $V_{th} = 1$ (blue line), $V_{th} = 2$ (green line) and $V_{th} = 4$ (red line). For the sake of comparison, we plotted the case of $\xi = 0.95$ and $\omega = 1.01$ with $V_{th} = 4$ (red dashed line).*

On the other hand, for a protocol with heterodyne detection the contribution of both the quadratures have to be taken into account. Therefore, the formula of the mutual information in that case is given by

$$I_{\text{het}} = \frac{1}{2}\log_2 \frac{V_{\hat{Q}_{B'}}}{V_{\hat{Q}_{B'}|Q_M}} + \frac{1}{2}\log_2 \frac{V_{\hat{P}_{B'}}}{V_{\hat{P}_{B'}|Q_M}} \tag{4.27}$$

$$= \log_2 \frac{\tau(V_M + V_{\text{th}} + V_s) + (1-\tau)\omega + 1}{\tau(V_{\text{th}} + V_s) + (1-\tau)\omega + 1} \tag{4.28}$$

where we have replaced from Eq. (4.21–4.24).

Firstly, we will calculate the von Neumann entropy of Eve's state with the help of the CV $\mathbf{V}_{E'e}$. This quantity is common for every version of the protocol. We first calculate its symplectic eigenvalues obtaining

$$\nu^{\pm}_{E'e} = \frac{1}{2}[\sqrt{(V_{E'} + \omega)^2 - 4\tau(\omega^2 - 1)} \pm (V_{E'} - \omega)] \tag{4.29}$$

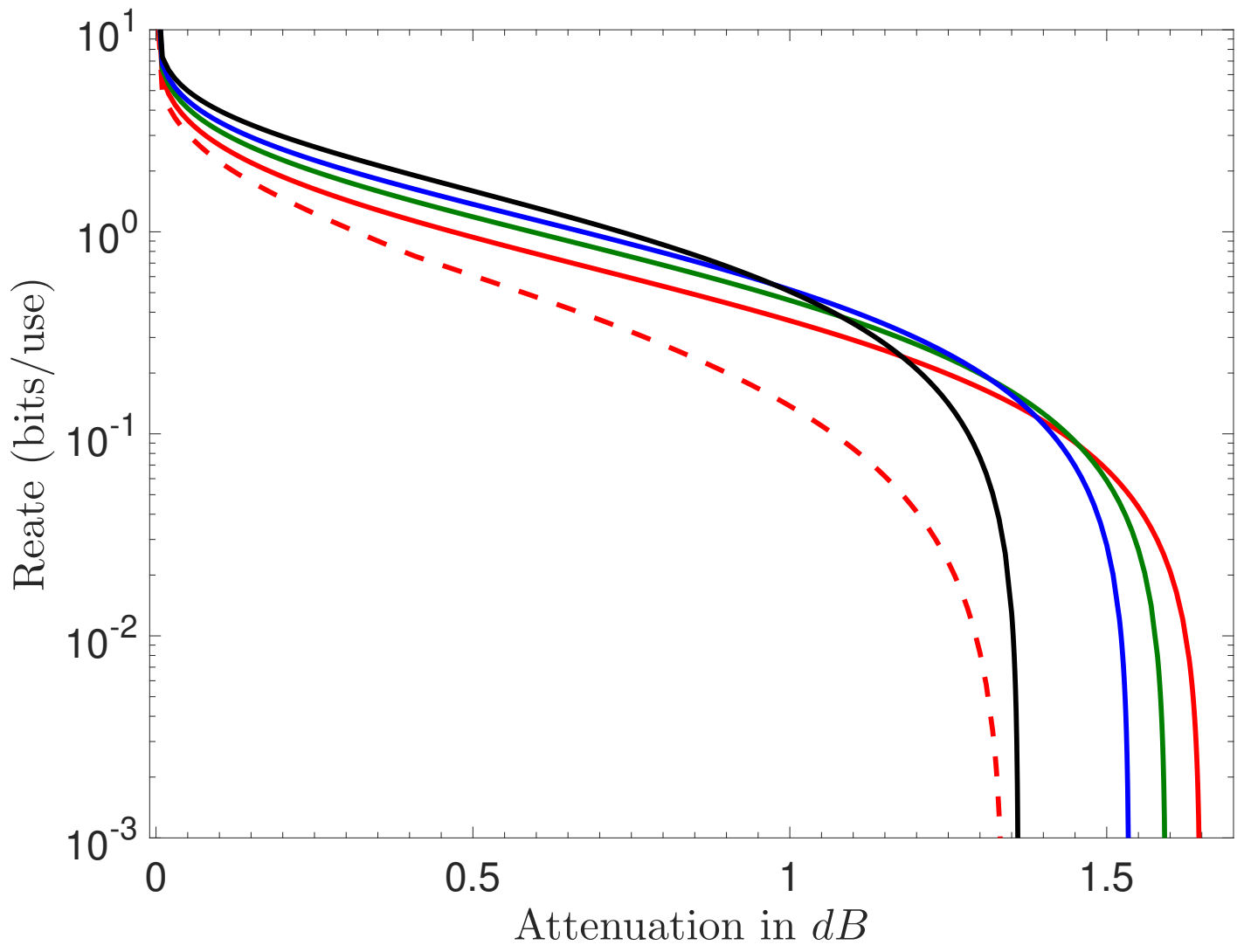

*Figure 4.3: The corresponding secret key rates of Fig. 4.2 for the direct reconciliation protocol with heterodyne detection.*

and the the von Neumann entropy will be given by

$$S(\rho_{E'e}) = h(\nu^{+}_{E'e}) + h(\nu^{-}_{E'e}). \tag{4.30}$$

In contrast, the conditional state for each of the four versions of the protocol is different with different entropies.

**Direct reconciliation with homodyne detection** Eve's CM of the conditional state on Alice's variable $Q_M$ is given by

$$\mathbf{V}_{E'e|Q_M} = \begin{pmatrix} \mathbf{V} & \sqrt{\tau\omega^2 - 1}\mathbf{Z} \\ \sqrt{\tau\omega^2 - 1}\mathbf{Z} & \omega\mathbf{I} \end{pmatrix}. \tag{4.31}$$

with $\mathbf{V} = \text{diag}\{V_{\hat{Q}_{E'}|Q_M}, V_{\hat{P}_{E'}}\}$.

**Direct reconciliation with heterodyne detection** Eve's CM of the conditional state on Alice's variable $Q_M$ is given by

$$\mathbf{V}_{E'e|Q_M P_M} = \begin{pmatrix} \mathbf{V} & \sqrt{\tau\omega^2 - 1}\mathbf{Z} \\ \sqrt{\tau\omega^2 - 1}\mathbf{Z} & \omega\mathbf{I} \end{pmatrix}. \tag{4.32}$$

with $\mathbf{V} = \text{diag}\{V_{\hat{Q}_{E'}|Q_M}, V_{\hat{P}_{E'|Q_M}}\}$.

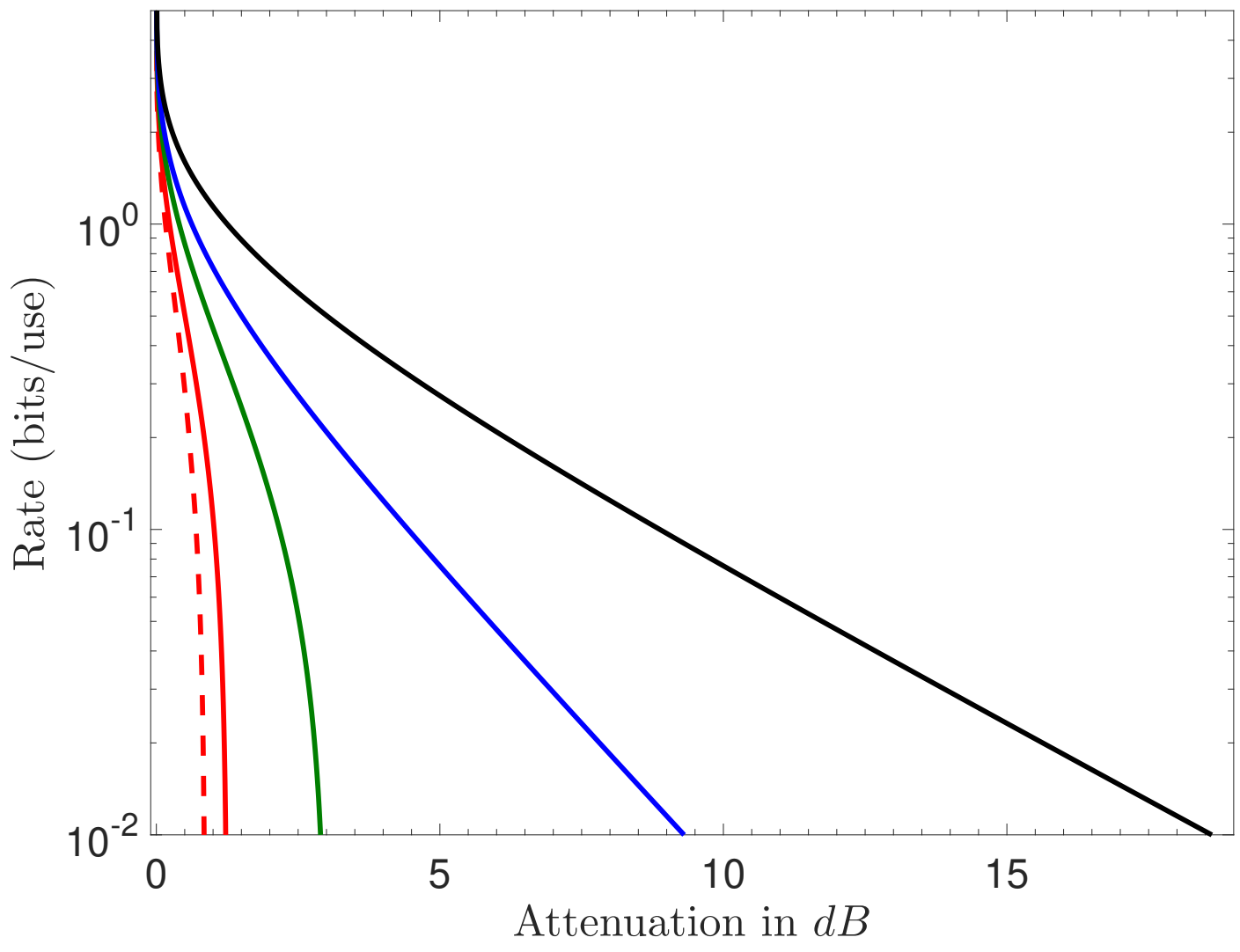

*Figure 4.4: The corresponding secret key rates of Fig. 4.2 for the reverse reconciliation protocol with homodyne detection.*

**Reverse reconciliation with homodyne detection** For the calculation of Eve's conditional matrix on Bob's variable $Q_B$, we apply the rule of homodyne measurement of mode $B$ to the CM

$$\mathbf{V}_{E'eB} = \begin{pmatrix} V_E\mathbf{I} & \sqrt{\tau(\omega^2-1)}\mathbf{Z} & \mathbf{D} \\ \sqrt{\tau(\omega^2-1)}\mathbf{Z} & \omega\mathbf{I} & \mathbf{d} \\ \mathbf{D} & \mathbf{d} & \mathrm{diag}\{V_{\hat{Q}_B}, V_{\hat{P}_B}\}\mathbf{I}, \end{pmatrix} \tag{4.33}$$

where the matrices $\mathbf{D}$ and $\mathbf{d}$ are describing the correlations of mode $B$ with mode $E'$ and $e$ respectively. By using Eq. (4.3–4.6) along with Eq. (4.9), we have

$$\mathbf{D} = \sqrt{\tau(1-\tau)}(V_M + V_{\mathrm{th}} + 1 - \omega)\mathbf{I} \tag{4.34}$$

$$\mathbf{d} = \sqrt{1-\tau}(\omega^2-1)\mathbf{Z}. \tag{4.35}$$

The resulting conditional CM will be

$$\mathbf{V}_{E'e|Q_B} = \begin{pmatrix} \mathbf{A} & \mathbf{C} \\ \mathbf{C} & \mathbf{B} \end{pmatrix}, \tag{4.36}$$

with

$$\mathbf{A} = \begin{pmatrix} \frac{(V_M+V_{\text{th}}+1)\omega}{\tau(V_M+V_{\text{th}}+1-\omega)+\omega} & 0 \\ 0 & \tau(V_M+V_{\text{th}}+1)+(1-\tau)\omega \end{pmatrix}, \tag{4.37}$$

$$\mathbf{B} = \begin{pmatrix} \frac{1-\tau+\tau(V_M+V_{\text{th}}+1)\omega}{\tau(V_M+V_{\text{th}}+1)+(1-\tau)\omega} & 0 \\ 0 & \omega \end{pmatrix}, \tag{4.38}$$

$$\mathbf{C} = \begin{pmatrix} \sqrt{\tau(\omega^2-1}\left(\frac{(V_M+V_{\text{th}}+1)}{\tau(V_M+V_{\text{th}}+1)+(1-\tau)\omega}\right) & 0 \\ 0 & -\sqrt{\tau(\omega^2-1} \end{pmatrix}. \tag{4.39}$$

**Reverse reconciliation with heterodyne detection** Correspondingly for the case of the heterodyne measurement we apply the rule of heterodyne measurement on $\mathbf{V}_{E'eB}$ and we obtain the conditional matrix

$$\mathbf{V}_{E'e|Q_{B'}P_{B'}} = \begin{pmatrix} a\mathbf{I} & c\mathbf{Z} \\ c\mathbf{Z} & b\mathbf{I} \end{pmatrix}, \tag{4.40}$$

with

$$a = \frac{(1-\tau)(V_M+V_{\text{th}}+1)+[\tau+(V_M+V_{\text{th}}+1)]\omega}{1+\tau(V_M+V_{\text{th}}+1)+(1-\tau)\omega}, \tag{4.41}$$

$$b = \frac{(1-\tau)+[1+\tau(V_M+V_{\text{th}}+1)]\omega}{1+\tau(V_M+V_{\text{th}}+1)+(1-\tau)\omega}, \tag{4.42}$$

$$c = \sqrt{\tau(\omega^2-1}\left(\frac{(V_M+V_{\text{th}}+2)}{\tau(V_M+V_{\text{th}}+1)+(1-\tau)\omega+1}\right). \tag{4.43}$$

The secret key rate now is a function of transmissivity $\tau$, modulation variance $\mu = V_M + 1$, channel thermal noise $\omega$ and thermal preparation noise $V_{\text{th}}$ and reconciliation efficiency $\xi$ so that the Eq. (3.45) is modified as follows

$$R(\xi, \mu, \tau, \omega, V_{\text{th}}) = \xi I_{AB} - I_E, \tag{4.44}$$

where the mutual information $I_{AB}$ is given in Eq. (4.25) and the Holevo information $I_E$ is calculated based on the symplectic spectra of the CM in Eq. (4.16) and the covariance matrices in Eq. (4.31), Eq. (4.32), Eq. (4.36) or Eq. (4.40) depending on the version of the protocol.

In Fig. 4.2–4.5, we have calculated numerically and plotted the secret key rates of the four protocols with respect the attenuation of the channel in $dB$ by setting $\tau = 10^{-dB/10}$ for different values of $V_{\text{th}}$. These rates have been calculated for a pure loss channel by setting $\omega = 1$ and assuming a very large modulation $V_M \sim 10^6$ and ideal reconciliation efficiency $\xi = 1$. For the last case of $V_{\text{th}} = 4$, we also plotted the non-ideal case with thermal channel noise, where we have set $\xi = 0.95$ and $\omega = 1.01$.

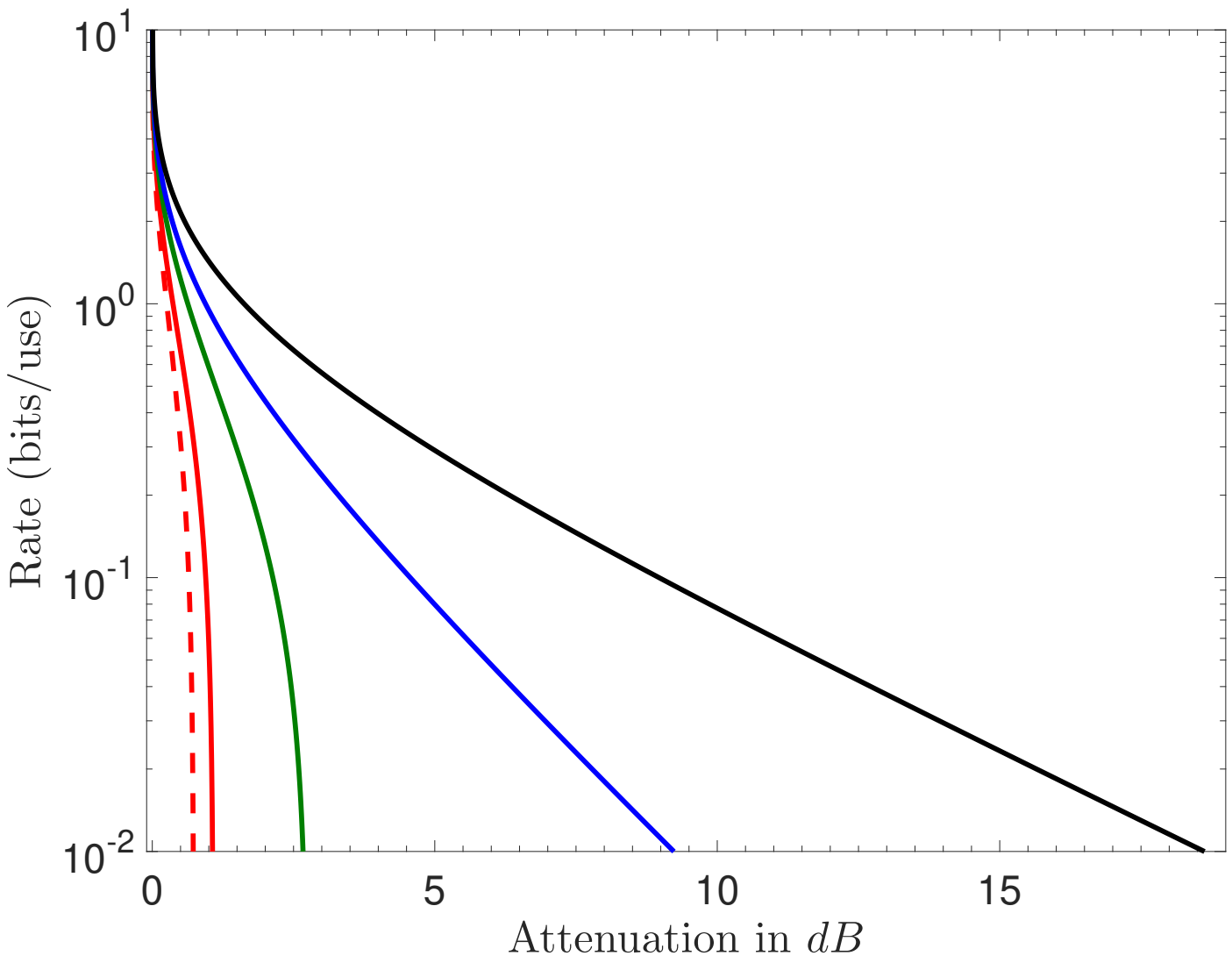

*Figure 4.5: The corresponding secret key rates of Fig. 4.2 for the reverse reconciliation protocol with heterodyne detection.*

## 4.4 Secret key rate for different frequencies

The average photon number of a thermal state is dependent on the frequency $f$ according to the following equation

$$\bar{n}(f) = \frac{1}{\exp[hf/k_B T] - 1}, \tag{4.45}$$

where $h$ is Planck's constant, $k_B$ is Bollzmann's constant and $T$ is the temperature which is set to $T := 300K$ (room temperature). This allows us to make the following substitution $V_{\text{th}} = 2\bar{n}(f)$. Therefore, we can consider secret key rates dependent on frequency and can evaluate the security performance of the protocol for different frequencies. Note that here the shot noise level is higher than this in the optical regime and equal to $V_{\text{th}} + 1$. Therefore the corresponding "pure loss attack" here will be an entangling cloner attack for $\omega = V_{\text{th}} + 1$. We have plotted the security thresholds of each version of the protocol with respect to frequency over transmissivity in Fig. 4.6, where we found that the most robust protocol with respect to lower frequencies in terms of channel transmissivity is the direct reconciliation with homodyne detection one.

## 4.5 Conclusion

In this chapter, we discussed about how the preparation noise manifested as thermal states normally modulated can affect the secret key rate of the one-way protocols. We noticed that the direct reconciliation protocol with homodyne detection is particularly robust to the addition of trusted thermal noise during the preparation of Alice's states. The asymptotic secret key rate calculated here will play an important role for the finite-size security analysis that follows in Chap. 7. By using the connection of the mean photon number of the preparation noise with the frequency of the electromagnetic field, this analysis can be extended to the microwave regime. I would like to mention here that the plots of this chapter is my contribution but for the sake of discussion I believe that are better to be presented here. Although they are my contribution they do not illustrate any recent advancements in the field.

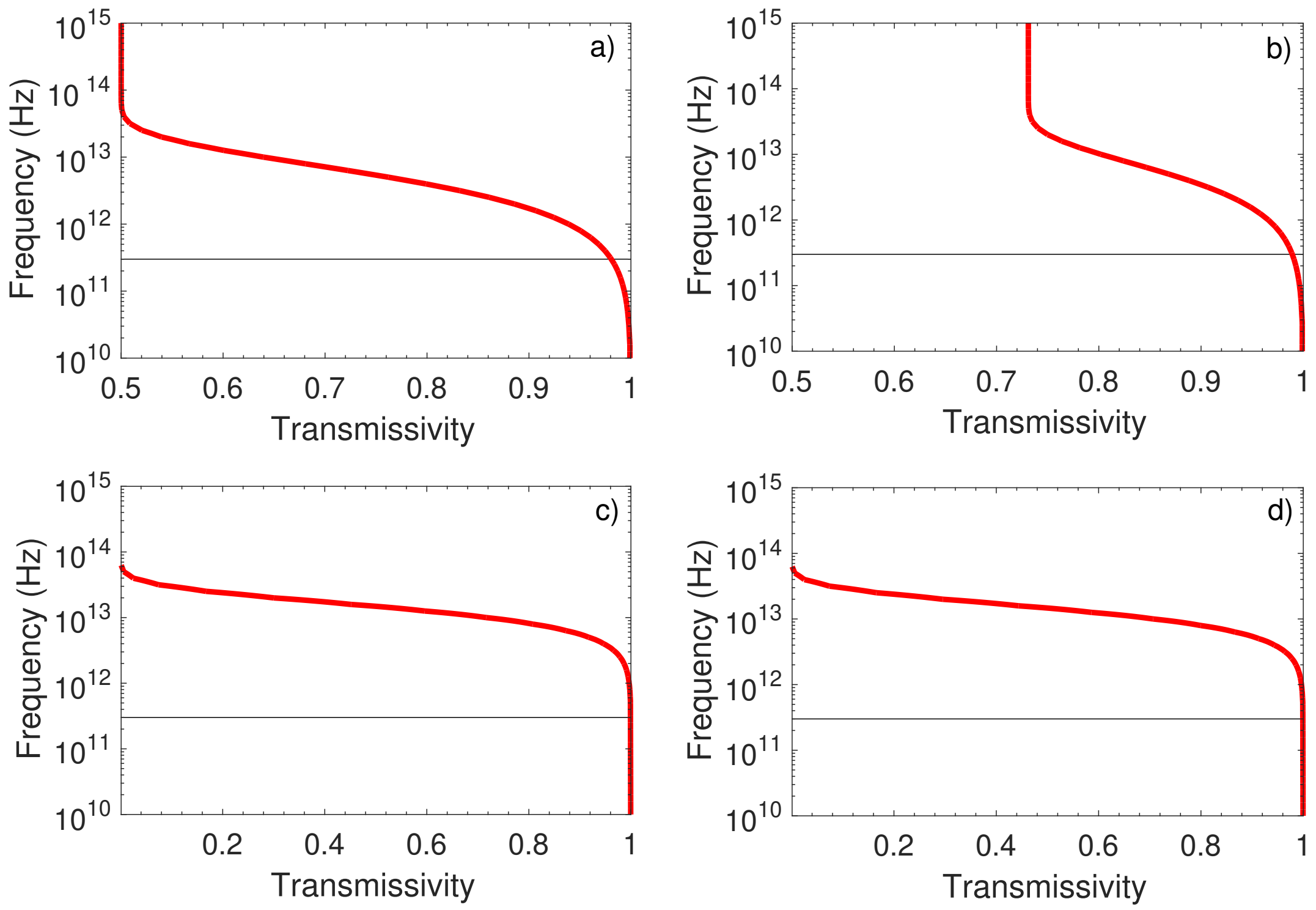

*Figure 4.6: Security thresholds for a) direct reconciliation with homodyne detection, b) direct reconciliation with heterodyne detection, c) reverse reconciliation with homodyne detection and d) direct reconciliation with heterodyne detection with respect the frequency over the transmissivity of the channel. For comparison we have plotted the line (black solid line) beneath we have microwave frequencies.*

# Chapter 5

# Measurement-device-independent quantum key distribution

## 5.1 Introduction

In this chapter we presents results from Ref. [29,30] about CV-QKD with continuous variables. This protocol uses an intermediate relay between Alice and Bob that performs the measurement for the parties and broadcasts the outputs. The realization of the measurements by a third party provides a counter measure on the most prominent of side-channel attacks [27], i.e., attacks on the measurement devices. Most interestingly, a secret key can be extracted even thought the intermediate relay is found under the control of Eve. Side-channels refer to degrees of freedom of the traveling light connected to the key variable that can hide behind any device imperfection [3,28]. They can be exploited by the eavesdropper to learn information about the key without the need of injecting any noise an thus making her untraceable. This protocol also provides with a scheme for network communications as its topology suggests, i.e., parties connect to a central node. Ref. [32,33] introduced such a multi-user protocol. I have done the calculations for Sec. 5.4 based on the references, however the plot presented there is my contribution.

## 5.2 Entanglement based representation of the protocol

Alice and Bob hold a TMSV state each with CM

$$\mathbf{V}_{aA} = \mathbf{V}_{Bb} = \begin{pmatrix} \mu\mathbf{I} & \sqrt{\mu^2-1}\mathbf{Z} \\ \sqrt{\mu^2-1}\mathbf{Z} & \mu\mathbf{I} \end{pmatrix} \tag{5.1}$$

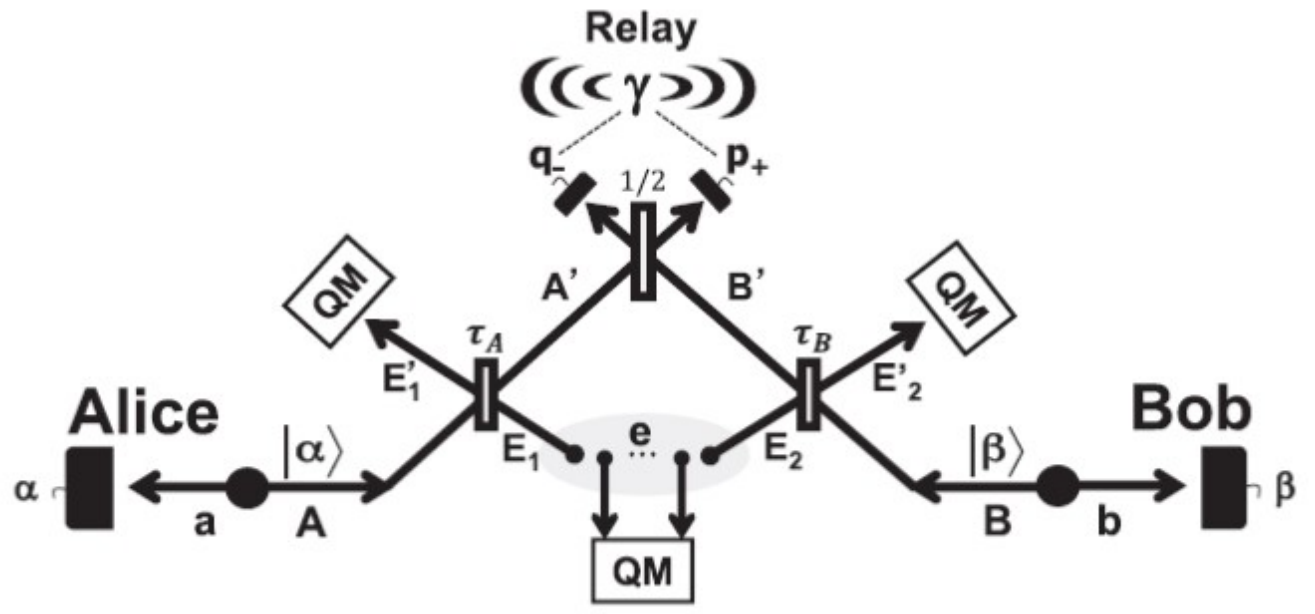

*Figure 5.1: The figure shows the EB representation of CV-MDI QKD. Alice and Bob have TMSV states with modes $(a, A)$ and $(b, B)$. Local modes $a$ and $b$ are kept by the parties, while $A$ and $B$ are sent to the relay through two channels with transmittance $\tau_A$ and $\tau_B$. When Alice and Bob heterodyne the local modes, the traveling ones $A$ and $B$ are projected onto coherent states $|\alpha\rangle$ and $|\beta\rangle$. The relay performs a Bell-measurement and broadcast the outcomes $\gamma$, creating correlation between the parties: for instance, Bob recovers Alice variable $\beta$ by subtracting his variable $\alpha$ from the relay's outputs $\gamma$. The Gaussian attack on the channels is simulated by Eve using ancillas $E_1$ and $E_2$, and thermal noise $\omega_A \geq 1$ and $\omega_B \geq 1$, respectively. These ancillary modes are, in general, two-mode correlated (see text for more details). The ancillary outputs are stored in a quantum memory for a later measurement.*

Then by applying a heterodyne detection to their local modes $a$ and $b$, with outcomes $\tilde{a} = \tilde{q}_\alpha + i\tilde{p}_\alpha$ and $\tilde{\beta} = \tilde{q}_\beta + i\tilde{p}_\beta$, they project the remote modes $A$ and $B$ to coherent states $|\alpha\rangle$ and $|\beta\rangle$. This is a method, called entanglement based representation [8, 55, 61], where a system is projected to a specific state by measuring part of a larger system including the initial one. The outcome variables are connected with the displacements of the coherent states by the following relations

$$\tilde{a} = \eta\alpha^\star,\ \eta = (\mu + 1)(\mu^2 - 1)^{\frac{1}{2}} \tag{5.2}$$

Then the two remote modes interact with Eve's modes $E_1$ and $E_2$ in the beam splitters with transmissivities $\tau_A$ and $\tau_B$ respectively. Eve's modes are described by the CM

$$\mathbf{V}_{E_1E_2} = \begin{pmatrix} \omega_A \mathbf{I} & \mathbf{G} \\ \mathbf{G} & \omega_B \mathbf{I} \end{pmatrix}, \quad \mathbf{G} = \begin{pmatrix} g & 0 \\ 0 & g' \end{pmatrix}. \tag{5.3}$$

where $\omega_A$ and $\omega_B$ quantify Eve's injected thermal noise. This attack is the extension of the entangling cloner attack to an attack including two communication channels. In fact, for $g = g' = 0$ this turns into an entangling cloner attack for each channel. However, for different values of $g$ and $g'$ the two ancillary modes are correlated. This is described more thoroughly in the Supplementary Information of Ref. [29], where are given constraints of the parameters $g$ and $g'$. Due to the fact that we use the Holevo information for quantifying Eve's information (see remarks after Def. 2.6.4), we need to take into account only Eve's modes that interact with the remote modes of Alice and Bob.

The output modes of the beam splitters are then $A'$, $B'$, $E_1'$ and $E_2'$ as it is shown in Fig. 5.1. Therefore, global input state is then described by the CM

$$\mathbf{V}_{aAE_1E_2Bb} = \mathbf{V}_{aA} \oplus \mathbf{V}_{E_1,E_2} \oplus \mathbf{V}_{Bb} \tag{5.4}$$

and the symplectic matrix of the action of the two beam splitters is described by

$$\mathbf{S}_1(\tau_A, \tau_B) = \mathbf{I} \oplus \mathbf{B}(\tau_A) \oplus \mathbf{B}(\tau_B)^T \oplus \mathbf{I}, \tag{5.5}$$

where $\mathbf{B}(\tau)$ is given in Eq. (2.67). Then the state of the output modes $aA'E_1'E_2'B'b$ has a CM

$$\mathbf{V}_{aA'E_1'E_2'B'b} = \mathbf{S}_1(\tau_A, \tau_B)\mathbf{V}_{aAE_1E_2Bb}\mathbf{S}_1(\tau_A, \tau_B)^T. \tag{5.6}$$

Then a balanced beam splitter action is applied on modes $A'$ and $B'$ . Thus we reordering the modes in the previous CM obtaining the CM $\mathbf{V}_{abA'B'E_1E_2}$ and then apply the symplectic transformation

$$\mathbf{S}_2 = \mathbf{I} \oplus \mathbf{I} \oplus \mathbf{B}(1/2) \oplus \mathbf{I} \tag{5.7}$$

from which we obtain

$$\mathbf{V}_{abA''B''E_1E_2} = \mathbf{S}_2\mathbf{V}_{abA'B'E_1E_2}\mathbf{S}_2^T. \tag{5.8}$$

We eliminate modes $E_1$ and $E_2$ from the previous CM and apply two conjugate homodyne detections on modes $A''$ and $B''$ given in Eq. (2.71) accounting for the Bell detection in the relay. This leads to a CM

$$\begin{aligned}\mathbf{V}_{ab|\gamma} = &\begin{pmatrix} \mu\mathbf{I} & \mathbf{0} \\ \mathbf{0} & \mu\mathbf{I} \end{pmatrix} - (\mu^2-1)\times \\ &\times \begin{pmatrix} \tau_A/\theta & 0 & -\sqrt{\tau_A\tau_B}/\theta & 0 \\ 0 & \tau_A/\theta' & 0 & \sqrt{\tau_A\tau_B}/\theta' \\ -\sqrt{\tau_A\tau_B}/\theta & 0 & \tau_B/\theta & 0 \\ 0 & \sqrt{\tau_A\tau_B}/\theta' & 0 & \tau_B/\theta' \end{pmatrix}\end{aligned} \tag{5.9}$$

with

$$\theta = (\tau_A + \tau_B)\mu + \lambda \quad \text{and} \quad \theta' = (\tau_A + \tau_B)\mu + \lambda', \tag{5.10}$$

where

$$\lambda =(1-\tau_A)\omega_A + (1-\tau_B)\omega_B - 2g\sqrt{(1-\tau_A)(1-\tau_B)}, \tag{5.11}$$

$$\lambda' =(1-\tau_A)\omega_A + (1-\tau_B)\omega_B + 2g'\sqrt{(1-\tau_A)(1-\tau_B)}. \tag{5.12}$$

Subsequently, we reorder the modes of the previous CM obtaining $\mathbf{V}_{ba|\gamma}$ and we apply a heterodyne measurement as in Eq. (2.72) in order to calculate the conditional CM on Alice's variable

$$\mathbf{V}_{b|\gamma\tilde{\alpha}} = \begin{pmatrix} \mu - \frac{\mu^2-1}{\tau_A+\tau_B\mu+\lambda} & 0 \\ 0 & \mu - \frac{\mu^2-1}{\tau_A+\tau_B\mu+\lambda'} \end{pmatrix} \tag{5.13}$$

A more detailed calculation of the previous covariance matrices is found in the Supplemental Material of Ref. [29].

## 5.3 Secret key rate

The mutual information between Alice and Bob are dependant on both the quadratures which have been modulated independently. Thus we have that

$$\begin{aligned} I(q_{\tilde{\alpha}}, p_{\tilde{\alpha}}, q_{\tilde{\beta}}, p_{\tilde{\beta}}|\gamma) =& \frac{1}{2}\log_2 \frac{V_{q_{\tilde{\beta}|\gamma}}}{V_{q_{\tilde{\beta}|\gamma\tilde{\alpha}}}} + \frac{1}{2}\log_2 \frac{V_{p_{\tilde{\beta}|\gamma}}}{V_{p_{\tilde{\beta}|\gamma\tilde{\alpha}}}} \\ =& \frac{1}{2}\log_2 \frac{\frac{1}{2}(V_{q_{b|\gamma}}+1)}{\frac{1}{2}(V_{q_{b|\gamma\tilde{\alpha}}}+1)} + \frac{1}{2}\log_2 \frac{\frac{1}{2}(V_{p_{b|\gamma}}+1)}{\frac{1}{2}(V_{p_{b|\gamma\tilde{\alpha}}}+1)} = \\ =& \frac{1}{2}\log_2 \frac{V_{q_{b|\gamma}}+1}{V_{q_{b|\gamma\tilde{\alpha}}}+1} + \frac{1}{2}\log_2 \frac{V_{p_{b|\gamma}}+1}{V_{p_{b|\gamma\tilde{\alpha}}}+1} \\ =& \frac{1}{2}\log_2 \Sigma. \end{aligned} \tag{5.14}$$

The first equality holds because the variables are Gaussian (see Eq. (2.76)) while the second emerges from the fact that the previous variables are outcomes of a heterodyne detection of the corresponding modes with variances given by $\mathbf{V}_{ab|\gamma}$ and $\Sigma = \frac{1+\det \mathbf{V}_{b|\gamma}+\mathrm{Tr}\mathbf{V}_{b|\gamma}}{1+\det \mathbf{V}_{b|\gamma\tilde{\alpha}}+\mathrm{Tr}\mathbf{V}_{b|\gamma\tilde{\alpha}}}$.

The Holevo function is calculated by

$$\chi(E_1E_2 : \tilde{\alpha}|\gamma) = S(\rho_{E_1E_2|\gamma}) - S(\rho_{E_1E_2|\gamma\tilde{\alpha}}) \tag{5.15}$$

Since Eve holds a purification of Alice and Bob (see remark after Def. 2.6.3) and the measurements applied for obtaining the conditional states are rank-one operations, the following relations should hold

$$S(\rho_{E_1E_2|\gamma}) = S(\rho_{ab|\gamma}) \quad \text{and} \quad S(\rho_{E_1E_2|\gamma\tilde{\alpha}}) = S(\rho_{b|\gamma\tilde{\alpha}}). \tag{5.16}$$

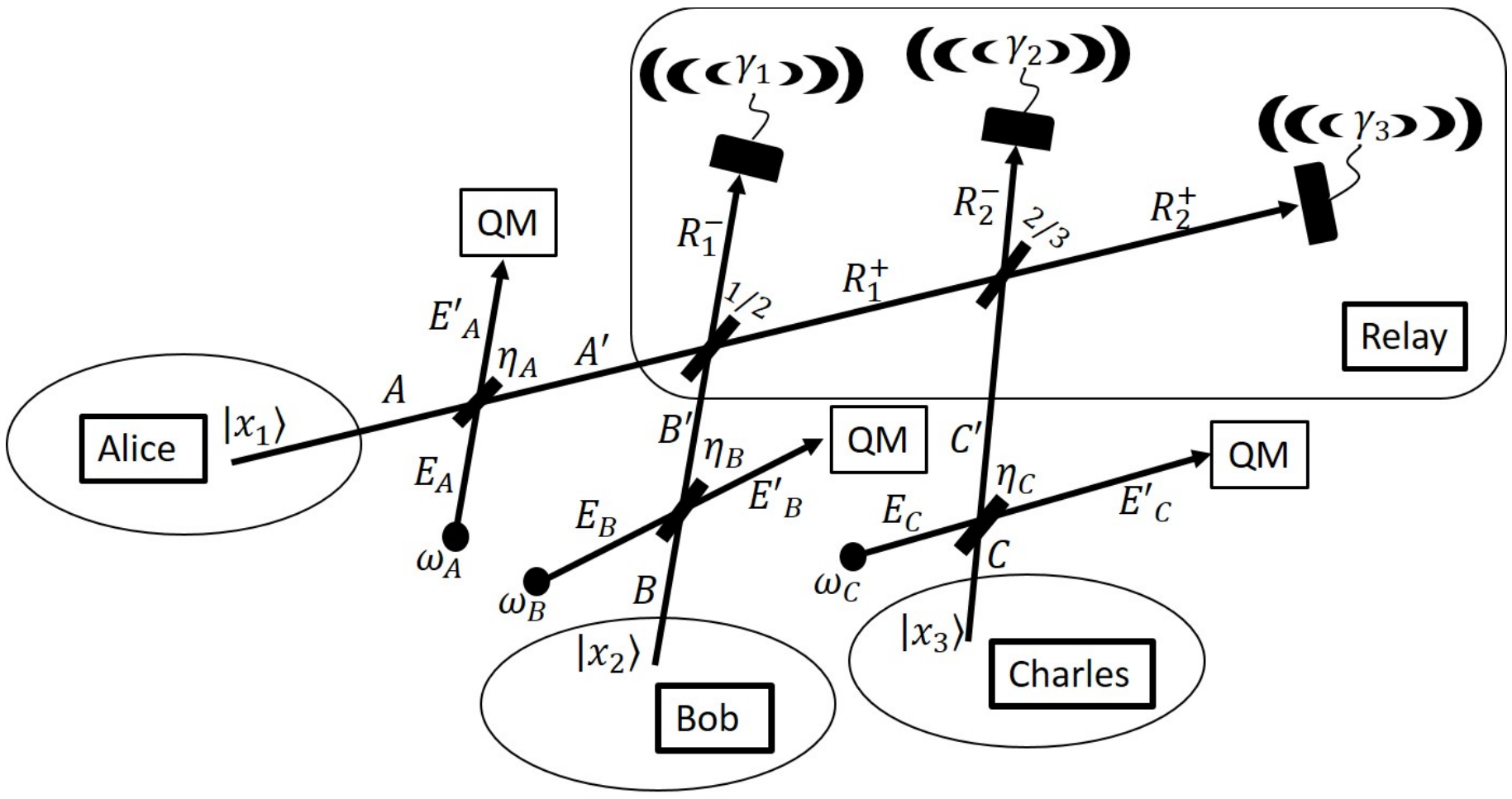

*Figure 5.2: parties prepare coherent states in the modes A, B and C whose mean values are Gaussian variables* $\mathbf{x}_1$, $\mathbf{x}_2$ *and* $\mathbf{x}_3$, *with variance* $V_M$. *Then the parties send these states to the relay using channels with transmissivities* $\eta_A$, $\eta_B$ *and* $\eta_C$ *respectively.After traveling through the channels the modes arrive at the relay as modes* $A'$, $B'$ *and* $C'$ *and processed by the relay. Although the relay is under the full control of the eavesdropper, we can assume without loss of generality that it operates consistently in each use of the channel: (a) firstly it mixes Alice's and Bob's modes in a beam splitter with transmissivity* $1/2$ *and measures the q-quadrature of mode* $R_1^-$ *with a homodyne detection, (b) subsequently mixes Charlie's mode with* $R_1^+$ *and then measures the q-quadrature and p-quadrature of modes* $R_2^-$ *and* $R_2^+$ *respectively, (c) finally the results of the measurements* $\gamma_1$, $\gamma_2$ *and* $\gamma_3$ *are broadcast. Any general attack affecting both the channels and the relay can be reduced to an attack tampering only with the channels. In this case, Eve is injecting thermal noise* $\omega_1$, $\omega_2$ *and* $\omega_3$ *in each of the channels by means of the modes* $E_A$, $E_B$ *and* $E_C$ *interacting with modes* $A$, $B$, $C$. *In a general Gaussian attack, Eve's modes are described by a correlated Gaussian state whose CM is specified in Eq.* (5.18). *The scheme is described in more detail in the Supplementary Information of Ref. [33].*

This allows us to calculate the Holevo information to based only on the covariance matrices $\mathbf{V}_{ab|\gamma}$ and $\mathbf{V}_{b|\gamma\tilde{\alpha}}$ and more specifically on their symplectic spectra (see Eq. (2.78)).

The secret key rate now is a function of the channel parameters such as transmissivity $\tau$, the attack parameters $\omega_A$, $\omega_B$, $g$, $g'$ quantifying the channel thermal noise and the modulation variance $\mu = V_M + 1$ as well as the reconciliation efficiency $\xi$ so that the relation of the secret key rate corresponding to that of Eq. (3.45) is expressed as follows

$$R(\xi, \mu, \tau_A, \tau_B, \omega_A, \omega_B, g, g') = \xi I_{AB} - I_E, \tag{5.17}$$

where $I_{AB}$ is the mutual information from Eq. (5.14) and the Holevo information $I_E$ is taken from Eq. (5.15). We have plotted the secret key rate for symmetric and asymmetric configuration in Fig. 8.1.

## 5.4 CV-MDI-QKD three-user network

In the three-user setting we assume three users Alice, Bob and Charlie using an intermediate relay in order to extract a common secret key. Each one prepares in a mode $A$, $B$ and $C$ respectively in a coherent state with amplitude modulated by the Gaussian random variables $x_1$, $x_2$, and $x_3$ with the same variance $V_m$. These modes propagate through three channels described by lossy channels with transmissivities $\tau_A$, $\tau_B$ and $\tau_C$ respectively and arrive to the relay as illustrated in Fig. 5.2. In each channel use, the relay is assumed to operate according to a certain way. Firstly, it mixes Alice's and Bob's modes in a beam splitter with transmissivity $\tau_1 = 1/2$. Then it homodynes the output mode $R_1^-$ with respect the $q$ quadrature. On the other hand, mode $R_1^+$ is mixed with Charlie's mode in a beam splitter with transmissivity $\tau_2 = 2/3$. Subsequently, the output modes $R_2^-$ and $R_2^+$ are conjugately homodyned. All the measurement outcomes $\boldsymbol{\gamma} = (\gamma_1, \gamma_2, \gamma_3)$ are then broadcast. Afterwards, the distributed classical correlations can be processed by the three parties to extract a common secret key for secure quantum conferencing Ref. [32,33]. The scheme is generalized for $N$ number of users and described in detail in the Supplemental Information of Ref. [33] along with the previous multi-mode Bell-detection.

Any attack that may include a relay that does not work as described above [29] can be broken to an attack on the channels plus a proper working relay. More specifically, the correlated attack among all the three channels can be described by the CM

$$\mathbf{V}_{E_A E_B E_C} = \begin{pmatrix} \omega_A \mathbf{I} & \mathbf{G}_1 & \mathbf{G}_3 \\ \mathbf{G}_1 & \omega_B \mathbf{I} & \mathbf{G}_2 \\ \mathbf{G}_3 & \mathbf{G}_2 & \omega_C \mathbf{I} \end{pmatrix}, \tag{5.18}$$

where $\omega_A$, $\omega_B$ and $\omega_C$ is the noise injected by the eavesdropper in each channel respectively. In addition, $\mathbf{G}_i = \mathrm{diag}(g_i, g_i')$ describes the correlations between the modes. When $g_i$ and $g_i'$ are equal to zero, the attack is reduced to an uncorrelated attack, which is the case that we investigate in this study.

In terms of the security analysis, we assume that the modes $A$, $B$ and $C$ are each one half of an TMSV state with parameter $\mu = V_M + 1$ (entanglement based representation). Heterodyne measurements are applied to the local TMSV modes $a$, $b$, and $c$ so that the traveling modes are projected onto modulated coherent states. In this representation, Eve's Holevo bound $I_E$ is given by the symplectic eigenvalues of the CM $\mathbf{V}_{abc|\boldsymbol{\gamma}}$ of Eve's average state and the conditional state CM $\mathbf{V}_{bc|\boldsymbol{\gamma},x_1}$ following the reasoning in Sec. 5.3. More specifically, the total CM is given by the CM of the parties' local modes after the application of the three relay measurements with outcome $\boldsymbol{\gamma}$. Since we assume reconciliation over Alice's variable the conditional CM is derived by the total CM after applying a heterodyne detection on mode $a$. The explicit expressions are found in Ref. [32].

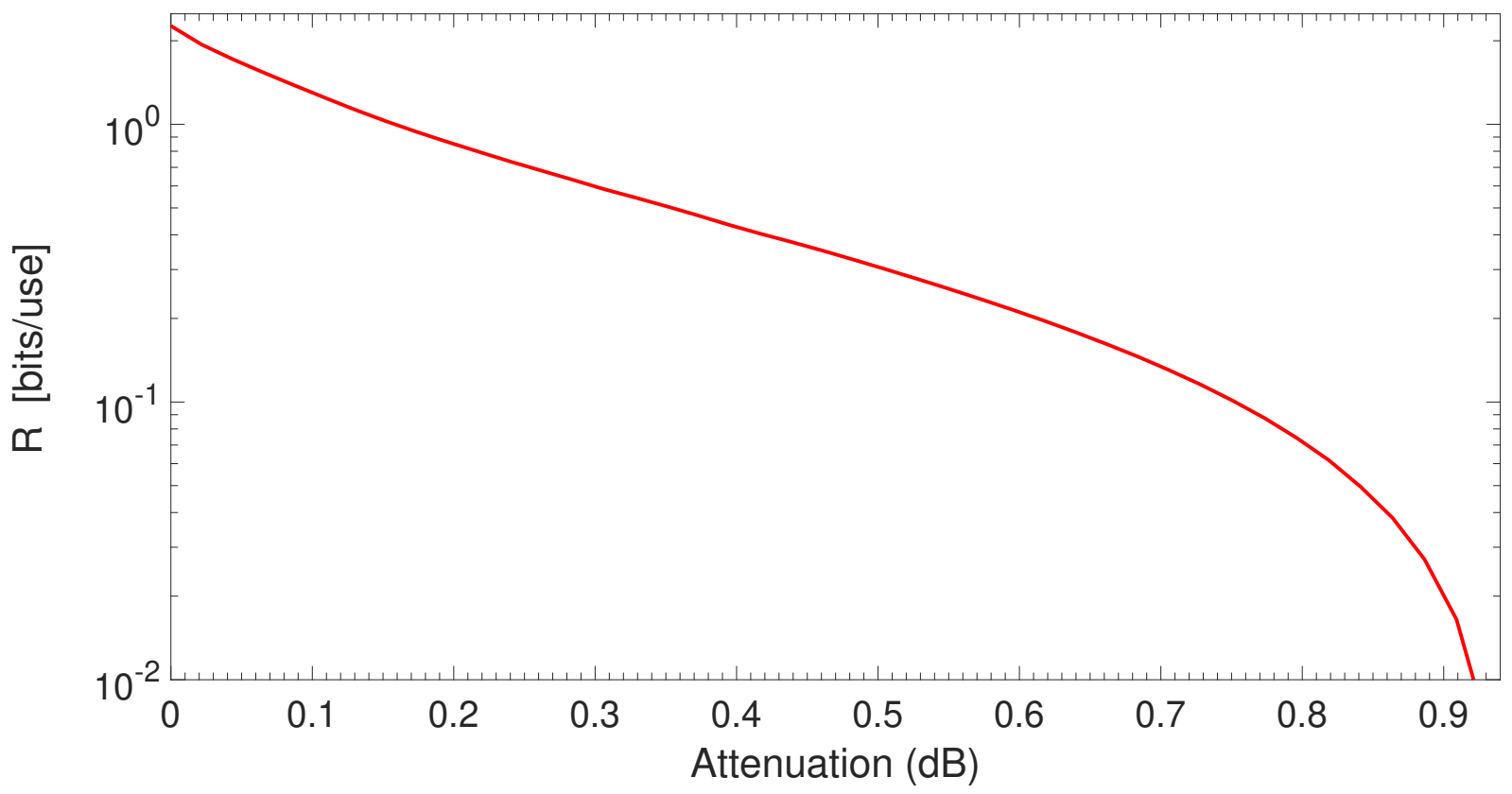

*Figure 5.3: Secret conferencing key rate versus $\eta$ for the three-party star configuration protocol based on the CV-MDI-QKD scheme, optimized over $\mu$.*

The mutual information in the case of the three parties is defined as the minimum of the mutual information between Alice-Bob and Alice-Charlie, i.e., $I_{\min} = \min\{I_{AB}, I_{AC}\}$. Each of the terms are evaluated by the formula $I_{AB(C)} = \frac{1}{2}\log_2 \boldsymbol{\Sigma}_{b(c)}$, where we have the following (for $m = b$ or $c$)

$$\boldsymbol{\Sigma}_m = \frac{\det(\mathbf{V}_{m|\boldsymbol{\gamma}}) + \mathrm{tr}(\mathbf{V}_{m|\boldsymbol{\gamma}}) + 1}{\det(\mathbf{V}_{m|\boldsymbol{\gamma},x_1}) + \mathrm{tr}(\mathbf{V}_{m|\boldsymbol{\gamma},x_1}) + 1}, \tag{5.19}$$

in terms of the covariance matrices of the local mode $m$.

As a result, the secret conferencing key rate is given by

$$R(\mu, \boldsymbol{\omega}, \boldsymbol{\eta}) = \xi I_{\min}(\mu, \boldsymbol{\omega}, \boldsymbol{\eta}) - I_E(\mu, \boldsymbol{\omega}, \boldsymbol{\eta}), \tag{5.20}$$

where $\boldsymbol{\omega} = (\omega_A, \omega_B, \omega_C)$ is the vector of the noise injected by the eavesdropper in each channel with corresponding transmissivity $\boldsymbol{\eta} = (\eta_A, \eta_B, \eta_C)$. Numerically, we have checked that, even for ideal reconciliation $\xi = 1$, the largest value of $\mu$ is not the optimal and we have therefore to optimize the rate over $\mu$.

Note that an asymmetric scenario where all the transmissivities and noises are different can be reduced to a symmetric scenario where we consider the lowest transmissivity and the highest noise for all the channels. This clearly gives a lower bound to the achievable key rate. In such a star configuration with identical channels the previous rate simplifies to

$$R^{\text{star}}(\mu, \omega, \eta) = \xi I_{\min}(\mu, \omega, \eta) - I_E(\mu, \omega, \eta), \tag{5.21}$$

where $\eta_A = \eta_B = \eta_C := \eta$ and $\omega_A = \omega_B = \omega_C := \omega$. This is plotted in Fig. 5.3 for passive eavesdropping ($\omega = 1$).

## 5.5 Conclusion

In this chapter, we presented the MDI protocol, where the two parties encode their variables to coherent states and send them to a central relay. The relay applies joint measurements to the aforementioned modes and broadcasts the outcome. The parties correlate their variables using the measurement outcomes of a central relay. This protects their communication from side-channel attacks on the detection apparatus. Due to the presence of the relay, the protocols configuration can be the basis for network settings with multiple users.

In the following chapters, we are going to present a security analysis incorporating finite-size effects and a composable security analysis assuming coherent attacks for the two-user CV-MDI scheme. Afterwards, we are going to extend the asymptotic security analysis of a three user instance of the multi-party CV-MDI scheme for the star configuration assuming fast fading channels with uniform distribution.

# Chapter 6

# Finite-size effects

## 6.1 Introduction

In the previous discussion we considered the case of the asymptotic secret key rate, i.e. assuming an infinite number of channel uses. This is the first approach that someone should adopt in order to assess the security of a protocol in the sense that if there is no security for an infinite number of signal exchange, there is no need to investigate the more realistic case of a finite exchange of signals.

Then the secret key rate in this case is modified in order to include the changes due to the previous assumption. These modifications are called finite-size effects and in general deteriorate the performance of a protocol. This is expected since these assumptions simulate realistic conditions and drive the investigation of the performance of a protocol away from an idealistic description as this of an infinite signal exchange between the honest parties. This deterioration is connected to the the number of signal states used for the communication between the honest parties. As this number is increasing and tending towards infinity, this decline disappears and one obtains the protocol performance of the asymptotic case.

## 6.2 Non-asymptotic security analysis

The secret key rate incorporating finite-size effects was firstly calculated in Ref. [53] and for the continuous variables protocol it turns out to be [54]

$$K = \frac{n}{n+m}\left(\xi \tilde{I}_{AB} - \tilde{I}_E - \Delta(n)\right). \tag{6.1}$$

The quantities $\tilde{I}_{AB}$ and $\tilde{I}_E$ are the same as for the asymptotic key rate (see Eq. (3.44)), namely the classical mutual information between Alice and Bob and the Holevo information for Eve's system on Alice's or Bob's variable depending on the reconciliation direction. However, there is a difference in the calculation of these quantities since they are dependent on the channel parameter estimation process which becomes relevant in the finite-size regime.

This is because the channel parameters, transmissivity $\tau$ and excess noise variance $V_\epsilon = (\tau - 1)(\omega - 1)$ (see entangling cloner attack in Sec. (2.5.1.1)), are not known and have to be estimated by sacrificing part of the signal states. In particular, part of the signal states is used for this estimation and has to be revealed and does not contribute to the secret key extraction (see Fig. (6.1)). By this process, Alice and Bob constrain the uncertainty of the channel parameters and thus can provide with a tighter bound for Eve's knowledge on their variables. This is in contrast with the asymptotic case, where in principle there is an infinite number of signals for the channel parameter estimation implying that the honest parties know the channel parameters with certainty.

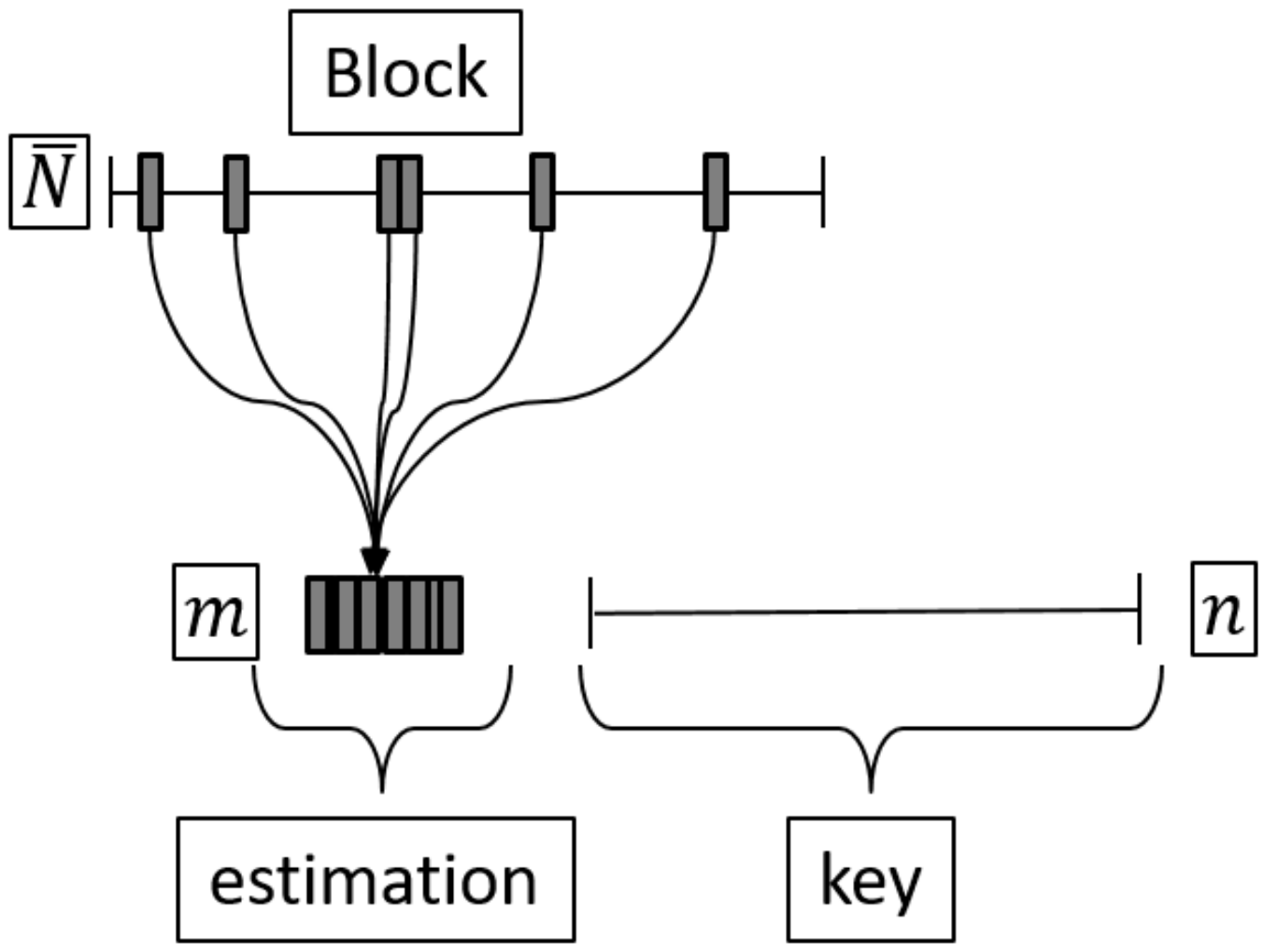

*Figure 6.1: This figure shows the procedure of channel parameter estimation. From a block of $\bar{N}$ signals states send through the channel Alice and Bob chose randomly and reveal the measurement outcomes of $m$ of them. These contribute to the channel parameter estimation, i.e., to statistically create confidence intervals for parameters such as the transmissivity $\tau$ and the excess noise variance $V_\epsilon$. The rest $n$ are not revealed and contribute to the extraction of the secret key. As $m$ increases the channel parameters are estimated more accurately and this results to a calculation of a tighter bound for the secret key rate meaning that the parties will end up with a larger key for a given number of $n$ states via the privacy amplification. On the other hand, if $\bar{N}$ is fixed, larger $m$ means smaller $n$ that results in a smaller key. In order to find the balance for this trade off we have to optimize over the ratio $r = m/\bar{N}$.*

In the finite-size effects regime, it is important that a balance exists between the number of signals used for key extraction and the channel parameter estimation. As $m$ increases the channel parameters are estimated more accurately and this results to a calculation of a tighter bound for the secret key rate meaning that the parties will end up with a larger key for a given number of $n$ states via the privacy amplification. On the other hand, if $\bar{N}$ is fixed, larger $m$ means smaller $n$ that results in a smaller key. An analysis taking into account this fact was presented in Ref. [15]. Here we give a summary of this study and in the following chapters of this thesis we modify it in an analogous so as to be suitable for other protocols.

In order to see the effect of the parameter estimation, we assume $m$ realizations of signal states that their measurement outcomes will be used in the channel parameter estimation. These outcomes follow Gaussian distributions since the protocols in discussion are the fully Gaussian one-way protocols (see Chap. 3) and the attack is the entangling cloner attack. These outcomes can be considered as samples from which we can define maximum likelihood estimators for the parameters of interest. For these estimators, we calculate their mean and variance and obtain confidence intervals for them. From these intervals, we choose the values that have the largest negative effect on the secret key rate, which accounts the for the worst case scenario for Alice and Bob in terms of security.

Furthermore, the relation of the secret key rate with finite size effects takes into consideration the fact that the Holevo information function is not working correctly in a finite size effects regime. Thus, there is an extra term that accounts for this correction, which is dependent on the number of signals used for key extraction. As presented in Ref. [14], this term is approximately calculated to be

$$\Delta(n) \sim (2 \times 2^d + 3)\sqrt{\frac{\log_2(2/\epsilon_{\rm sm})}{n}}, \tag{6.2}$$

where $\epsilon_{\rm sm}$ is the smoothing parameter that has to do with the use of Holevo information in an non-asymptotic analysis, and $d$ is the discretization parameter corresponding to $2^d$ intervals used by the parties to discretize their variables. In the most cases, we assume that the key is given in terms of bits and set $d = 1$. In theory, the parameter $\epsilon_{\rm sm}$ is assumed to be optimized for maximizing the rate. However, this makes a negligible difference and thus we set it to be approximately $10^{-10}$. In the following section, we give the explicit steps for the channel parameter estimation.

## 6.3 Channel parameter estimation

In order to achieve the channel parameter estimation we first describe the channel output variable in terms of signal and noise

$$y = \sqrt{\tau}x + z,$$

where $\tau$ is the transmissivity of the channel $x$ is the variable describing the signal and $z$ the variable describing the noise. Both variables are independent and in our case follow Gaussian distributions. In fact, the actual relation is given by equation Eq. (3.11), where have we set $y \to Q_B$, $x \to Q_M$ and $z \to Q_N$. Then we can define estimators for the covariance of the signal variable $x$ with the channel output variable $y$ and the variance of the variable $z = y - \sqrt{\tau}x$. We then use these estimators to define estimators for the transmissivity and the variance of the excess noise.

### 6.3.1 Variance of the transmissivity

Let us assume $m$ realizations of the Alice's variable $Q_M$ denoted by $M_i$ for $i = 1, 2, \dots, m$. These correspond to $m$ realization of Bob's quadrature variable $Q_B$ given in Eq.(3.11). Both $Q_M$ and $Q_B$ follow normal distributions with variances $V_M$ and $V_B$ and zero mean and can be seen as members of a bivariate normal distribution. Thus, we can define the maximum likelihood estimator [51] of the covariance $\mathrm{Cov}(Q_M, Q_B) = \sqrt{\tau}V_M := C_{MB}$ with the help of the previous realizations by

$$\tilde{C}_{MB} = \frac{1}{m}\sum_{i=1}^{m} M_i B_i \tag{6.3}$$

This estimator is normally distributed as the sample mean of the variable $Q_M Q_B$ according to the central limit theorem. Its mean value is

$$\mathbb{E}(\tilde{C}_{MB}) = \frac{1}{m}\sum_{i=1}^{m} \mathbb{E}(M_i B_i) = \mathbb{E}(Q_M, Q_B) = \sqrt{\tau}V_M = C_{MB} \tag{6.4}$$

where the second equality holds because each of the realizations $M_i$ ($B_i$) follows the same distribution as $Q_M$ ($Q_B$). The variance of the estimator is given by

$$\mathrm{Var}(\tilde{C}_{MB}) = \frac{1}{m^2}\sum_{i=1}^{m} \mathrm{Var}(\mathbb{E}(M_i B_i)) = \frac{1}{m}\mathrm{Var}(Q_M \hat{Q}_B) \tag{6.5}$$

$$= \frac{1}{m}\mathrm{Var}(Q_M, Q_B) = \frac{1}{m}(2\tau V_M^2 + V_M V_N) := V_{C_{MB}}, \tag{6.6}$$

where we have replaced $Q_B$ and used the moments of the Gaussian distribution.

The relation $\sqrt{\tau}V_M := C_{MB}$ can also be expressed as

$$\tau = \frac{1}{V_M^2}(C_{MB})^2. \tag{6.7}$$

We assume that the variance $V_M$ is known to the parties since it describes Alice's state preparation and thus we only replace with the estimators of $\tau$ and $C_{MB}$ in order to obtain an expression for the estimator of $\tau$ given by

$$\tilde{\tau} = \frac{1}{V_M}(\tilde{C}_{MB})^2 = \frac{V_{C_{MB}}}{V_M^2}\left(\frac{\tilde{C}_{MB}}{\sqrt{V_{C_{MB}}}}\right)^2. \tag{6.8}$$

From this equation, we can calculate the variance and the mean value of the transmissivity estimator by using the fact that $\frac{\tilde{C}_{MB}}{\sqrt{V_{C_{MB}}}}$ is following a standard normal distribution. According to Eq. (A.3.4), this means that its square is chi-squared distributed

$$\left(\frac{\tilde{C}_{MB}}{\sqrt{V_{C_{MB}}}}\right)^2 \sim \chi^2\left(1, \frac{C_{MB}^2}{V_{C_{MB}}}\right) \tag{6.9}$$

with mean $(1+\frac{C_{MB}^2}{V_{C_{MB}}})$ and variance $2(1+2\frac{C_{MB}^2}{V_{C_{MB}}})$. In fact, for very large $m >> 1$ we have that $V_{C_{MB}} << 1$. Therefore $\frac{C_{MB}^2}{V_{C_{MB}}} >> 1$ implying that the previous distribution can be approximated by a normal distribution with the same variance and mean (see Remark in Appendix A). Then we calculate that

$$\mathbb{E}(\tilde{\tau}) = \tau + \mathcal{O}(1/m) \tag{6.10}$$

and

$$\mathrm{Var}(\tilde{\tau}) = \frac{4}{m}\tau^2(2 + \frac{V_N}{\tau V_M}) + \mathcal{O}(1/m^2) := \sigma^2. \tag{6.11}$$

### 6.3.2 Variance of the excess noise variance

We can have a maximum likelihood estimator for the noise variance $V_N$ of $Q_N = Q_B - \sqrt{\tau}Q_M$ associated with realizations $B_i - \sqrt{\tilde{\tau}}M_i$. This estimator is given by

$$\tilde{V}_N = \frac{1}{m}\sum_{i=1}^{m}(B_i - \sqrt{\tilde{\tau}}M_i)^2 \tag{6.12}$$

According to Eq. (3.15), we have the estimator for $V_\epsilon$

$$\tilde{V}_\epsilon = \tilde{V}_N - 1 \tag{6.13}$$

The uncertainty of this estimator consists of two levels of uncertainty. The first one and most important is created from the realizations $B_i$ and $M_i$. The second is created due to the variable $\sqrt{\tilde{\tau}}$, which in principle has to be replaced from Eq.(6.8). However, here we replace the variable $\sqrt{\tilde{\tau}}$ with $\sqrt{\tau}$ since we assume its uncertainty is negligible. This is justified by the fact that for large values of $m$ the variance $\sigma^2$ vanishes. Therefore we obtain

$$\mathbb{E}(\tilde{V}_\epsilon) = \mathbb{E}\left(\frac{1}{m}\sum_{i=1}^{m}(B_i - \sqrt{\tau}M_i)^2 - 1\right) \tag{6.14}$$

where we have that $\sum_{i=1}^{m}(\frac{B_i-\sqrt{\tau}M_i}{\sqrt{V_N}})^2$ is chi-squared distributed with mean $m$ and variance $2m$ as the sum of squares of random variables follow the standard normal distribution (see Eq. A.3.4). Therefore we have that the mean value is given by

$$\mathbb{E}(\tilde{V}_\epsilon) = \mathbb{E}\left(V_N\frac{1}{m}\sum_{i=1}^{m}\left(\frac{B_i - \sqrt{\tau}M_i}{\sqrt{V_N}}\right)^2 - 1\right) \tag{6.15}$$

$$= V_N - 1 = V_\epsilon \tag{6.16}$$

while the variance is

$$\mathrm{Var}(\tilde{V}_\epsilon) = \mathrm{Var}\left(V_N\frac{1}{m}\sum_{i=1}^{m}\left(\frac{B_i - \sqrt{\tau}M_i}{\sqrt{V_N}}\right)^2 - 1\right) = \frac{2}{m}V_N^2 := s^2. \tag{6.17}$$

## 6.4 Secret key rate with finite size effects

Using Eq. (A.4.7), we then calculate the confidence intervals of the estimators

$$\tilde{\tau} = \{\tau - 6.5\sigma, \tau + 6.5\sigma\}, \quad \tilde{V}_\epsilon = \{V_\epsilon - 6.5s, V_\epsilon + 6.5s\}, \tag{6.18}$$

given that the error is $\sim 10^{-10}$. The worst case scenario is satisfied by using the values

$$\tau^{\mathrm{low}} = \tau - 6.5\sigma, \quad \text{and} \quad V_\epsilon^{\mathrm{up}} = V_\epsilon + 6.5s. \tag{6.19}$$

Then we express the secret key rate function of Eq. (3.45) as

$$R(\xi, V_M, \tau, V_\epsilon) = \xi I_{AB} - I_E, \tag{6.20}$$

where we have set $V_\epsilon = \tau\epsilon$ and $\epsilon = \frac{(1-\tau)(\omega-1)}{\tau}$ and we replace with $V_\epsilon \to V_\epsilon^{\mathrm{up}}$ and $\tau \to \tau^{\mathrm{low}}$ so that the secret key rate with channel estimated parameters to be

$$\tilde{R}(\xi, V_M, \tau, V_\epsilon, m) = R(\xi, V_M, \tau^{\mathrm{low}}, V_\epsilon^{\mathrm{up}}) = \xi\tilde{I}_{AB} - \tilde{I}_E. \tag{6.21}$$

The secret key rate with finite size effects is given by replacing the the asymptotic key rate with estimated channel parameters of Eq. (6.21) into Eq. (6.1)

$$K = \frac{n}{n+m}\left(\tilde{R}(\xi, V_M, \tau, V_\epsilon, m) - \Delta(n)\right) \tag{6.22}$$

and if $\bar{N} = n + m$ and $r = m/\bar{N}$, then

$$K = (1 - r)\left(\tilde{R}(\xi, V_M, \tau, V_\epsilon, r\bar{N}) - \Delta((1 - r)\bar{N})\right) \tag{6.23}$$

so that for a given block size number $\bar{N}$ we can optimized over the ratio $r$.

## 6.5 Conclusion

In this chapter, we summarized the procedure of channel parameter estimation for some of the fully Gaussian protocols. This method is based on the Central Limit Theorem and on the Gaussian nature of the variables describing the given protocol and applies changes to the asymptotic key rate in order to include finite-size effects. We are going to modify it appropriately and used it for the non-asymptotic security analysis of the CV-MDI and thermal state protocols presented in the following chapters.

# Part II

# Contribution to recent advancements in CV-QKD

# Chapter 7

# Finite-size security analysis for one-way protocols using thermal states

## 7.1 Introduction

In this chapter we present results from Ref. [26] about the secret key rate of the protocols using normally distributed thermal states (see Chap. 4) incorporating finite-size effects by modifying appropriately the method of Ref. [15] summarized in Chap 6. We constrain our study in the protocol with direct reconciliation and homodyne detection since this is the most robust in terms of additional thermal trusted noise during Alice's state preparation (see Fig. 4.6). We then study of the finite-size secret key rate for different trusted thermal noise variances with respect to the block size needed for having a performance close to the asymptotic one. Finally, we connect the secret key rate with the frequency of the electromagnetic filed in order to investigate the behaviour of the protocol in the microwave regime with respect different block sizes.

## 7.2 Channel parameter estimation

For the channel parameter estimation, we use the maximum likelihood estimator of the covariance of $Q_M$ and $Q_B$ (see Eq. (6.3)) to obtain the estimator for transmissivity $\tilde{\tau}$. Subsequently, we use a maximum likelihood estimator for the variance of the variable $Q_N = Q_B - \sqrt{\tau} Q_M$ (see Eq.(6.12)) for an the estimator of the excess noise variance $V_\epsilon$.

Both of them are calculated by assuming $m$ realizations $M_i$ and $B_i$ of Alice's and Bob's variables, where $i = 1, 2, \ldots, m$. Each of the realizations $M_i$ ($B_i$) are following the same normal distribution as $Q_M$ ($Q_B$) with given modulation variance $V_M$

### 7.2.1 Variance of the transmissivity estimator

The definition of the estimator of covariance $\sigma_{MB} = \mathrm{Cov}(Q_M Q_B)$, between variables $Q_M$ and $Q_B$, is then easy to define as follows

$$\tilde{\sigma}_{MB} = \frac{1}{m}\sum_{i=1}^{m} M_i B_i. \tag{7.1}$$

From Eq. (7.1) we can compute both the expectation value and variance of $\tilde{\sigma}_{MB}$. This is done by assuming $M_i$ and $B_i$ are independent and normally distributed Gaussian variables. Therefore, we obtain the expectation value

$$\mathbb{E}\left[\tilde{\sigma}_{MB}\right] = \sqrt{\tau} V_M = \sigma_{MB}, \tag{7.2}$$

and the variance

$$V_{\tilde{\sigma}_{MB}} = \frac{\tau V_M^2}{m}\left(2 + \frac{V_N}{\tau V_M}\right). \tag{7.3}$$

According to Eq. (7.2), we can obtain the expectation value and variance of the estimator, of the transmissivity

$$\tilde{\tau} = \frac{\tilde{\sigma}_{MB}^2}{V_M^2} = \frac{V_{\tilde{\sigma}_{MB}}}{V_M^2}\left(\frac{\tilde{\sigma}_{MB}}{\sqrt{V_{\tilde{\sigma}_{MB}}}}\right)^2, \tag{7.4}$$

where

$$\left(\frac{\tilde{\sigma}_{MB}}{\sqrt{V_{\tilde{\sigma}_{MB}}}}\right)^2 \sim \chi^2(1, \frac{\sigma_{MB}^2}{V_{\tilde{\sigma}_{MB}}})$$

is chi-squared distributed with mean $(1+\frac{\sigma_{MB}^2}{V_{\tilde{\sigma}_{MB}}})$ and variance $2(1+2\frac{\sigma_{MB}^2}{V_{\tilde{\sigma}_{MB}}})$ (see Eq. (A.3.4)). Note that for $m >> 1$ the variance $V_{\tilde{\sigma}_{MB}} << 1$, which implies that the parameter $\lambda = \frac{\sigma_{MB}^2}{V_{\tilde{\sigma}_{MB}}} >> 1$. Therefore, according to the remark in Appendix A, this distribution can be approximated by a normal distribution with the same mean and variance. Eq. (7.4) allows us to compute the following expectation value

$$\mathbb{E}\left[\tilde{\tau}\right] = \frac{V_{\tilde{\sigma}_{MB}}}{V_M^2}\mathbb{E}\left[\left(\frac{\tilde{\sigma}_{MB}}{\sqrt{V_{\tilde{\sigma}_{MB}}}}\right)^2\right] = \tau + \mathcal{O}(1/m), \tag{7.5}$$

and variance

$$\sigma_{\tilde{\tau}}^2 = \frac{4\tau^2}{m}\left(2 + \frac{V_N}{\tau V_M}\right) + \mathcal{O}(1/m^2). \tag{7.6}$$

### 7.2.2 Variance of the excess noise variance estimator

We follow the same steps in order to obtain the variance $V_N$ starting from the statistical sampling $B_i$ and $M_i$. Using the relations $Q_N = Q_B - \sqrt{\tau}Q_M$, we can write the estimator of $V_N$ as follows

$$\tilde{V}_N = \frac{1}{m}\sum_{i=1}^{m}\left(B_i - \sqrt{\tilde{\tau}}M_i\right)^2. \tag{7.7}$$

Eq. (7.5) and Eq. (7.6) imply that the standard deviation $\sigma_{\tilde{\tau}}$ becomes rapidly negligible as $m \gg 1$. For reasons of simplicity, one can then securely substitute the estimator $\tilde{\tau}$ with its actual value $\tau$ in Eq. (7.7) since we can assume that the main source of uncertainty stems only from $B_i$ and $M_i$. Note that variable $B_i - \sqrt{\tau}M_i$ is normally distributed with variance $V_N$, which means that $\sum_{i=1}^{m}\left(\frac{B_i - \sqrt{\tau}M_i}{\sqrt{V_N}}\right)^2$ is also $\chi^2$-distributed with expectation values $m$ and variance $2m$ (see Eq. (A.3.4)). For degrees of freedom $m >> 1$, this distribution can be approximated by a normal distribution with the same mean and variance. Thus we can write

$$\tilde{V}_N = \frac{V_N}{m}\sum_{i=1}^{m}\left(\frac{B_i - \sqrt{\tau}M_i}{\sqrt{V_N}}\right)^2.$$

The estimator for the variance $V_\varepsilon$, can now be expressed using Eq. (4.14) leading to the following formula

$$\tilde{V}_\varepsilon = \tilde{V}_N - \tilde{\tau}V_{\text{th}} - 1,$$

with expectation value

$$\mathbb{E}(\tilde{V}_\varepsilon) = V_N - \tau V_{\text{th}} - 1, \tag{7.8}$$

and variance

$$\sigma^2_{\tilde{V}_\varepsilon} = \frac{2V_N^2}{m} + V_{\text{th}}^2\sigma^2_{\tilde{\tau}}. \tag{7.9}$$

We note that these equations are similar to the corresponding ones from Chap 6. The only but crucial difference in this case is the contribution of thermal noise $V_{\text{th}}$ in $V_N$ . This has as a result an extra term to appear in Eq. (7.9) which is dependent on the variance of $\tilde{\tau}$ as well.

## 7.3 Secret key rate with finite size effects

Assuming an error probability for the parameter estimation of the order of $\varepsilon_{PE} = 10^{-10}$ and using Eq. (A.4.7) the corresponding confidence intervals of channel parameters are

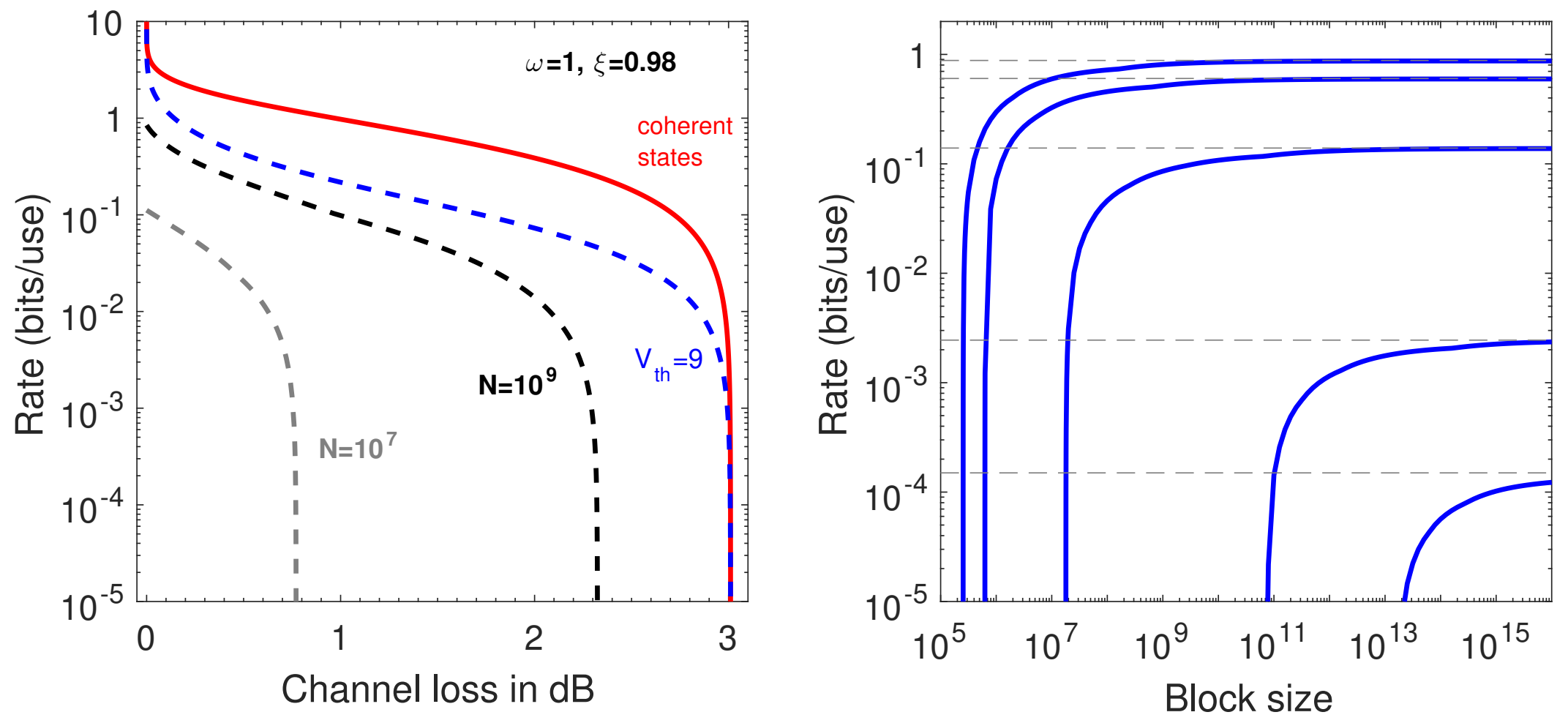

*Figure 7.1: (Color online) This figure focuses on the key-rate in the optical regime. The left panel describes the key rate versus channel attenuation given in dB. The red solid curve describes the ideal key-rate, using just coherent states. The blue-dashed curve describes the ideal key rate assuming a preparation noise with variance $V_{\rm th} = 9$ SNU. Then, we keep the same $V_{\rm th}$ and plot the finite-size rate for block size with $\bar{N} = 10^9$ (black dashed line) and $\bar{N} = 10^7$ (gray dashed) with $\xi = 0.98$ and $\omega = 1$. The key rate is optimized over the Gaussian modulation $V_M$. The right panel presents the key rate as a function of the block size ($\bar{N}$). We fix channel attenuation to 1 dB, and we assume pure loss attack $\omega = 1$ while $\xi = 0.98$. The plot shows the convergence of the key rates toward the asymptotic values (dashed curves) for different values of the preparation noise $V_{\rm th} = 0, 1, 10, 100, 150$ SNU, from top to bottom.*

given by

$$\tilde{\tau} =\{\tau - 6.5\sigma_{\tilde{\tau}}, \tau + 6.5\sigma_{\tilde{\tau}}\}, \tag{7.10}$$

$$\tilde{V}_\varepsilon =\{V_\varepsilon - 6.5\sigma_{\tilde{V}_\varepsilon}, V_\varepsilon + 6.5\sigma_{\tilde{V}_\varepsilon}\} \tag{7.11}$$

from which the worst case scenario parameter values are chosen

$$\tau^{\rm low} := \tau - 6.5\sigma_{\tilde{\tau}}, \quad V_\varepsilon^{\rm up} := V_\varepsilon + 6.5\sigma_{\tilde{V}_\varepsilon}. \tag{7.12}$$

The quantities in Eq. (7.12) are then used to compute the finite-size key rate, which is given by the following expression

$$K = \frac{n}{\bar{N}}\left[\tilde{R}(\xi, V_M, V_{\rm th}, V_\varepsilon^{\rm up}, \tau^{\rm low}) - \Delta\right], \tag{7.13}$$

where $\bar{N} = n+m$, is the total number of signals points, $n$ is the number of signals used to build the key, and the correction term $\Delta(n, d)$ given in Eq. (6.2) accounts for the penalty for using the Holevo bound in the key rate of Eq. (7.13) using a finite number of signals.

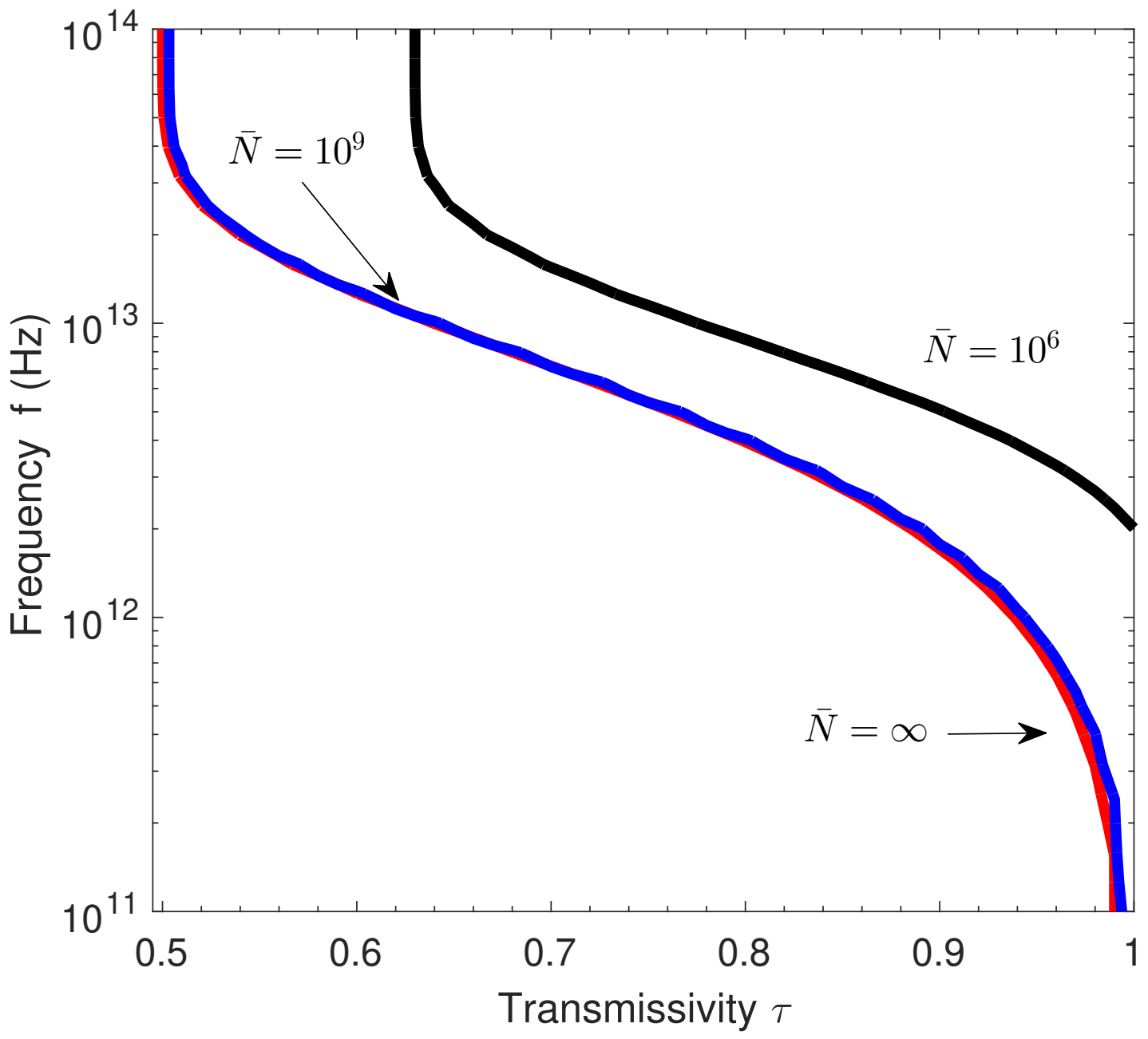

*Figure 7.2: (Color online) The red line shows the security threshold (frequencies vs channel's transmissivity) for a shot-noise level attack $\omega = V_{th} + 1$ without finite-size effects, and assuming infinite Gaussian modulation. Then we have the case for block size with $\bar{N} = 10^6$ signal points (black) and $\bar{N} = 10^9$ (blue).*

Here, we assume that $d = 4$ since thermal states are using a larger phase-space so that we should have a more refined discretization. The secret key rate $\tilde{R}$ comes from the asymptotic key rate of Eq.(4.44) by expressing it with respect the parameter $V_\epsilon = (1-\tau)(\omega-1)$. Then we set the values of $V_\varepsilon^{\text{up}}$ and $\tau^{\text{low}}$ computed previously.

By virtue of Eq. (7.13), we quantifying relevant quantities like achievable distance and block-size needed to obtain a positive secret key rate. More specifically, we concentrate on the size of the signal blocks needed in order to achieve a positive key rate as the thermal noise is increasing. By using $V_\varepsilon = \tau\varepsilon$, with $\varepsilon = \left[(1-\tau)(\omega-1)\right]/\tau$ for $\omega$ being the variance of thermal noise of Eve's ancillary states used in the attack, we express the key rate as follows

$$K' = (1-r)\left[\tilde{R}'(\xi, V_M, V_{\text{th}}, \omega, dB, r, \bar{N}) - \Delta\right], \tag{7.14}$$

where the transmissivity $\tau$ is given in terms of $dB$ of attenuation by using $\tau = 10^{-\frac{dB}{10}}$ and $r := m/\bar{N}$. The asymptotic key rate $\tilde{R}'$ is given by the corresponding key rate $\tilde{R}$ of Eq. (7.13), where we have made the previous replacements. From Eq. (7.14) we can plot the key rate as a function of the channel attenuation, fixing the values of $V_{\text{th}}$, reconciliation efficiency $\xi$, thermal noise $\omega$. Then we can optimize over the remaining parameters.

The results for pure-loss attacks are shown in Fig. 7.1. In the left panel, we plot the key rate for different values of the block-size and preparation noise. In particular, the red solid line describes the asymptotic key rate when Alice send coherent states, i.e., $V_{\text{th}} = 0$, while the blue-dashed line is for $V_{\text{th}} = 9$ SNU. Then, we compare the previous curves with the key rate of Eq. (7.14) for $\bar{N} = 10^9$ (black dashed line) and $\bar{N} = 10^7$ (gray dashed line). In Fig. 7.1 (right panel), we quantify the block-size needed to achieve a positive key rate for increasing values of the preparation noise. We fix the attenuation to 1 dB and assume pure loss attack ($\omega = 1$ SNU). We then plot the key-rate as a function of the block-size, for preparation noise $V_{\text{th}} = 0,\ 1,\ 10,\ 100,\ 150$ SNU from top to bottom and efficiency $\xi = 0.98$ [62]. Our results show that, by an increase in $V_{\text{th}}$, the block-size need to be increased in order to match the asymptotic value of the key rate (dashed lines). This stems from the fact that the confidence intervals are dependent on Alice's thermal noise as for large amount of $V_{\text{th}}$ they widen resulting in estimating lower transmissivity and higher channel thermal noise.

## 7.4 Secret key rate in different frequencies

Here we follow the reasoning of Sec. 4.4 and we set for Eve's thermal variance in here ancillary modes $\omega = V_{\text{th}} + 1$ and $V_{\text{th}} = 2\bar{n}$, where $\bar{n}$ is taken from Eq. (4.45) for room temperature $T = 300K$, in the rate of Eq. (7.13). Then we can plot the corresponding thresholds, illustrated in Fig. 7.2, in terms of frequency versus transmissivity for different block sizes. We see that in the microwave region security is achieved only for transmissivities very close to $\tau = 1$ for a moderately high block size number of $N = 10^9$. Note that the rapid decrease of the protocols performance due to the additional preparation noise affecting the channel parameter estimated values is now deteriorating due to considering an entangling cloner attack even in the shot-noise level which is as high as the preparation noise.

## 7.5 Conclusion

In this chapter, we expanded the a non-asymptotic secret key rate analysis to the one-way protocols using thermal states. This extra trusted noise in Alice's preparation allows us to investigate the performance of such a protocol in the microwave regime. The context of this analysis was focused on the parameter estimation step based on Gaussian distribution assumption and on the central limit theorem. We notice that the estimated values of the channel parameters are dependent on the presence of the preparation noise, which as it increases gives a more pessimistic secret key rate. This reflects on the limitations of this protocol in terms of using a small block size as Alice's thermal noise is increasing for achieving key rates near to the asymptotic ones.

As a later step, we would like to investigate a protocol of thermal states using a discrete modulation instead of a Gaussian (see Sec. 3.6). Also protocols with post selection or two-way communication might be more appropriate of handling the extra preparation noise. It would be interesting to proceed with their non-asymptotic analysis comparing their robustness in terms of high preparation noise with respect block size.

# Chapter 8

# Finite-size security analysis for CV-MDI-QKD

## 8.1 Introduction

In this chapter, we present the work of Ref. [34]. We base our finite size analysis in Chap. 6, where we have seen that such an analysis is based on the asymptotic secret key rate calculation (see Chap. 5). We assume a two-mode Gaussian attack equivalent to the entangling cloner attack of the conventional one-way protocols. Our statistical estimation of the channel parameters is assuming a Gaussian framework and is based on the central limit theorem. We start with separating the signal and noise variables of the relay outputs and use these variables in order to estimate the channel parameters and create confidence intervals for them. Then we choose the most pessimistic values of the intervals for these parameters so as to simulate the worst case scenario for our security analysis. We finally use these values in the function of the secret key rate incorporating finite size effects in order to evaluate the performance of the CV-MDI protocol in a more realistic situation. We notice then that, as we increase the block size, the protocol achieves a performance closer to the asymptotic one as it was expected. In addition to this, block sizes in the range of $10^6 - 10^9$ can produce a secret key rate of about $10^{-2}$ bit/use even when the excess noise is as high as 0.01 vacuum shot noise units (SNU) for an attenuation compatible with metropolitan distances when using standard optical-fibre equipment.

## 8.2 Relay outputs: losses and excess noise

The output variables of the relay are denoted with $Q_R := q_-$ and $P_R = p_+$, where the output $\gamma = q_- + \mathrm{i}p_+$ and are dependent on the evolution of Alice's and Bob's traveling modes $A$ and $B$ with quadratures $\alpha = (Q_A, P_A)$ and $\beta = (Q_B, P_B)$. In terms of these input field quadratures, one can then write the following relations

$$\begin{aligned} Q_R &= \frac{1}{\sqrt{2}}(\sqrt{\tau_B}(Q_{M,B} + Q_{0,B}) - \sqrt{\tau_A}(Q_{M,A} + Q_{0,A})) \\ &+ \frac{1}{\sqrt{2}}(\sqrt{1-\tau_B}Q_{E_2} - \sqrt{1-\tau_A}Q_{E_1}) \\ &= \frac{1}{\sqrt{2}}(\sqrt{\tau_B}(Q_{M,B} - \sqrt{\tau_A}Q_{M,A}) + Q_N, \end{aligned} \tag{8.1}$$

$$\begin{aligned} P_R &= \frac{1}{\sqrt{2}}(\sqrt{\tau_B}(P_{M,B} + P_{0,B}) + \sqrt{\tau_A}(P_{M,A} + P_{0,A})) \\ &+ \frac{1}{\sqrt{2}}(\sqrt{1-\tau_B}P_{E_2} + \sqrt{1-\tau_A}P_{E_1}) \\ &= \frac{1}{\sqrt{2}}(\sqrt{\tau_B}(P_{M,B} - \sqrt{\tau_A}P_{M,A}) + P_N, \end{aligned} \tag{8.2}$$

where

$$Q_N = \frac{1}{\sqrt{2}}(\sqrt{\tau_B}Q_{0,B} - \sqrt{\tau_A}Q_{0,A}) + \frac{1}{\sqrt{2}}(\sqrt{1-\tau_B}Q_{E_2} - \sqrt{1-\tau_A}Q_{E_1}), \tag{8.3}$$

$$P_N = \frac{1}{\sqrt{2}}(\sqrt{\tau_B}P_{0,B} + \sqrt{\tau_A}P_{0,A}) + \frac{1}{\sqrt{2}}(\sqrt{1-\tau_B}P_{E_2} + \sqrt{1-\tau_A}P_{E_1}), \tag{8.4}$$

with $Q_N$ and $Q_N$ are noise terms accounting for both excess noise and quantum shot noise coming form the signal modes as well as Eve's ancillary modes. Their variances are given by

$$V_{Q_N} = 1 + V_{Q,\epsilon}, \quad V_{P_N} = 1 + V_{P,\epsilon}, \tag{8.5}$$

with

$$V_{Q,\epsilon} = k - gu, \quad V_{P,\epsilon} = k + g'u, \tag{8.6}$$

and

$$k = \frac{(1-\tau_B)(\omega_B - 1) + (1-\tau_A)(\omega_A - 1)}{2}, \tag{8.7}$$

$$u = \sqrt{(1-\tau_B)(1-\tau_A)}, \tag{8.8}$$

where $g$ and $g'$ have been defined in Eq. (5.3). Therefore, we can express the variances of $Q_R$ and $P_R$ in terms of signal and noise

$$V_{Q_R} = \frac{1}{2}(\tau_B + \tau_A)V_M + V_{Q_N}, \tag{8.9}$$

$$V_{P_R} = \frac{1}{2}(\tau_B + \tau_A)V_M + V_{P_N}. \tag{8.10}$$

## 8.3 Channel parameter estimation

### 8.3.1 Variance of the transmissivity

The parameters estimation procedure is described in more detail as follows: Alice's and Bob's Gaussian modulation $V_M$ is assumed to be a known parameter while the channel parameters transmissivity $\tau_A$, $\tau_B$ and variance of the excess noises $V_{Q,\epsilon}$ and $V_{Q,\epsilon}$ need to be estimated and confidence intervals to be defined for them. Assuming that $m$ Gaussian distributed signals are used for this task, we associate to $A_{Q,i}$ ($A_{P,i}$) and $B_{Q,i}$ ($B_{P,i}$), for $i \in \{1, 2, \ldots, m\}$, the empirical realizations of the field quadrature of the traveling modes. By contrast, we denote by $R_{Q,i}$ and $R_{P,i}$ the realizations of the relay outputs. We first present the procedure for output $Q_R$ of the relay. From Eq. (8.1) one can write the estimator of transmissivity $\tau_A$ as follows

$$\tilde{\tau}_{A_Q} = \frac{2}{V_M^2} \tilde{C}_{AR_Q}^2,$$

where the covariance $C_{AR_q} = \mathrm{Cov}(Q_{M,A} Q_R) = \sqrt{\tau_A/2} V_M$ has maximum likelihood estimator given by

$$\tilde{C}_{AR_Q} = \frac{1}{m} \sum_{i=1}^{m} A_{Q,i} R_{Q,i},$$

normally distributed as the sample mean of variable $Z = Q_{M,A} Q_R$. We can compute the expectation value by

$$\mathbb{E}(\tilde{C}_{AR_Q}) = \mathbb{E}(Q_{M,A} Q_R) = \sqrt{\frac{\tau_A}{2}} V_M = C_{AR_Q}, \tag{8.11}$$

and the variance can be defined as

$$V_{\mathrm{Cov}} := \mathrm{Var}(\tilde{C}_{AR_Q}) \tag{8.12}$$

with

$$\begin{aligned}
\mathrm{Var}(\tilde{C}_{AR_Q}) &= \frac{1}{m} \mathrm{Var}(Q_{M,A} Q_R) = \\
&= \frac{1}{2m} \left[ \tau_A \mathrm{Var}(Q_{M,A}^2) + \tau_B \mathrm{Var}(Q_{M,A} Q_{M,B}) \right] \\
&+ \mathrm{Var}(Q_{M,A} Q_N), && (8.13) \\
&= \frac{1}{m} \left( \tau_A V_M^2 + \frac{\tau_B}{2} V_M^2 + V_M V_{Q_N} \right) \\
&= \frac{1}{m} \left( \tau_A + \frac{\tau_B}{2} \right) V_M^2 \left[ 1 + \frac{V_{Q_N}}{\left( \tau_A + \frac{\tau_B}{2} \right) V_M} \right], && (8.14)
\end{aligned}$$

where we have considered the independence of the variables and the second order moments of the normal distribution.

Therefore, we can derive the mean and variance for the estimator of $\tau_A$. We rewrite the estimator as

$$\tilde{\tau}_{A_Q} = \frac{2V_{\text{Cov}}}{V_M^2}\left(\frac{\tilde{C}_{AR_Q}}{\sqrt{V_{\text{Cov}}}}\right)^2. \tag{8.15}$$

Note that the variable $\left(\tilde{C}_{AR_Q}/\sqrt{V_{\text{Cov}}}\right)^2$ is $\chi^2-$distributed (see Eq.(A.3.4)), i.e.,

$$\left(\frac{\tilde{C}_{AR_Q}}{\sqrt{V_{\text{Cov}}}}\right)^2 \sim \chi^2\left[1, \left(\frac{C_{AR_Q}}{\sqrt{V_{\text{Cov}}}}\right)^2\right], \tag{8.16}$$

with expectation value

$$\mathbb{E}\left[\left(\frac{\tilde{C}_{AR_Q}}{\sqrt{V_{\text{Cov}}}}\right)^2\right] = 1 + \left(\frac{C_{AR_Q}}{\sqrt{V_{\text{Cov}}}}\right)^2, \tag{8.17}$$

and variance

$$\text{Var}\left[\left(\frac{\tilde{C}_{AR_Q}}{\sqrt{V_{\text{Cov}}}}\right)^2\right] = 2\left[1 + 2\left(\frac{C_{AR_Q}}{\sqrt{V_{\text{Cov}}}}\right)^2\right]. \tag{8.18}$$

For $m >> 1$, this distribution can be approximated by a normal distribution with the same variance and mean since then $V_{\text{Cov}} << 1$ and $\left[\frac{C_{AR_Q}}{\sqrt{V_{\text{Cov}}}}\right]^2 >> 1$ (see remark in Appendix A). The expectation value of $\tilde{\tau}_{A_Q}$ is then given by

$$\begin{aligned}\mathbb{E}(\tilde{\tau}_{A_Q}) &= \frac{2V_{\text{Cov}}}{V_M^2}\left[1 + \left(\frac{C_{AR_Q}}{\sqrt{V_{\text{Cov}}}}\right)^2\right] = \\ &= \frac{2C_{AR_Q}^2}{V_M^2} + \mathcal{O}(1/m) = \tau_A + \mathcal{O}(1/m)\end{aligned} \tag{8.19}$$

and its variance is

$$\begin{aligned}\text{Var}(\tilde{\tau}_{A_Q}) &= \frac{4V_{\text{Cov}}^2}{V_M^4}2\left[1 + 2\left(\frac{C_{AR_Q}}{\sqrt{V_{\text{Cov}}}}\right)^2\right] = \\ &= \frac{16V_{\text{Cov}}C_{AR_Q}^2}{V_M^4} + \mathcal{O}(1/m^2).\end{aligned} \tag{8.20}$$

By replacing from Eq. (8.11) and Eq. (8.14), we obtain

$$\begin{aligned}\text{Var}(\tilde{\tau}_{A_Q}) &= \frac{16}{mV_M^4}\left(\tau_A + \frac{\tau_B}{2}\right) \\ &\frac{V_M^4\tau_A}{2}\left[1 + \frac{V_{Q_N}}{\left(\tau_A + \frac{\tau_B}{2}\right)V_M}\right] + \mathcal{O}(1/m^2), \\ &= \frac{8\tau_A}{m}\left(\tau_A + \frac{\tau_B}{2}\right)\left[1 + \frac{V_{Q_N}}{\left(\tau_A + \frac{\tau_B}{2}\right)V_M}\right] \\ &+ \mathcal{O}(1/m^2).\end{aligned} \tag{8.21}$$

The estimator of the transmissivity $\tilde{\tau}_{A_Q}$ is only asymptotically unbiased. In fact the standard deviation $\sqrt{\mathrm{Var}(\tilde{\tau}_{A_Q})}$ is of order $1/\sqrt{m}$ while the bias goes as $1/m$. As we consider $m > 10^5$ in our analysis, the value of the bias becomes rapidly negligible as $m \gg 1$.

We also can have an estimator of $\tau_A$ considering the other output of the relay, $P_R$ as well for which hold very similar relations. We can write the estimator of the covariance $C_{AR_P}$, which is given by

$$\tilde{C}_{AR_P} = \frac{1}{m}\sum_{i=1}^{m} A_{P,i}R_{P,i},$$

then using Eq. (8.2) one can write the estimator of the transmissivity $\tau_A$

$$\tilde{\tau}_{A_P} = \frac{2}{V_M^2}\tilde{C}_{AR_P}^2,$$

having variance

$$\mathrm{Var}(\tilde{\tau}_{A_P}) = \frac{8}{m}\tau_A\left(\tau_A + \frac{\tau_B}{2}\right)\left[1 + \frac{V_{P_N}}{\left(\tau_A + \frac{\tau_B}{2}\right)V_M}\right]. \tag{8.22}$$

We notice that it differs from the formula of Eq. (8.21) in the expression of $V_{p_N}$, given in Eq. (8.5). Now, from Eq. (8.21) and Eq. (8.22) we calculate the optimum linear combination of the variances of the two estimators (see Eq.(A.5.11))

$$\mathrm{Var}(\tilde{\tau}_A) = \frac{\mathrm{Var}(\tilde{\tau}_{A_q})\mathrm{Var}(\tilde{\tau}_{Ap})}{\mathrm{Var}(\tilde{\tau}_{A_q}) + \mathrm{Var}(\tilde{\tau}_{Ap})} := \sigma_A^2. \tag{8.23}$$

The same steps can be performed to obtain the relevant estimators for transmissivity $\tau_B$ and the corresponding variance $\mathrm{Var}(\tilde{\tau}_B) = \sigma_B^2$.

### 8.3.2 Variance of the excess noise variance

The estimator of the variance of the excess noise $V_{Q,\epsilon}$ present on the communication channel estimator can be derived from the maximum likelihood estimator of $V_{Q,N}$,

$$\tilde{V}_{Q,\epsilon} = \frac{1}{m}\sum_{i=1}^{m}\left[R_{Q,i} - \frac{\sqrt{\tilde{\tau}_B}B_{Q,i} - \sqrt{\tilde{\tau}_A}A_{Q,i}}{\sqrt{2}}\right]^2 - 1, \tag{8.24}$$

We can replace the estimator of $\tau_A$ ($\tau_B$) with its value as in Chap. 6 assuming that any uncertainty in the estimator of the excess noise comes only from variables $R_{Q,i}$, $A_{Q,i}$ and $B_{Q,i}$. We then have that the expression inside square brackets is normally distributed with zero mean and variance $V_{q,N}$. In addition to this, one also has that the following expression

$$Y := \sum_{i=1}^{m}\left(\frac{R_{Q,i} - \left(\sqrt{\tau_B}B_{Q,i} - \sqrt{\tau_A}A_{Q,i}\right)/\sqrt{2}}{\sqrt{V_{Q,N}}}\right)^2 \sim \chi^2(m,0), \tag{8.25}$$

is $\chi^2$-distributed (see Eq. (A.3.4)), and has mean $\mathbb{E}(Y) = m$ and variance $\mathrm{Var}(Y) = 2m$. Note that for $m >> 1$ this distribution can be approximated by a normal distribution with the same mean and variance see remark in (Appendix A). This allows to approximate the sum of Eq. (8.24) with $V_{Q,N}Y$ when we assume large values for $m$, obtaining the expectation value

$$\mathbb{E}(\tilde{V}_{Q,\epsilon}) \approx \frac{1}{m} V_{Q_N} \mathbb{E}(Y) - 1 = V_{Q,\epsilon} \tag{8.26}$$

and the variance

$$\mathrm{Var}(\tilde{V}_{Q,\epsilon}) \approx \frac{1}{m^2} V_{Q_N}^2 \mathrm{Var}(Y) = \frac{2}{m} V_{Q_N}^2 := s_Q^2. \tag{8.27}$$

Correspondingly, we obtain an estimator for $V_{P,\epsilon}$ expressed as

$$\tilde{V}_{P,\epsilon} = \frac{1}{m} \sum_{i=1} \left[ R_{P,i} - \frac{1}{\sqrt{2}} (\sqrt{\tilde{\tau}_B} B_{P,i} + \sqrt{\tilde{\tau}_A} A_{P,i}) \right]^2 - 1 \tag{8.28}$$

and variance

$$\mathrm{Var}(\tilde{V}_{P,\epsilon}) \approx \frac{2}{m} V_{P,N}^2 := s_P^2. \tag{8.29}$$

### 8.3.3 Secret key rate with finite size effects

Subsequently, from Eq. (8.27) and Eq. (8.29) we compute the confidence intervals from Eq. (A.4.7) by assuming an error of $10^{-10}$ and select the pessimistic values given by the following choice of parameters

$$\tau_A^{\mathrm{low}} = \tau_A - 6.5\sigma_A, \quad \tau_B^{\mathrm{low}} = \tau_B - 6.5\sigma_B, \tag{8.30}$$

$$V_{q,\epsilon}^{\mathrm{up}} = V_{q,\epsilon} + 6.5 s_Q, \quad V_{p,\epsilon}^{\mathrm{up}} = V_{p,\epsilon} + 6.5 s_P. \tag{8.31}$$

Once we have obtained the estimation of the transmissivities of the channels and the corresponding excess noises, we write the secret key rate of Eq. (5.17) in terms of the excess noise variances $V_{Q,\epsilon}$ and $V_{P,\epsilon}$ and replace with the relations of Eq. (8.30–8.30) and $V_M = \mu - 1$ in order to obtain the asymptotic key rate with channel parameter estimated values

$$\tilde{R}(\xi, V_M, \tau_A^{\mathrm{low}}, \tau_B^{\mathrm{low}}, V_{q,\epsilon}^{\mathrm{up}}, V_{p,\epsilon}^{\mathrm{up}}, g, g'). \tag{8.32}$$

Then we use this rate into the relation of the secret key rate with finite size effects, which yields

$$K = \frac{n}{\bar{N}} \left( \tilde{R}(\xi, V_M, \tau_A^{\mathrm{low}}, \tau_B^{\mathrm{low}}, V_{q,\epsilon}^{\mathrm{up}}, V_{p,\epsilon}^{\mathrm{up}}) - \Delta(n) \right), \tag{8.33}$$

where $n = \bar{N} - m$ is the number of signals used to prepare the key and $\bar{N}$ the total number of signal exchanged.

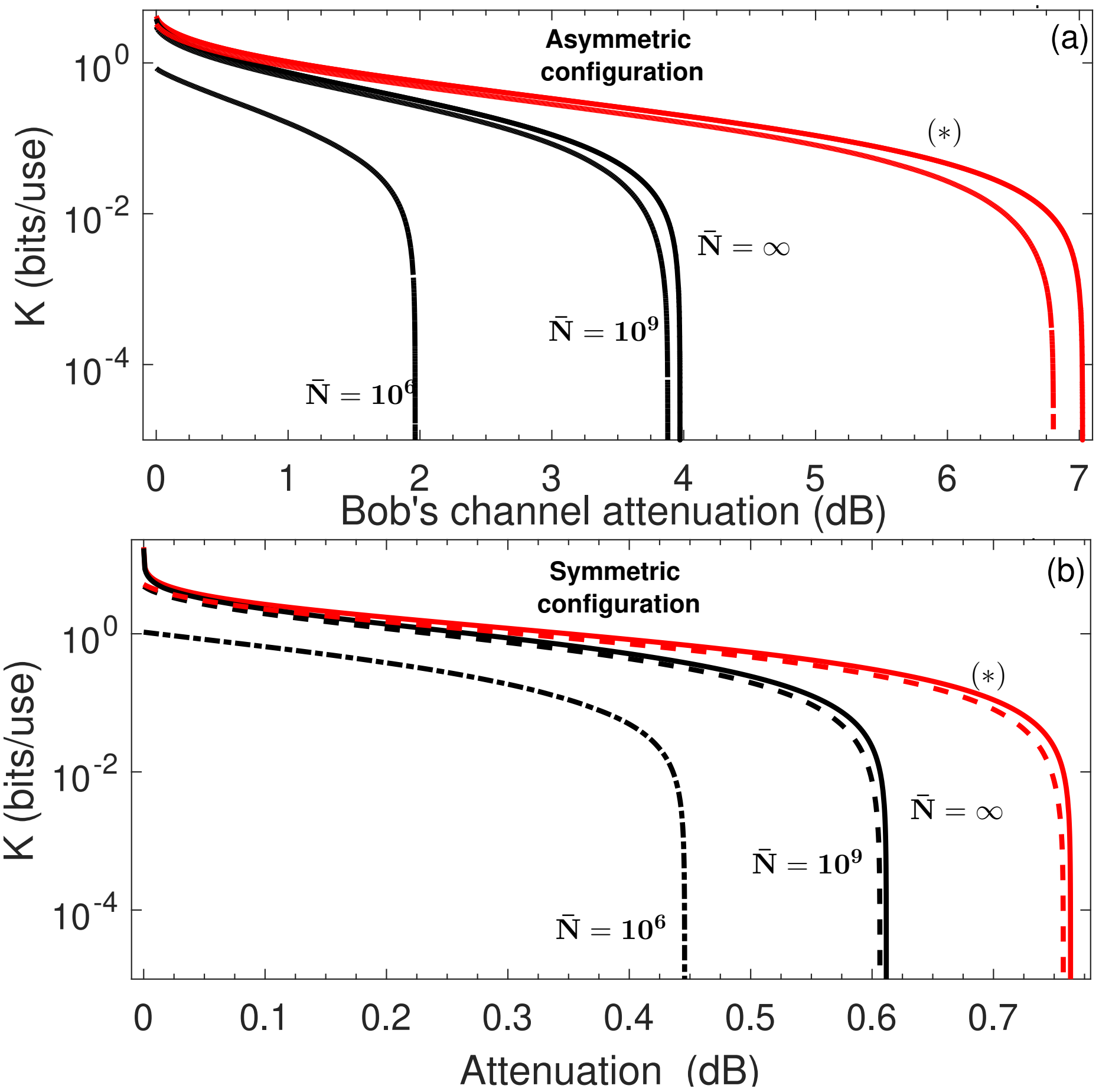

*Figure 8.1: (Color online)The figure summarize the impact of finite size effects on the performance of CV-MDI QKD for both asymmetric (panel a) and symmetric (panel b) configuration of the relay, in the presence of optimal two-mode attack. In panel (a) the key rate is plotted as a function of the dB of attenuation on Bob's channel, with the relay placed near Alice $\tau_A = 0.98$. From top to bottom, the black curves describe: the rate for $\bar{N} \gg 1$ with $\xi = 0.98$, and optimizing over $V_M$ (solid line). Then we have the cases with finite block-size. The dashed line is for $\bar{N} = 10^9$ while the dot-dashed curve is obtained for $\bar{N} = 10^6$. In all cases we have assumed excess of noise $\varepsilon = 0.01$ SNU. The red curves (*) describe the case obtained for pure loss and assuming $\xi = 1$, $V_M \to \infty$, $\bar{N} \to \infty$ (solid line) and $\bar{N} = 10^9$ (dashed line). Panel (b) focuses on the symmetric configuration of the relay. The curves are obtained using the same parameters as in panel (a), but setting $\tau_A = \tau_B = \tau$.*

## 8.4 Results

We use Eq. (8.33) of the non-asymptotic secret key rate in order to quantify the finite size effects and compare it with the performance of the protocol in the asymptotic regime. To do so, for the asymmetric configuration of the relay, we plot in the top-panel (a) of Fig. 8.1 the key rate as a function of Bob's channel transmissivity, expressed in terms of $dB$ of attenuation by replacing $\tau_B = 10^{-\frac{dB}{10}}$ while the transmissivity of Alice's channel is set to $\tau_A = 0.98$. The attack considered here is a two-mode optimal one, for which $g = -g'$ with $g = \min\left[\sqrt{(\omega_A - 1)(\omega_B + 1)}, \sqrt{(\omega_B - 1)(\omega_A + 1))}\right]$ and $\omega_A \sim \omega_B \sim 1.01$ [29]. The reconciliation efficiency is set to $\xi = 0.98$, and the final key rate is optimized over the variance of the Gaussian modulation $V_M$ and the ratio $r = n/\bar{N}$. The black-solid line gives the asymptotic key rate for very large $\bar{N}$ ($>> 10^9$), while the dashed line is for block-size $\bar{N} = 10^9$ and the dot-dashed line is obtained for $\bar{N} = 10^6$.

The symmetric case [30], where the parties have the same distance from the relay ($\tau_A = \tau = B$), is presented in the bottom panel of Fig. 8.1 (b). We set the same values for the all the parameters as in Fig. 8.1 (a) and optimizing as before. The black solid line describes again the asymptotic case $\bar{N} \to \infty$ of the symmetric configuration while the dashed lines is obtained for $\bar{N} = 10^9$ and the dotted one for $\bar{N} = 10^6$.

We notice that the performance of finite-size CV-MDI-QKD converges to the ideal one if the number of signals exchanged increases. We observe also that for block size $10^9$ the performance of the protocol is not so different from the asymptotic case showing that the protocol has a sufficiently robust behaviour with respect the finite size effects. In terms of achievable distance, we can see for example that for the asymmetric case Alice is considered to be roughly 0.4 km away from the relay and Bob can be 20 km away from the relay for $\bar{N} = \infty$ and 19.4 km away from the relay for $\bar{N} = 10^9$ approximately.

## 8.5 Conclusion

In this chapter, we followed the steps of Chap. 6 in order to present a finite-size security analysis of the CV-MDI protocol in its CV version focusing on the parameter estimation. This analysis can be seen as an intermediate step towards a final goal that of obtaining a composable security analysis (see Chap. 9) taking into consideration all the subroutines of a protocol. In other words, achieving a secret key rate expression sensitive to phenomena appearing in a realistic situation. The method of parameter estimation is based on a Gaussian framework and on the central limit theorem. However, it is made in such a way so as to take into consideration the optimal amount of signals from the block for the secret key extraction given a specific setting for the protocol.

Although we have assumed a non-asymptotic analysis, our results show that the protocol can still achieve high enough secret key rates for metropolitan communication using state-of-the-art equipment. In particular a range of $\bar{N} = 10^6 \div 10^9$ block size is sufficient so as to have a performance of $10^{-2}$ bits/use under channel excess noise as high as $\varepsilon = 0.01$ (SNU) and an attenuation corresponding to previous distances. Future works, can include phenomena of preparation noise in Alice's and Bob's labs and use discrete encoding instead of a Gaussian. The later approach might be able to manifest phenomena appearing on communication with repeaters by using the relay configuration as such.

# Chapter 9

# Composable security analysis for CV-MDI-QKD

## 9.1 Introduction

The secret key rate in a composable security framework is dependent on the effectiveness of all assumed subtasks during the protocol procedure (see remark in Sec. 3.3) and incorporates already finite size effects. This was presented for the one-way protocols assuming collective attacks in Ref. [17] and expanded for general attacks in Ref. [18]. A similar security analysis has been given for the CV-MDI protocol (see Chap. 5) in Ref. [35]. Based on this study, we will explain the adopted channel parameter estimation scheme which is not based on the central limit theorem assumption and give some insight on how this secret key rate works. The channel parameter estimation analysis estimates focuses on the estimation of the CM of the protocol variables instead of the channel parameters separately and makes use of the optimality of Gaussian attacks [18]. Finally, we present some numerical results for the secret key rate performance with respect different block sizes.

## 9.2 Channel parameter estimation without the Central Limit Theorem

The parameter estimation we present here follows a different direction with respect to this in Chap. 8. First of all, we assume that Alice and Bob prepare coherent states by displacing vacuum states according to Gaussian distributed variables $(Q'_A, P'_A)$ and $(Q'_B, P'_B)$ respectively with variance $V_M$. After the relay output $\gamma = (Q_Z, P_Z)$ is broadcast, the authenticated parties apply displacements to their variables obtaining

$$Q_A = Q'_A - g_{Q'_A}(\gamma)\,, \tag{9.1}$$

$$P_A = P'_A - g_{P'_A}(\gamma)\,, \tag{9.2}$$

$$Q_B = P'_B - g_{Q'_B}(\gamma)\,, \tag{9.3}$$

$$Q_B = P'_B - g_{P'_B}(\gamma)\,, \tag{9.4}$$

The estimation procedure is applied on the entries of the CM of the variables above which is given by

$$\mathbf{V}_{AB} = \langle \begin{pmatrix} Q_A^2 & Q_A P_A & Q_A Q_B & Q_A P_B \\ p_A Q_A & P_A^2 & P_A Q_B & P_A P_B \\ Q_B Q_A & Q_B P_A & Q_B^2 & Q_B P_B \\ P_B Q_A & P_B Q_B & P_B Q_B & P_B^2 \end{pmatrix} \rangle = \begin{pmatrix} x\mathbf{I} & z\mathbf{I} \\ z\mathbf{I} & y\mathbf{I} \end{pmatrix}\,, \tag{9.5}$$

where $\mathbf{I} = \mathrm{diag}(1,1)$, and

$$x = \frac{\langle Q_A^2 \rangle + \langle P_A^2 \rangle}{2}\,, \tag{9.6}$$

$$y = \frac{\langle Q_B^2 \rangle + \langle P_B^2 \rangle}{2}\,, \tag{9.7}$$

$$z = \frac{\langle Q_A Q_B \rangle + \langle P_A P_B \rangle}{2}\,. \tag{9.8}$$

Note that $g_\star$, where $\star$ stands for any of $(Q'_A, P'_A, Q'_B, P'_B)$, is an affine functions of $\gamma$ that can be optimally chosen as in Ref. [36] to be

$$g_\star(\gamma) = u_\star\, Q_Z + v_\star\, P_Z\,, \tag{9.9}$$

with

$$u_\star = \frac{\langle \star\, Q_Z \rangle \langle P_Z^2 \rangle - \langle \star\, P_Z \rangle \langle Q_Z P_Z \rangle}{\langle P_Z^2 \rangle \langle Q_Z^2 \rangle - \langle Q_Z P_Z \rangle^2}\,, \tag{9.10}$$

$$v_\star = \frac{\langle \star\, P_Z \rangle \langle Q_Z^2 \rangle - \langle \star\, Q_Z \rangle \langle Q_Z P_Z \rangle}{\langle Q_Z^2 \rangle \langle P_Z^2 \rangle - \langle Q_Z P_Z \rangle^2}\,. \tag{9.11}$$

for $\langle Q_z \rangle = \langle P_z \rangle = 0$ for the sake of simplicity. We notice that the parameters $u_\star$ and $v_\star$ can be computed directly from the estimated CM.

The diagonal entries $x$ and $y$ of the CM $\mathbf{V}_{AB}$ can be calculated by the local data either of Alice or Bob. The off-diagonal entries corresponding to $z$ can be computed based on Eq. (9.1–9.4)

$$\begin{aligned} z &= \frac{\langle Q_A Q_B \rangle + \langle P_A P_B \rangle}{2} = \\ &= w_1 \langle Q_Z^2 \rangle + w_2 \langle P_Z^2 \rangle + w_3 \langle Q_Z P_Z \rangle\,, \end{aligned} \tag{9.12}$$

where we have set

$$w_1 := \frac{1}{2}\left(u_{q'_A}u_{q'_B} + u_{p'_A}u_{p'_B}\right)\,, \tag{9.13}$$

$$w_2 := \frac{1}{2}\left(v_{q'_A}v_{q'_B} + v_{p'_A}v_{p'_B}\right)\,, \tag{9.14}$$

$$w_3 := \frac{1}{2}\left(u_{q'_A}v_{q'_B} + v_{q'_A}u_{q'_B} + u_{p'_A}v_{p'_B} + v_{p'_A}u_{p'_B}\right)\,. \tag{9.15}$$

Note that the authenticated parties can locally compute the variances $\langle Q_Z\rangle$, $\langle P_Z\rangle$ and the covariance $\langle Q_Z P_Z\rangle$. This indicates that the parties may exploit the option of completing the parameter estimation without disclosing their local data. In other words, all their raw data can be used for both parameter estimation and secret key extraction.

In order to create confidence intervals for the variances of $Q_A, P_A, Q_B, P_B, Q_Z, P_Z$, we associate with them empirical variances for estimators. For example, for the variance $\langle Q_A^2\rangle$ we have that $n^{-1}\Sigma_{j=1}^{n} Q_{Aj}^2$ for $n$ empirical realizations $Q_{Aj}$ of the variable $Q_A$ independently and identically distributed. Then these estimators can be considered chi-squared distributed after the appropriate rescaling. This holds because the variables $Q_A, P_A, Q_B, P_B$ are assumed to follow a Gaussian distribution (see Eq. A.3.4). This is a consequence of the fact that we assume a Gaussian attack on the channels due to the Gaussian optimality in addition to the $Q'_A, P'_A, Q'_B, P'_B$ being normally distributed as it is required by the protocol. This is also true for the variables $Q_Z, P_Z$ since they can be seen as a linear combination with respect the variables $Q_A, P_A, Q_B, P_B$ and $Q'_A, P'_A, Q'_B, P'_B$.

Furthermore, the covariance $\langle Q_Z P_Z\rangle$ can be expresses as a combination of two terms

$$\langle Q_Z P_Z\rangle = \frac{1}{4}\langle (Q_Z + P_Z)^2\rangle - \frac{1}{4}\langle (Q_Z - P_Z)^2\rangle\,, \tag{9.16}$$

each one corresponding in a chi-squared distributed estimator

$$\frac{1}{n}\sum_{j=1}^{n} q_{Zj}p_{Zj} = \frac{1}{4n}\sum_{j=1}^{n}(q_{Zj}+p_{Zj})^2 - \frac{1}{4n}\sum_{j=1}^{n}(q_{Zj}-p_{Zj})^2 \tag{9.17}$$

since the variables $Q_{Zj}+P_{Zj}$ and $Q_{Zj}-Q_{Zj}$ are normally distributed as linear combination of independent Gaussian variables.

To each of the estimators, which follows a chi-squared distribution, we can apply tail bounds coming from its corresponding cumulative function. For $\langle Q_A^2\rangle$ these are given by

$$\Pr\left\{\langle Q_A^2\rangle < \frac{n^{-1}\sum_j Q_{Aj}^2}{1+t}\right\} \le e^{-nt^2/8}\,, \tag{9.18}$$

$$\Pr\left\{\langle Q_A^2\rangle > \frac{n^{-1}\sum_j Q_{Aj}^2}{1-t}\right\} \le e^{-nt^2/8}\,, \tag{9.19}$$

and hold similar relations for the other cases. These bounds are calculated with the help of Chernoff bound [1] using the sub-exponential behaviour of the chi-squared distribution. Subsequently, for the the covariance term we have that

$$\begin{aligned}
&\Pr\left\{\langle Q_Z Q_Z\rangle > \frac{n^{-1}\sum_j (Q_{Zj}+P_{Zj})^2}{4(1-t)} - \frac{n^{-1}\sum_j (Q_{Zj}-P_{Zj})^2}{4(1+t)}\right\} \\
&\leq \Pr\left\{\langle (Q_Z+P_Z)^2\rangle > \frac{n^{-1}\sum_j (Q_{Zj}+P_{Zj})^2}{(1-t)}\right\} \\
&+ \Pr\left\{\langle (Q_Z-P_Z)^2\rangle < \frac{n^{-1}\sum_j (Q_{Zj}-P_{Zj})^2}{(1+t)}\right\} \\
&\leq 2e^{-nt^2/8} 
\end{aligned} \tag{9.20}$$

and

$$\Pr\left\{\langle Q_Z P_Z\rangle < \frac{n^{-1}\sum_j (Q_{Zj}+P_{Zj})^2}{4(1+t)} - \frac{n^{-1}\sum_j (Q_{Zj}-P_{Zj})^2}{4(1-t)}\right\} \leq 2e^{-nt^2/8}\,. \tag{9.21}$$

Finally, we have that

$$\Pr\{x > x_{\max}\} \leq 2e^{-nt^2/8}\,,\ \Pr\{y > y_{\max}\} \leq 2e^{-nt^2/8}\,,\ \Pr\{z < z_{\min}\} \leq 4e^{-nt^2/8}\,, \tag{9.22}$$

with

$$x_{\max} = \frac{1}{1-t}\sum_j \frac{Q_{Aj}^2+P_{Aj}^2}{2n}\,,\ y_{\max} = \frac{1}{1-t}\sum_j \frac{Q_{Bj}^2+P_{Bj}^2}{2n}\,, \tag{9.23}$$

and

$$\begin{aligned}
z_{\min} = \min_{s_1,s_2,s_3\in\{-1,1\}} \Bigg| & w_1 \frac{n^{-1}\sum_j Q_{Zj}^2}{1+s_1 t} + w_2 \frac{n^{-1}\sum_j P_{Zj}^2}{1+s_2 t} \\
& + w_3 \left(\frac{n^{-1}\sum_j (Q_{Zj}+P_{Zj})^2}{4(1+s_3 t)} - \frac{n^{-1}\sum_j (Q_{Zj}-P_{Zj})^2}{4(1-s_3 t)}\right)\Bigg|\,,
\end{aligned} \tag{9.24}$$

where $w_1$, $w_2$ and $w_3$ are defined in Eq. (9.13-9.15). We set

$$t = \sqrt{n^{-1}\, 8 \ln(8/\epsilon_{\mathrm{PE}})}$$

so that

$$\Pr\{x > x_{\max} \vee y > y_{\max} \vee z < z_{\min}\} \leq \epsilon_{\mathrm{PE}}\,. \tag{9.25}$$

Next we investigate the case of an entangling cloner attack for two settings:

1. symmetric attacks in which both communication channels from Alice to the relay and from Bob to the relay are simulated by a beam-splitter with equal transmissivity $\tau_A = \tau_B = \tau$;

2. asymmetric attacks where the relay is assumed very close to Alice's lab, $\tau_A \simeq 1$.

We assume that Alice and Bob's variables have modulation variance $\langle {Q'_A}^2 \rangle = \langle {P'_A}^2 \rangle = \langle {Q'_B}^2 \rangle = \langle {P'_B}^2 \rangle = V_M$, so that the following covariance values are

$$\langle Q'_A Q_Z \rangle = -\sqrt{\frac{\tau_A}{2}} V_M \,, \tag{9.26}$$

$$\langle P'_A P_Z \rangle = \sqrt{\frac{\tau_A}{2}} V_M \,, \tag{9.27}$$

$$\langle Q'_B Q_Z \rangle = \langle P'_B q_Z \rangle = \sqrt{\frac{\tau_B}{2}} V_M \,, \tag{9.28}$$

and the covariances of mutually conjugate quadratures vanish. It also holds that $\langle Q_Z P_Z \rangle = 0$ and

$$\langle Q_Z^2 \rangle = \langle P_Z^2 \rangle = \frac{\tau_A + \tau_B}{2} V_M + 1 + \frac{\xi_A + \xi_B}{2} =: \nu \,, \tag{9.29}$$

where $\xi_A = (1 - \tau_A)(\omega_A - 1)$, $\xi_B = (1 - \tau_B)(\omega_B - 1)$ are the excess noise variances and $\omega_{A,B}$ are the thermal noise that Eve injects in the channels respectively. Then the only non-zero displacement coefficients are

$$u_{q'_A} = -\sqrt{\frac{\tau_A}{2}} \frac{V_M}{\nu} \,, \tag{9.30}$$

$$v_{p'_A} = \sqrt{\frac{\tau_A}{2}} \frac{V_M}{\nu} \,, \tag{9.31}$$

$$u_{q'_B} = v_{p'_B} = \sqrt{\frac{\tau_B}{2}} \frac{V_M}{\nu} \,, \tag{9.32}$$

that means

$$w_1 = w_2 = -\frac{\sqrt{\tau_A \tau_B}}{4} \frac{V_M^2}{\nu^2} \,, \tag{9.33}$$

and $w_3 = 0$. Finally, applying Eq. (9.24) we obtain

$$z_{\min} = \frac{\sqrt{\tau_A \tau_B}}{2(1+t)} \frac{V_M^2}{\nu} \,, \tag{9.34}$$

and similarly, from Eq. (9.23),

$$x_{\max} = \frac{V_M}{1-t} \left( 1 - \frac{\tau_A}{2} \frac{V_M}{\nu} \right) \,, \tag{9.35}$$

$$y_{\max} = \frac{V_M}{1-t} \left( 1 - \frac{\tau_B}{2} \frac{V_M}{\nu} \right) \,. \tag{9.36}$$

## 9.3 Secret key rate in a composable framework

The secret key rate in a composable framework is given by

$$\begin{aligned} r_n^{\epsilon'} \geq r_n^0 &- \frac{1}{\sqrt{n}} \Delta_{\text{AEP}} \left( \frac{2}{3} p \epsilon_s, d \right) \\ &+ \frac{1}{n} \log \left( p - \frac{2}{3} p \epsilon_s \right) + \frac{1}{n} 2 \log (2\epsilon) \,, \end{aligned} \tag{9.37}$$

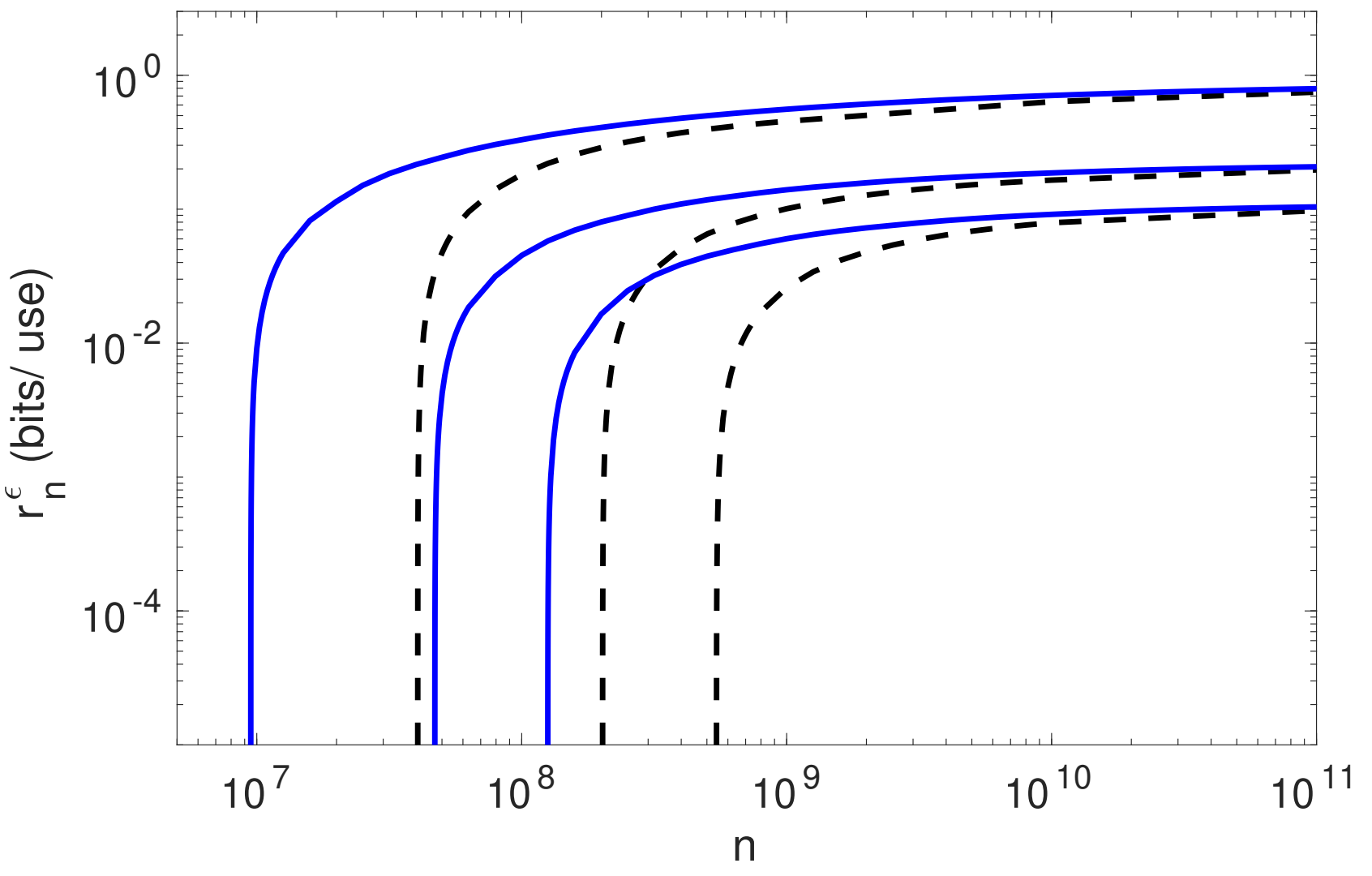

*Figure 9.1: Secret key rate vs block size for asymmetric attacks: $\tau_A = 0.99$ and different values of $\tau_B$ (from top to bottom the attenuation of the communication channel from Bob to the relay is of 1dB, 2dB, and 4dB). The excess noise is $\xi_A = 0$ and $\xi_B = 0.01$ (in shot noise unit). Solid lines are for collective Gaussian attacks, and dashed lines are for coherent attacks. For both kinds of attack, the overall security parameter is smaller than $10^{-20}$.*

where $\epsilon' = \epsilon + \epsilon_s + \epsilon_{\text{EC}} + \epsilon_{\text{PE}}$ and $n$ is the block size. The main term connected to the asymptotic secret key rate is

$$r_n^0 = \xi \tilde{I}_{AB} - \tilde{I}_E, \tag{9.38}$$

where $\tilde{I}_{AB}$ and $\tilde{I}_E$ are the classical mutual information and Holevo information calculated by the estimated CM given in the previous paragraph and accepting an error for $\epsilon_{PA}$ for parameter estimation. We assume that this estimated CM corresponds to a Gaussian state that minimizes the secret key rate. The parameter $\xi \leq 1$ stands for the reconciliation efficiency and theoretically is dependent on $n$ and on the error $\epsilon_{EC}$ of the error correction procedure (for plotting the secret key rate later in this section we assumed that is constant and $\xi = 0.95$).

The last term $\frac{1}{n} 2 \log(2\epsilon)$ accounts for the privacy amplification procedure and is dependent on $n$ and on the privacy amplification parameter $\epsilon_{PA}$. The term $\frac{1}{n} \log\left(p - \frac{2}{3} p \epsilon_s\right)$ is connected to the fact that the state after error correction with probability of success $p$ is not a product state as it was initially assumed for the calculation of the rate. The term

$$\Delta_{\text{AEP}}(\delta, d) \leq 4(d+1)\sqrt{\log(2/\delta^2)}$$

with $\delta = \frac{2}{3} p \epsilon_s$ compensates for the fact that we use the Holevo information for calculating a secret key rate in the non-asymptotic regime. It is also dependent on $d$, the number of bits after discretizing the variables of Alice and Bob. Finally, the smoothing parameter $\epsilon_s$ is a technical parameter that is connected with the fact that initially the number of the secret bits between Alice and Bob are quantified by the conditional smooth min-entropy of their variables as is shown in Ref. [53].

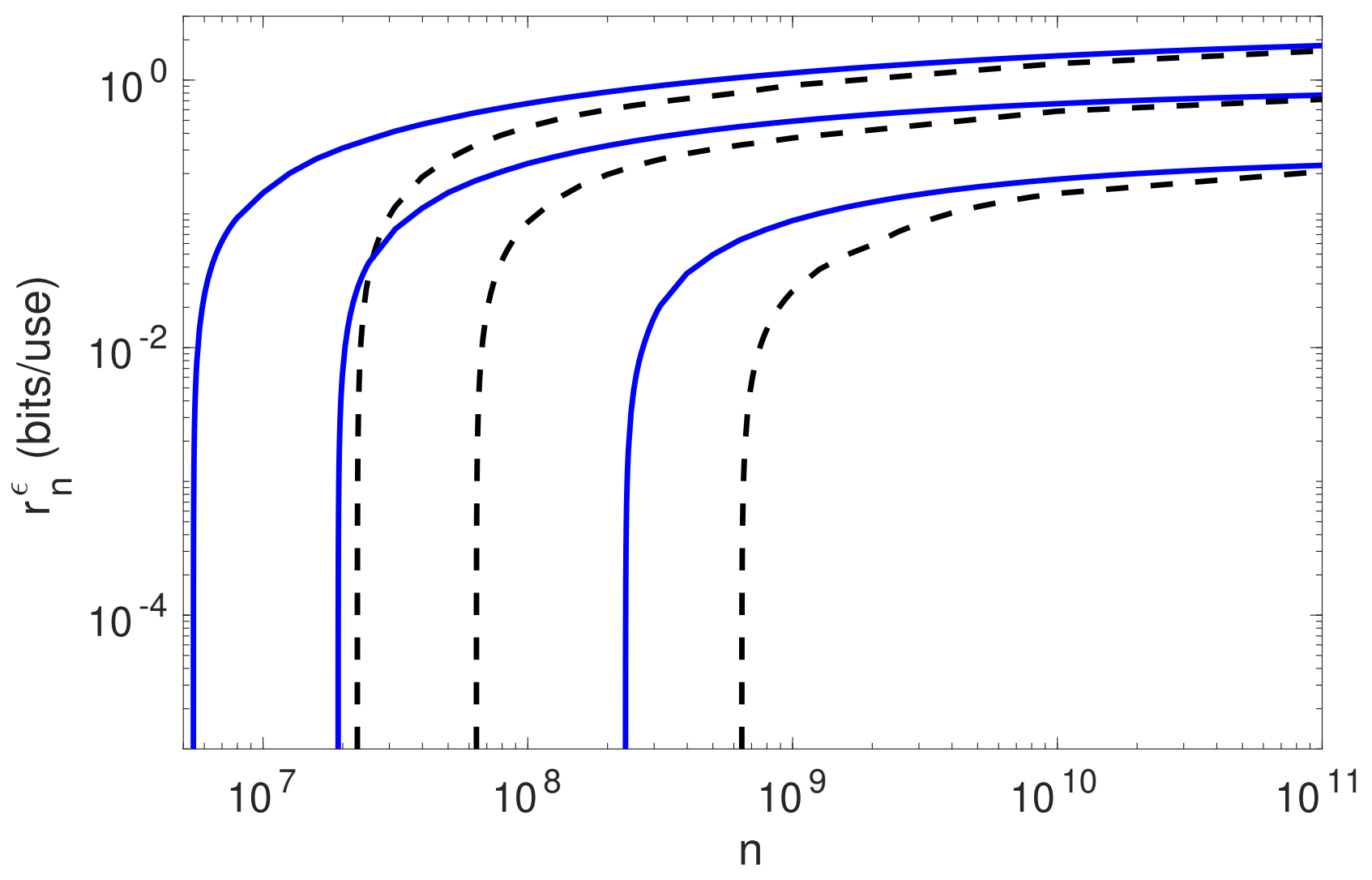

*Figure 9.2: Secret key rate vs block size for symmetric attacks and different values of $\tau_A = \tau_B$ (from top to bottom the symmetric attenuation is of $0.1dB$, $0.3dB$, $0.5dB$, and $0.55dB$). The excess noise is $\xi_A = \xi_B = 0.01$ (in shot noise unit). Solid lines are for collective Gaussian attacks, and dashed lines are for coherent attacks. For both kinds of attack, the overall security parameter is smaller than $10^{-20}$.*

For coherent attacks, by applying the results of Ref. [18] we obtain

$$\begin{aligned} r_n^{\epsilon''} \geq & \frac{n-k}{n} r_n^0 - \frac{\sqrt{n-k}}{n} \Delta_{\mathrm{AEP}} \left( \frac{2}{3} p \epsilon_s, d \right) \\ & + \frac{1}{n} \log \left( p - \frac{2}{3} p \epsilon_s \right) + \frac{1}{n} 2 \log (2\epsilon) \\ & - \frac{1}{n} 2 \log \binom{K+4}{4}, \end{aligned} \tag{9.39}$$

where $k$ is the number of signals used for the energy test that allows us to use to reduce any general coherent attacks to a Gaussian collective one, $K \sim n$ is the correction term that decreases the length of the secret key calculated for collective attacks in return for extending the security to coherent attacks and $\epsilon'' = \frac{K^4}{50} \epsilon'$.

In Fig. 9.1 and Fig. 9.2, this rate is plotted versus the block size $n$, for different values of the transmissivities and excess noise, for error correction efficiency of $\beta = 95\%$. The plots are obtained for $\epsilon = \epsilon_s = \epsilon_{\rm EC} = \epsilon_{\rm PE}$ chosen in such a way to obtain $\epsilon'' < 10^{-20}$. The rate is then obtained by maximizing over $k$ and the modulation $V_M$ and for $p = 0.99$.

## 9.4 Conclusion

In this chapter we presented numerical examples of the performance of the CV-MDI protocol taking into consideration finite-size effects in a composable framework. The statistical analysis of the channel parameter estimation has been given without using a central limit theorem assumption and thus is more rigorous mathematically for being used in a non-asymptotic regime. For this reason, it does not give as tight bounds as it may be manifested by other methods (see Chap. 8). However, it is still in accordance with the results of the asymptotic analysis of the CV-MDI protocol. In particular, it provides with almost the same level of performance even for the case of coherent attacks for moderate block sizes of about $10^9$ confirming the potential of this scheme for the use of metropolitan telecommunications.

# Chapter 10

# Uniform fast-fading channels

## 10.1 Introduction

Apart from the finite size effects (see Chap. 6) that should be taken into consideration in a realistic description of a protocol, there are also other assumptions that can contribute to a simulation of a practical situation. One of them is about the temporal variations of the communication channel between the two authenticated remote users. This is described by the so-called fading of the channel. In this case, its transmissivity $\tau$ is not fixed and can vary consistent with a given probability distribution [37]. This phenomenon can exist due to the use of a free-space channel [38] affected by environmental conditions, e.g., atmospheric turbulence [39–45].

In other studies, e.g., Ref. [46], both the parties as well as the eavesdropper are susceptible to variations in the channel transmissivity due the environment. In this section, based on Ref. [47], we present a basic model considering a uniform distribution for the variation of the transmissivities and describe an asymmetric situation between the users and the Eavesdropper encapsulating the worst-case scenario. Here the fading of the channel is under the complete control of the eavesdropper who can prepare different values of the transmissivity spontaneously for each channel use.

We present results for the one-way protocol in reverse reconciliation and homodyne detection (see Chap. 3), the measurement device independent protocol and a setting for three users connected via a relay (expanding the later in a network configuration as in Sec. 5.4).

## 10.2 Secret key rate with for a fast-fading channel

Under the previous assumption of fading, the transmissivity is quite fast meaning that the parties have access to its statistical distribution and have no knowledge of its instantaneous values. This brings an asymmetry between the eavesdropper and the authorized parties. More rigorously, this can be expressed by calculating in a different way the two asymptotic secret key rate terms (see Eq. 3.44) expressing a more conservative perspective. In particular, given a distribution for the transmissivity we will compute the mutual information term between the parties $I_{AB}$ with respect its minimum value and we will replace with the average of the term $I_E$ accounting for Eve's knowledge on the parties shared data string over it. This will give us the secret key rate with fast-fading

$$R_{\text{fast}} = \xi I_{AB}|_{\tau_{\min}} - \bar{I}_E \tag{10.1}$$

with $\xi \in [0, 1]$ the reconciliation efficiency. On the other hand, in slow- fading both parts of the asymptotic secret key rate are averaged over the transmissivity distribution. This is because a slow change of the values of the transmissivity is taken under consideration, which stays constant for a large number of channel uses resulting in its accurate estimation. We base our calculations in a simple model for the fading of the transmissivity assuming a uniform distribution, which, however, encapsulates the main concept of the eavesdropper tampering with the transmissivity of the channel.

## 10.3 One-way QKD

Let us assume that we have a fast uniform fading channel with minimum value $\tau_{\min}$ a maximum and a maximum value $\tau_{\max} = \tau_{\min} + \Delta\tau$, where $\Delta\tau$ is the variance of the fading. We calculate the fast-fading rate by using the asymptotic key rate for fixed transmissivity values in Eq. (3.44). More specifically, we set in the mutual information $I_{AB}$ the value $\tau_{\min}$ and integrate the Holevo function, which is dependent on the transmissivity over the fast-fading distribution. This results in the following formula

$$R_{\text{fast},\Delta\tau}(\tau_{\min}) = \xi \ I_{AB}|_{\tau=\tau_{\min}} - \frac{1}{\Delta\tau} \int_{\tau_{\min}}^{\tau_{\max}} d\eta I_E(\eta). \tag{10.2}$$

In contrast, for slow-fading, both Alice and Bob's mutual information and Eve's Holevo information need to be averaged with respect the uniform fading probability distribution so that

$$R_{\text{slow},\Delta\tau}(\tau_{\min}) = \frac{1}{\Delta\tau} \int_{\tau_{\min}}^{\tau_{\max}} d\eta [\xi I_{\text{AB}}(\eta) - I_E(\eta)]. \tag{10.3}$$

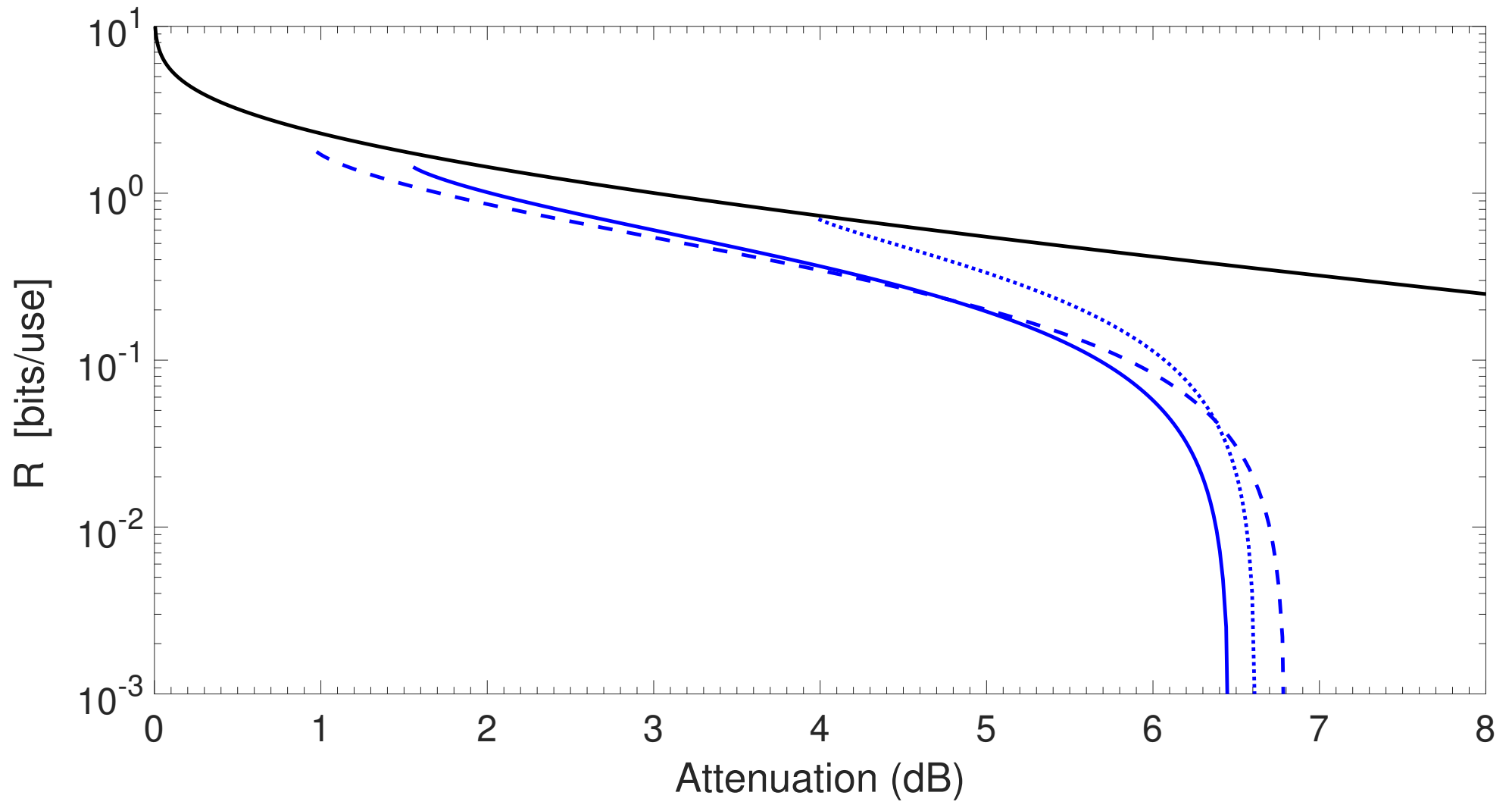

*Figure 10.1: Secret key rate is plotted in terms of $\eta_{min}$ for a fast-fading channel with $\Delta\eta = 0.2$ (dashed blue line), $\Delta\eta = 0.5$ (solid blue line) and $\Delta\eta = 0.6$ (dotted blue line). We set $\omega = 1$ (passive eavesdropping), $\beta = 1$ (ideal reconciliation) and $\mu = 10^6$. We compare the results with the PLOB bound for repeaterless private communications over a lossy channel (black line) [63]. We can see that high rates can be achieved up to losses of about $6 - 7dB$, where the rates start to rapidly decrease. Note that some curves start from non-zero dBs because we need to enforce $\eta_{min} + \Delta\eta \leq 1$, otherwise the fading channel may become an amplifier.*

The fast-fading secret key rate is illustrated in Fig. 10.1 for $\Delta\eta = 0.2$, $\Delta\eta = 0.5$ and $\Delta\eta = 0.6$ over the minimum of the transmissivities of the fading channel $\tau_{\rm min}$. We compare these rates with the PLOB bound plotted with respect to the minimum transmissivity $\tau_{\rm min}$, which sets the limit for repeaterless private communication over a lossy channel [63]. The fast- and slow- fading secret key rates are presented in Fig. 10.2 over the attenuation in dB by setting $\tau_{\rm min} = 10^{-\frac{dB}{10}}$ and $\Delta\tau = 0.1$ for the ideal case of $\xi = 1$, classical modulation variance $V_M = 10^6$ and pure loss channel $\omega = 1$. In Fig. 10.3, the same cases are illustrated for $\xi = 0.98$ and $\omega = 1.01$ after optimizing over the classical modulation variance $V_M$. We can see that the two key rates are comparable for losses approximately less than 6 dB.

## 10.4 Continuous-variables measurement-device-independent quantum key distribution

In this section, we assume that the two parties are found in the same distance with respect the relay (symmetric configuration). Each channel connecting them with the relay is assumed to be a fading channel whose transmissivity follows a uniform probability distribution. Eve's thermal noise is the same for both channels $\omega_A = \omega_B = \omega$ and fixed. We modify Eq. (5.17) appropriately in order to simulate the fast-fading of the channels. In more detail, we calculate the mutual information $I_{AB}$ setting $\tau_A = \tau_B = \tau$ and then we replace with minimum of the transmissivities $\tau_{\min}$.

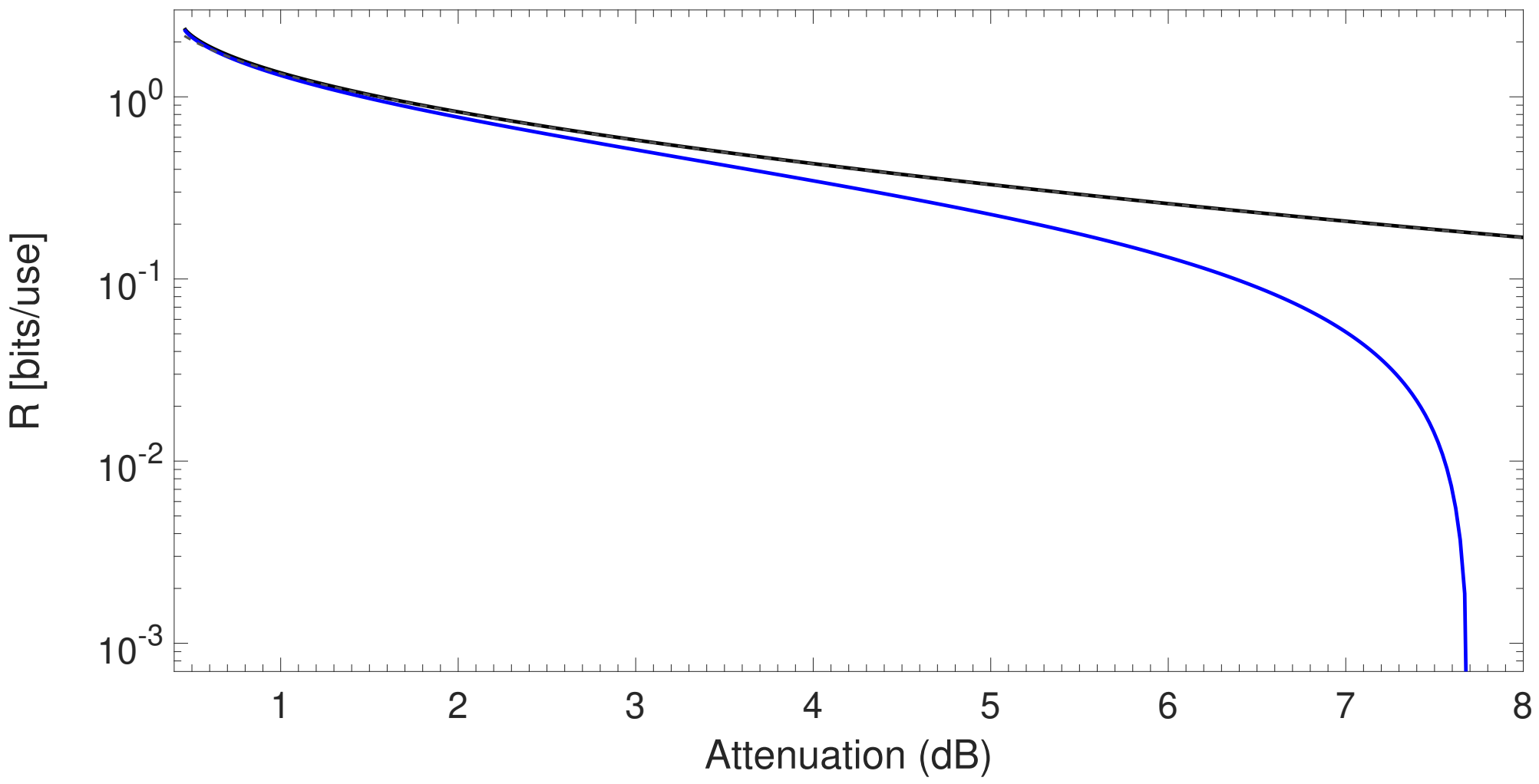

*Figure 10.2: Comparison between the key rates for fast and slow fading versus $\tau_{min}$. We plot the secret key rate for the fast-fading channel (lower blue line) and slow-fading channel (upper black line) for $\Delta\tau = 0.1$. We also set $V_M = 10^6$, $\omega = 1$ (passive eavesdropping) and $\beta = 1$ (ideal reconciliation). Performances are comparable within the range between* 0 *and* 6 dB.

Subsequently, we integrate the Holevo information $I_E$ depended on the transmissivities $\tau_A$ and $\tau_B$ over their uniform fading probability distribution of both transimmisivities. Each one of them is described by the same $\tau_{\min}$ and $\tau_{\max} = \tau_{\min} + \Delta\tau$, where $\Delta\tau$ is the fading variance. We finally obtain the following secret key rate expression

$$\begin{aligned} R^{\text{MDI}}_{\text{fast}}(\tau_{\min}) =& \xi \; I_{\text{AB}}|_{\tau=\tau_{\min}} - \frac{1}{(\Delta\tau)^2} \times \\ & \int_{\tau_{\min}}^{\tau_{\max}} \int_{\tau_{\min}}^{\tau_{\max}} d\eta_A d\eta_B I_E(\eta_A, \eta_B). \end{aligned} \tag{10.4}$$

In Fig. 10.4, we present the secret key rates for $\beta = 1$, $\omega = 1$ and very large modulation $\mu \simeq 10^6$, while we set $\Delta\eta = 0.1$ (solid lines). We see that the performance for fast-fading is not so far from that related to slow-fading and that is achievable with a standard lossy channel. We also present the same instances but for $\beta = 0.98$, optimizing over $\mu$ and setting $\omega = 1.01$ (dashed lines). In the latter case, the eavesdropper may also optimize her attack by exploiting correlations in the injected environmental state.

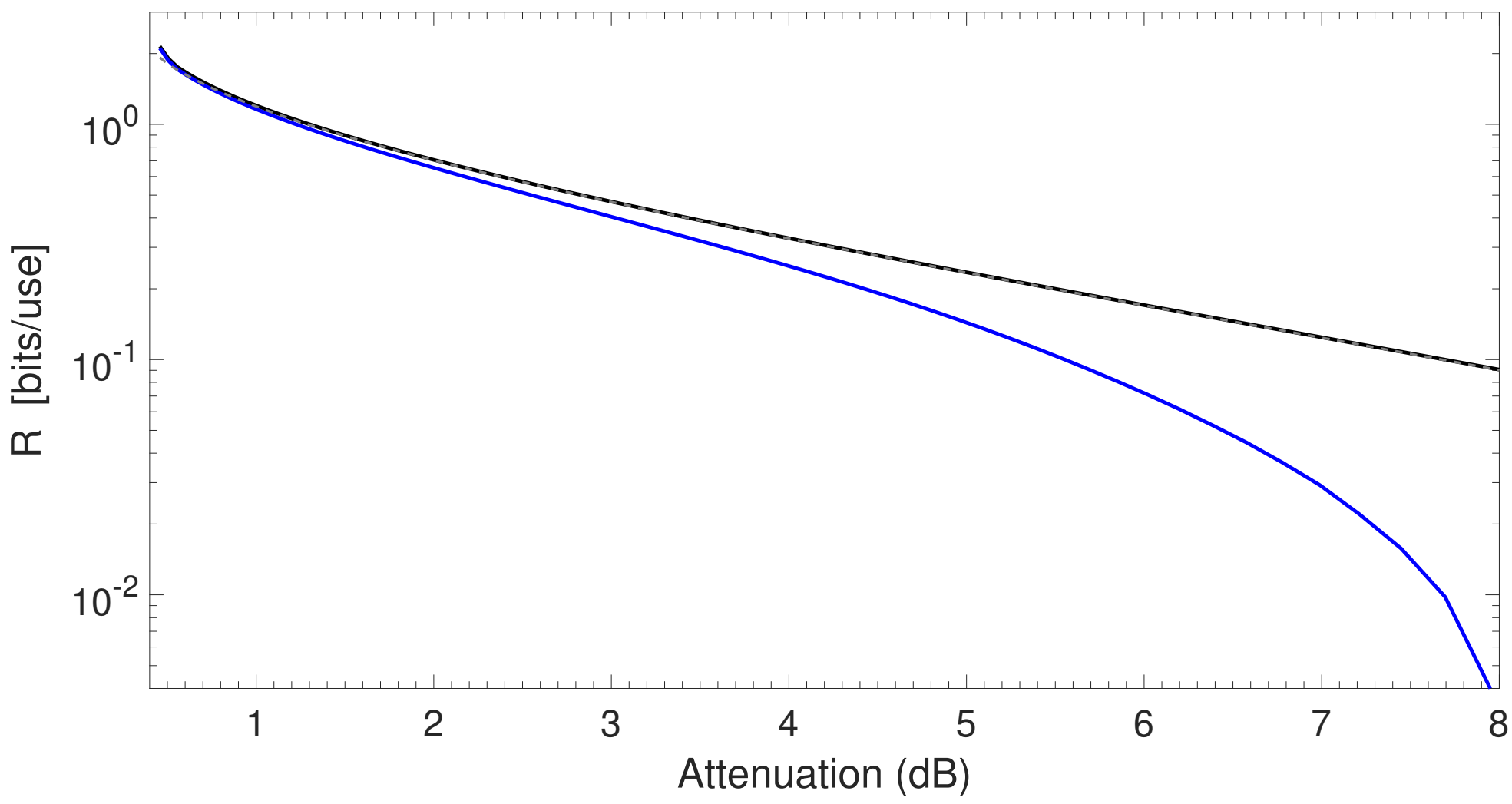

*Figure 10.3: Comparison between fast and slow-fading. As in Fig. 10.2 but for $\omega = 1.01$, $\beta = 0.98$, and optimized over $V_M$.*

## 10.5 CV-MDI-QKD three-user network

Let us assume the star configuration of Eq. (5.20) with fading affecting the channels, where the transmissivities follow uniform distribution between $\tau_{\min}$ and $\tau_{\max} = \tau_{\min} + \Delta\eta$. For obtaining an expression with the uniform fast-fading we need to integrate the Holevo bound over the distribution of the three channel transmissivities and compute the mutual information in the worst-case scenario by replacing with the minimum transmissivity. Thus, we write

$$R^{\text{star}}_{\text{fast},\Delta\tau}(\tau_{\min}) = \xi \; I_{\min}|_{\tau=\tau_{\min}} - \iiint_{\tau_{\min}}^{\tau_{\max}} \frac{I_E(\boldsymbol{\eta})}{(\Delta\tau)^3} d\boldsymbol{\eta}, \tag{10.5}$$

The rate for $\Delta\eta = 0.05$ and $\omega = 1$ is optimized over $\mu$ and shown in Fig. 10.5. From the figure, we can see that the performance is comparable to the case of slow-fading, where the parties' mutual information is averaged over the statistical distribution.

## 10.6 Conclusion

In this chapter, we have studied the one-way protocol, the measurement device independent protocol and a three user variation of it in terms of fast-fading. Fading of a channel is the phenomenon associated with the non-constant transmissivity of the channels connecting the parties during the communication process. This can be a consequence of changes in environmental conditions or in a more pessimistic scenario can result from of the eavesdropper's tampering with the channels transmissivity. The first case can sometimes be considered slow in the sense that the transmissivty can remain constant for sufficient number of channel uses part of which the parties can use so as to estimate it. However these changes may follow a more complicated distribution dependent on different aspects of the setting and the experimental equipment.

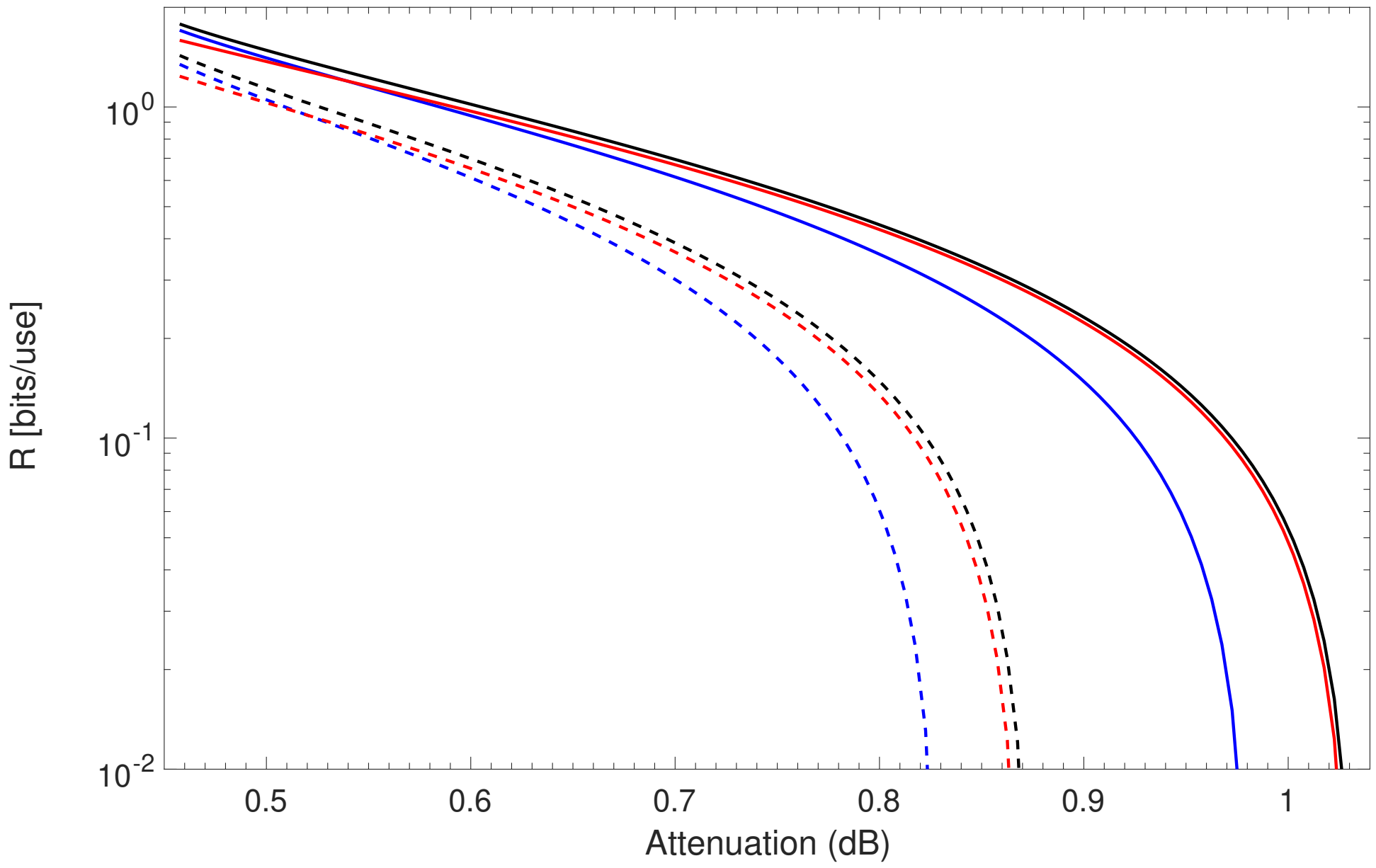

*Figure 10.4: Performance of the CV-MDI-QKD protocol in symmetric configuration assuming two fading channels in the channels with $\Delta\eta = 0.1$ and no excess noise ($\omega = 1$). We plot the secret key rate versus $\eta_{min}$ for fast fading (lower blue solid line) and slow-fading (upper black solid line). We also show the standard case of a non-fading lossy channel (middle red solid line) which is plotted in terms of the expectation value of $\eta$, i.e., $\bar{\eta} = \eta_{\min} + \frac{\Delta\eta}{2}$. Here we set $\beta = 1$ and $\mu = 10^6$. Then we plot the same rates but for $\beta = 0.98$, $\omega = 1.01$ and optimizing over $\mu$ (see the corresponding dashed lines).*

On the other hand, the distribution on the second case can be based on a simple model of the uniform distribution since here we assume a mechanical process executed by an antagonist. In this case, we consider is considering a fast change of the transmissivity randomly chosen from the eavesdropper for each channel use. This results in a disadvantage for the users, who have access to the distribution only in the end of the protocol and thus are forced to assume its minimum for their variable correlations. However, the eavesdropper's information is averaged over the distribution.

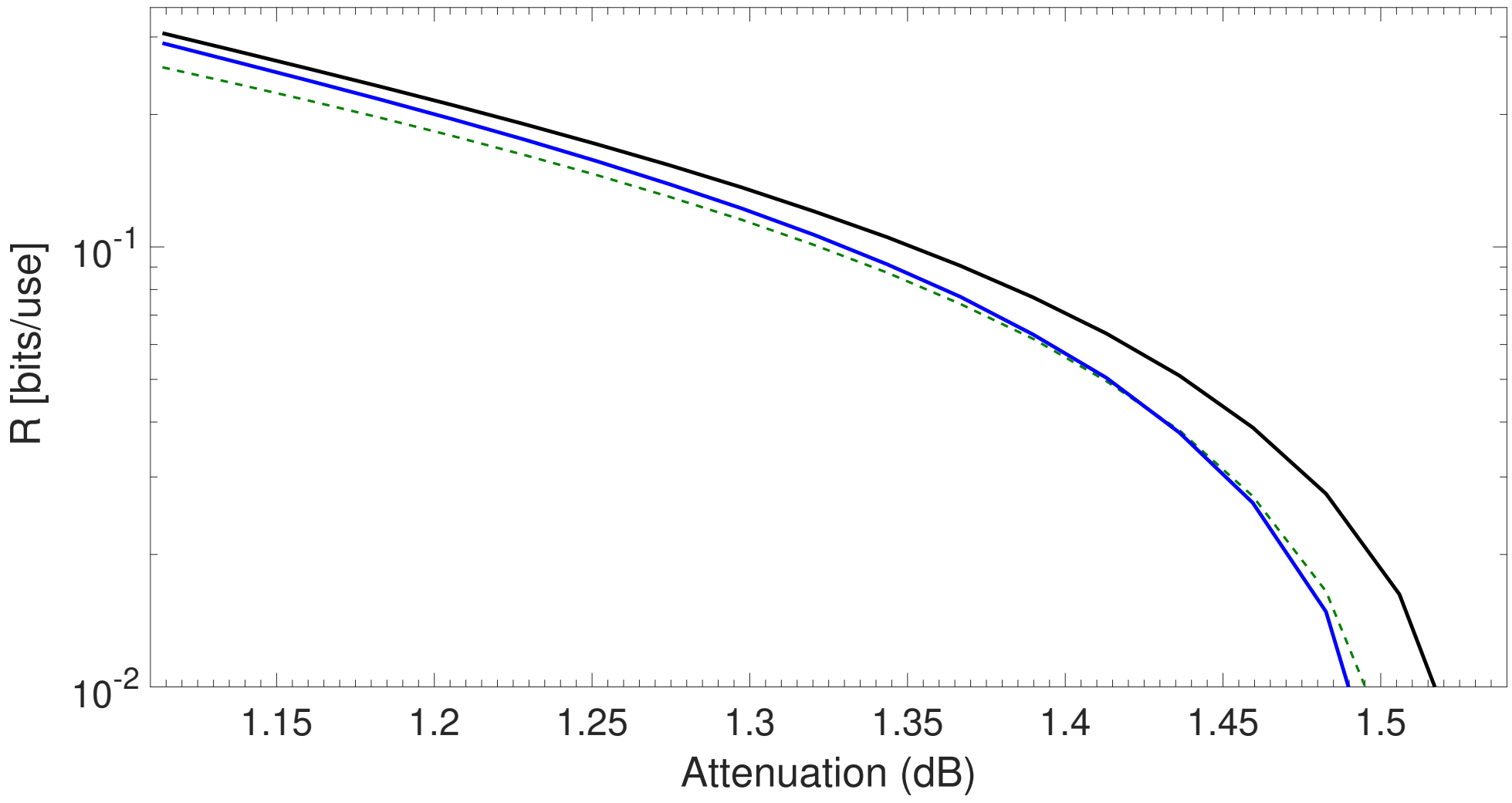

*Figure 10.5: The secret conferencing key rate versus $\eta_{min}$ in a star configuration of the three-party CV-MDI-QKD network assuming fast-fading channels (blue solid line) and slow-fading channels (black solid line) with the same variance $\Delta\eta = 0.05$. We also include the rate of the protocol in the presence of lossy channels with transmissivity $\bar{\eta} = \eta_{min} + \frac{\Delta\eta}{2}$. For all the plots we have optimized over $\mu \in [2, 20]$ and set $\omega = 1$.*

Although the last situation describes a more pessimistic secret key rate, our results show that the performance of the aforementioned protocols with fast-fading are comparable with the corresponding slow-fading cases. In fact, for typical distance ranges for each protocol, we notice high enough performances which implies a robustness to such channel fading assumptions.

# Chapter 11

# Discrete phase-encoding in coherent states: asymptotic security analysis in thermal-loss channels

## 11.1 Introduction

In this chapter we present results from Ref. [64], where we have extended the security analysis from considering a pure loss channel (see Sec. 3.6) to the thermal loss case. In order to estimate the excess noise of the channel in this case, we assume that the parties have send a sufficient amount of coherent states modulated with a Gaussian distribution that are not participating in the secret key extraction. Since the protocol is not a fully Gaussian protocol we have to use the density matrix notation in the Fock basis which we have to truncate for the numerical calculations.

## 11.2 Thermal-loss channel assumption

The calculations are based on Eve's system. According to the entangling cloner attack, Eve's input states is a TMSV state described by the density matrix (see Eq.2.53)

$$\rho_{Ee}(\lambda) = (1-\lambda^2)\sum_{n=0}^{\infty}(-\lambda)^{(k+l)}|k\rangle\langle l| \otimes |k\rangle\langle l|, \tag{11.1}$$

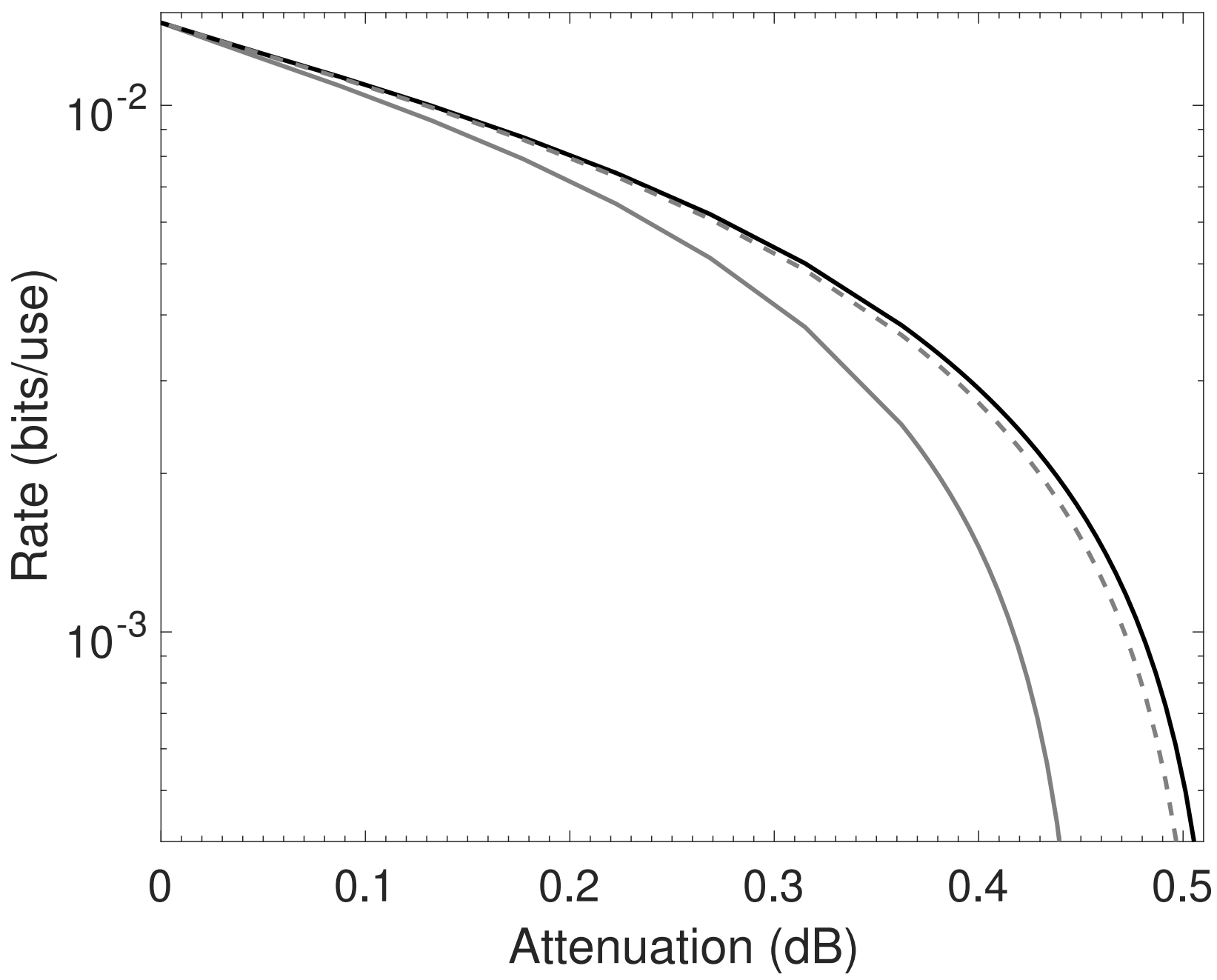

*Figure 11.1: Realistic secret-key rate (bits/use) over the attenuation (decibels) in direct reconciliation for $N = 4$ and $z = 0.1$. We plot the rate for a pure-loss channel (upper solid line) and a thermal-loss channel with mean photon number $\bar{n} = 0.01$ (middle dashed line) and $\bar{n} = 0.1$ (lower solid line).*

with $\lambda = \tanh\left[\frac{1}{2}\text{arcosh}(2\bar{n}+1)\right]$, where $\bar{n}$ is the mean number of thermal photons. The beam spliter operation that is applied on Alice's and Eve's modes is given by Eq. (2.66)

$$U(\theta) = \exp\left[\theta\left(\hat{a}_A^\dagger \hat{a}_E - \hat{a}_A \hat{a}_E^\dagger\right)\right], \tag{11.2}$$

where $\theta = \text{arcos}(\sqrt{\tau})$. According to Fig. 3.8, we notice that the global output state of Bob (mode $B$) and Eve (modes $e$ and $E'$) for a given $k$ is given by

$$\rho_{BE'e}(\theta, a_k, \lambda) = U(\theta)\Pi_{\text{A}}(a_k)\rho_{Ee}(\lambda)U^\dagger(\theta), \tag{11.3}$$

where Alice had prepared the state$\Pi_{\text{A}}(a_k) := |a_k\rangle\langle a_k|$. In order to find Eve's state after the propagation of the channel we have to trace with respect to Bob's mode $B$ obtaining

$$\rho_{\text{Eve}|k} := \rho_{\text{E'e}}(\theta, a_k, \lambda) = \text{Tr}_{\text{B}}[\rho_{BE'e}(\theta, a_k, \lambda)]. \tag{11.4}$$

As a result, Eve's average state can be obtained as the convex sum of Eve's conditional states from Eq. (11.4)

$$\rho_{\text{Eve}}(\theta, z, \lambda) = \frac{1}{N}\sum_{k=0}^{N} \rho_{\text{Eve}|k}. \tag{11.5}$$

Therefore, based on the von Neumann entropy of the previous states we can evaluate the Holevo information defined in Eq.(2.82)

$$\chi(\text{Eve} : X_A) = S(\rho_{\text{Eve}}) - \frac{1}{N}\sum_{k=0}^{N} S(\rho_{\text{Eve}|k}). \tag{11.6}$$

The entropy of the state $\rho_{\text{Eve}|k}$ does not depend on $k$, i.e., the phase of the amplitude of the coherent state that Alice has sent (see remarks after Def. 2.6.4). Thus Eq. (11.6) can be simplified to

$$\chi(\text{Eve} : X_A) = S(\rho_{\text{Eve}}) - S(\rho_{\text{Eve}|k}) \tag{11.7}$$

for any $k$.

The calculation of the mutual information between the authorized parties is based on the discussion of Sec. 3.6.2. The main difference is that Bob's distribution is dependent on the channel thermal mean photon number described by the following relation

$$p(b|a_k)(\bar{n}) = \text{Tr}[\Pi(b)\rho(\sqrt{\tau}a_k, (1-\tau)\bar{n})\Pi^\dagger(b)], \tag{11.8}$$

where $\Pi(b) := |b\rangle\langle b|$ and $\rho(\sqrt{\tau}a_k, (1-\tau)\bar{n})$ is a displaced thermal state with amplitude $\sqrt{\tau}a_k$ and mean photon number $(1-\tau)\bar{n}$. By the calculation in the Appendix B.3, we obtain

$$p(b|a_k)(\bar{n}) = \frac{\exp\left[\frac{|b-\sqrt{\tau}a_k|^2}{1+(1-t)\bar{n}}\right]}{\pi(1+(1-\tau)\bar{n})}. \tag{11.9}$$

Subsequently, we apply Bayes' rule and calculate the probability distribution $p(a_k|b)(\bar{n})$. Then by replacing it in Eq. (2.75), we obtain the mutual information.

Finally, we compute the secret key rate

$$R(\bar{n}) = I(X_A : X_B)(\bar{n}) - \chi(\text{Eve} : X_A) \tag{11.10}$$

with $I(X_A : X_B)(\bar{n})$ being the mutual information between Alice and Bob calculated from Eq. (11.9) and $\chi(\text{Eve} : X_A)$ is given by Eq. (11.7). In Fig. 11.1, we plot this secret key rate over the attenuation for a protocol with $N = 4$ and $z = 0.1$. In particular, we see that the performance obtained in the presence of thermal noise $\bar{n} = 0.01$ is not so far from the performance achievable in the presence of a pure-loss channel. In other words, the four-state protocol is sufficiently robust to the presence of excess noise. In Fig. 11.2, we optimize the rate with respect to the radius $z$ for both pure loss and thermal loss channels. In the lower panel, we can see the difference between the two optimal radii corresponding to the two cases. In Fig. 11.3, we also see incremental changes in the security threshold of the protocol by increasing the number of encoding states, e.g. from $N = 4$ to 10, assuming a thermal loss channel. In fact, we can see that after $N > 7$ the threshold saturates showing that there is no advantage by using a larger number of states.

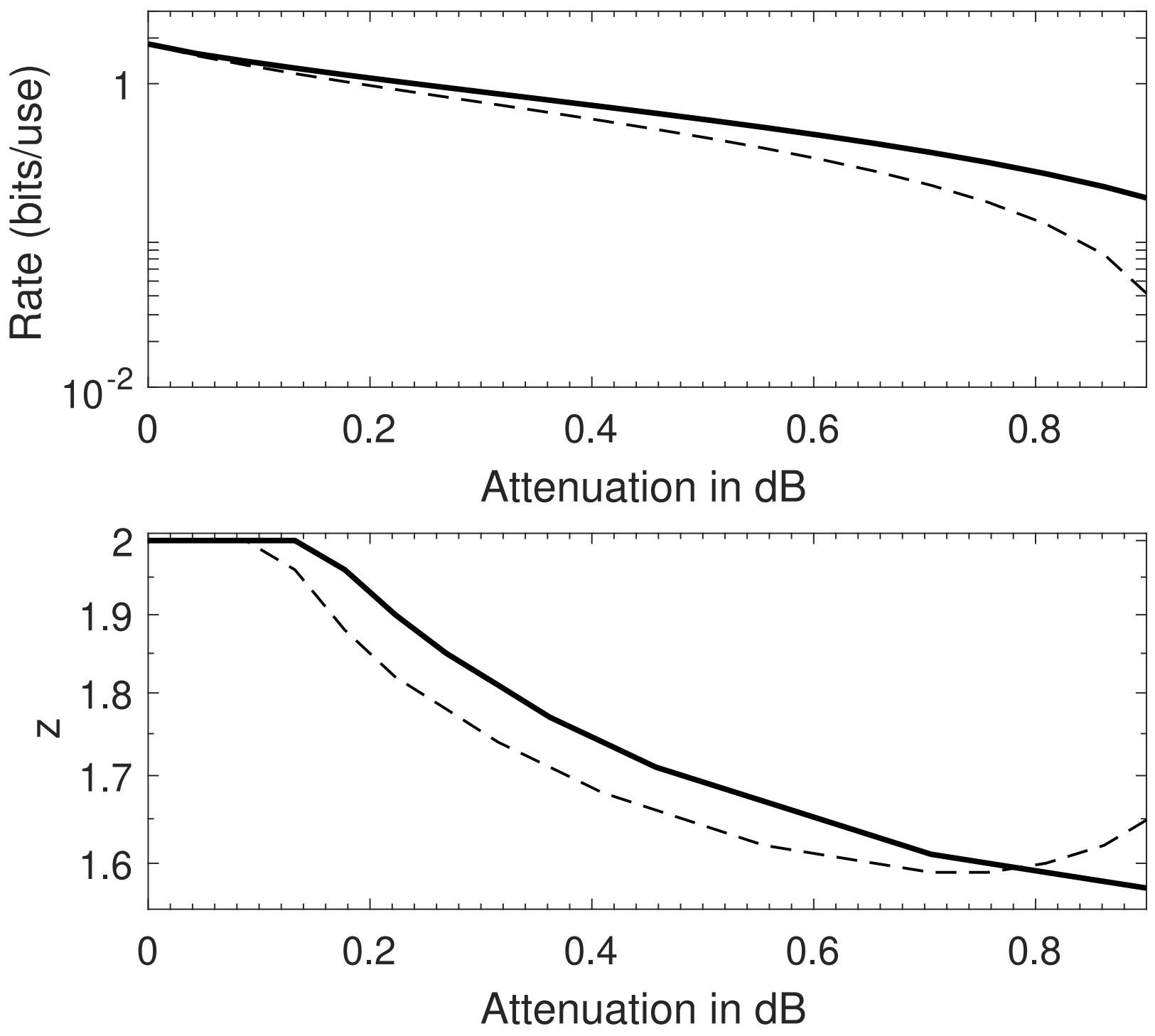

*Figure 11.2: Protocol for $N = 4$ in direct reconciliation. Upper panel: Key rate is optimized over $z \in (0, 2)$ and plotted versus attenuation. We assume a pure loss channel (solid line) and a thermal loss channel with $\bar{n} = 0.1$ (dashed line). Lower panel: We plot the optimal value of $z$ versus attenuation, for the pure loss case (solid line) and the thermal loss case (dashed line).*

In order to estimate the excess noise $\epsilon$ we assume a variation of the original protocol: Alice may send Gaussian-modulated decoy coherent states to Bob mixed with coherent states used for the key. At the end of all quantum communication, Alice informs Bob in which instances she was sending decoys so that they can use their data to compute $\epsilon$. As expected, we also have that direct reconciliation restricts the use of the protocol to low loss. The case is different for reverse reconciliation that we study below.

Let us now consider the reverse reconciliation case. Let us assume that the probability of Alice sending a state with amplitude $a_k$ given that Bob measured $b$ is given by $p(a_k|b)(\bar{n})$. Then Eve's conditional state can be given by the average state of an ensembles of states $\rho_{\mathrm{Eve}|k}$ each one associated with the previous probability. Therefore, Eve's conditional state

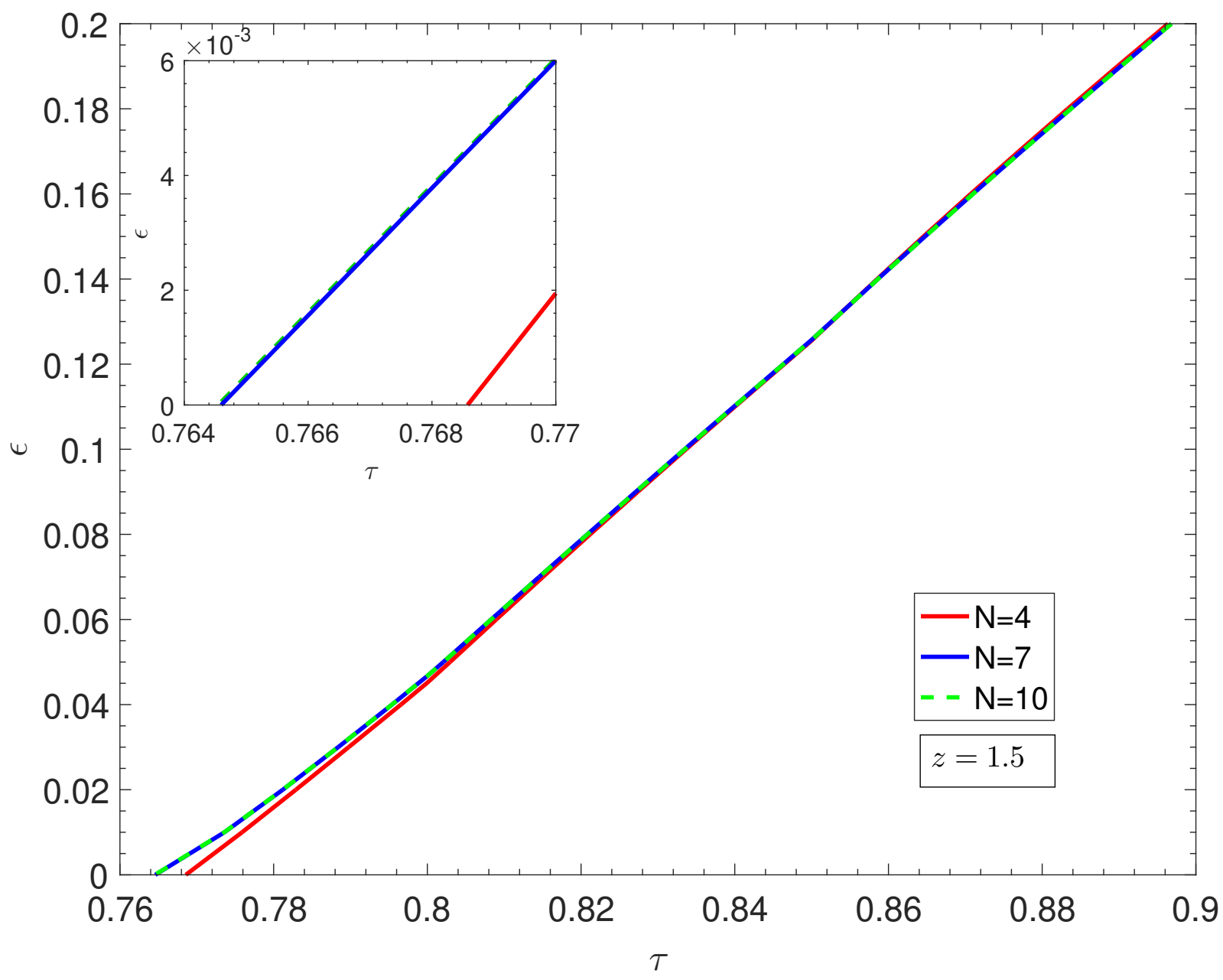

*Figure 11.3: We plot the security thresholds in direct reconciliation, i.e., maximum tolerable excess noise $\epsilon$ versus transmissivity $\tau$ for $N = 4$ (red line), $N = 7$ (blue dashed line) and $N = 10$ (green line) with $z = 1.5$. The threshold saturates for $N > 7$ showing that there is no advantage in terms of achievable distance by increasing the number of states. We have also included a figure containing a magnified area so as to clarify this saturation tendency.*

is given by a convex sum as follows

$$\rho_{E'e|b} = \sum_{k=0}^{N-1} p(a_k|b)(\bar{n})\rho_{\mathrm{Eve}|k}. \tag{11.11}$$

The conditional states $\rho_{\mathrm{Eve}|k}$ is given in Eq. (11.4) and $p(a_k|b)(\bar{n})$ comes from Eq. (11.9). Therefore, we may derive $S(\rho_{E'e|b})$ by using the definition of Eq. (2.77). Then we replace in the Holevo information function of Eq. (2.82) and calculate the secret-key rate

$$R(\bar{n}) = I(X_A : X_B)(\bar{n}) - S(\rho_{\mathrm{Eve}}) + \int d^2b\ p(b)(\bar{n})\rho_{E'e|b}, \tag{11.12}$$

where $p(b)(\bar{n}) := \frac{1}{N}\sum_{k=0}^{N-1} p(b|a_k)(\bar{n})$, $I(X_A : X_B)(\bar{n})$ is calculated based on Eq. (11.9) and $\rho_{\mathrm{Eve}}$ is given in Eq. (11.5). Numerically, we compute this rate by truncating the Hilbert space to a suitable number of photons, which is of the order of $\simeq 10-15$ photons for the specific regime of parameters considered.

In Fig. 11.4, we plot the reverse reconciliation secret key rate over the attenuation for the four-state protocol $N = 4$ with radius $z = 0.1$ and excess noise $\epsilon = 0.001$. We can see that the protocol is sufficiently robust to excess noise, achieving a rate of $6 \times 10^{-4}$ bits per channel use for attenuation values of about 20 dB. In this regime of energy, the performance of the protocol coincides with that of a Gaussian protocol modulating coherent state with modulation variance $V_M = 2z^2$ (and performing heterodyne detection on the channel output). On the contrary, for larger energies, e.g., for a constellation radius $z = 1$, the rate of the four-state protocol does not coincide with its Gaussian counterpart, as also illustrated in Fig. 11.5. Note that here we have not optimized over $z$ so as to be able to compare the actual potential of a discrete modulation protocol(in principle meaning ideal reconciliation) against a protocol with Gaussian modulation and $\xi < 1$ (see also Fig. 3.12). Here the four-state protocol can achieve a rate of the order of $4 \times 10^{-3}$ bits per channel use for attenuation values of about 15 dB and excess noise $\epsilon = 0.01$. So for this order of noise, which is a realistic consideration coming from experiments [24, 29], the protocol can achieve distances up to 75 km.

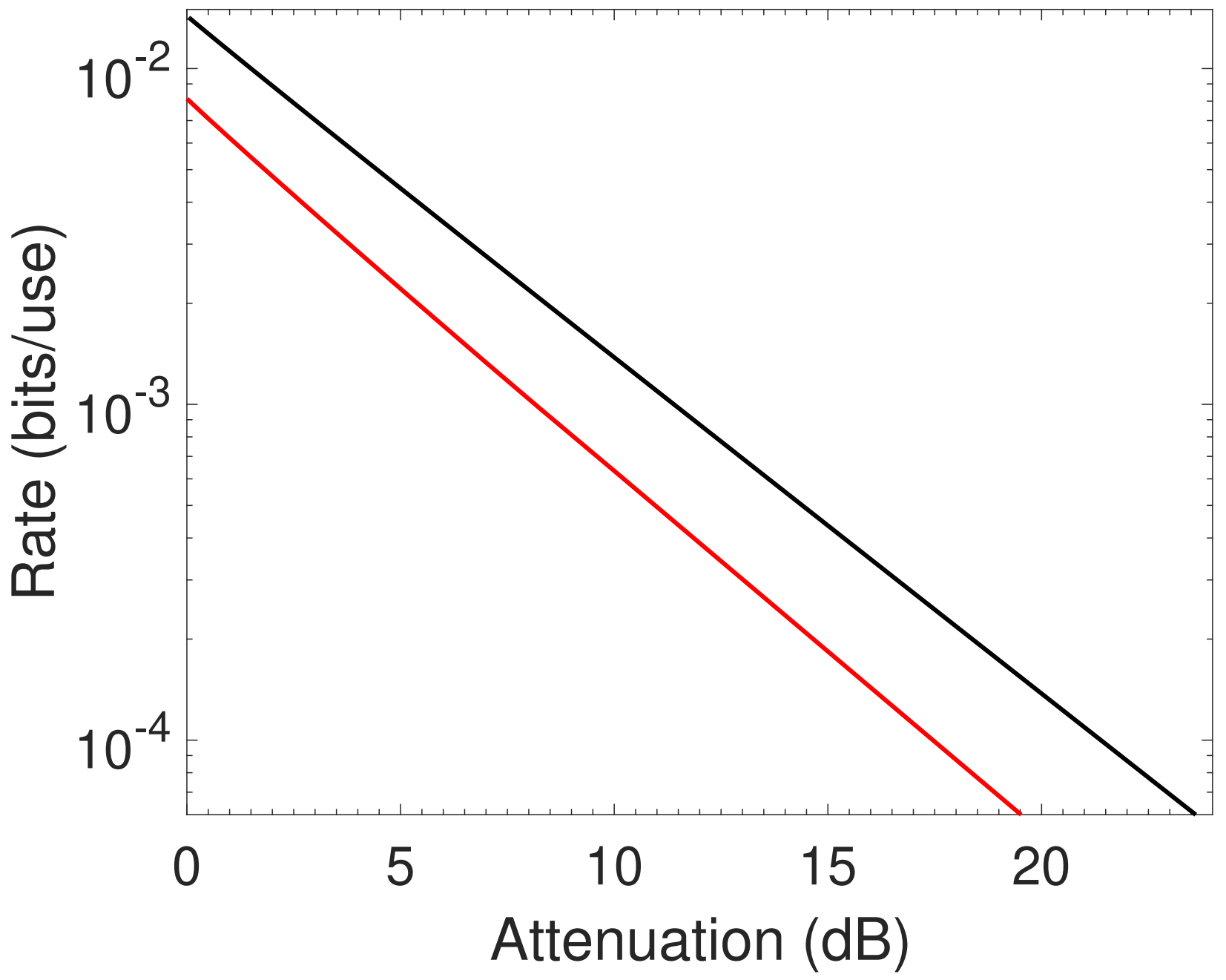

*Figure 11.4: Realistic secret key rate (bits/use) over the attenuation (decibels) in reverse reconciliation for $N = 4$ and $z = 0.1$. We have plotted the rate for a pure-loss channel (upper black line) and a thermal-loss channel with excess noise $\epsilon = 0.001$ (lower red line). Both these rates coincide with the corresponding rates achievable by a Gaussian protocol modulating coherent states with variance $V_M = 0.02$.*

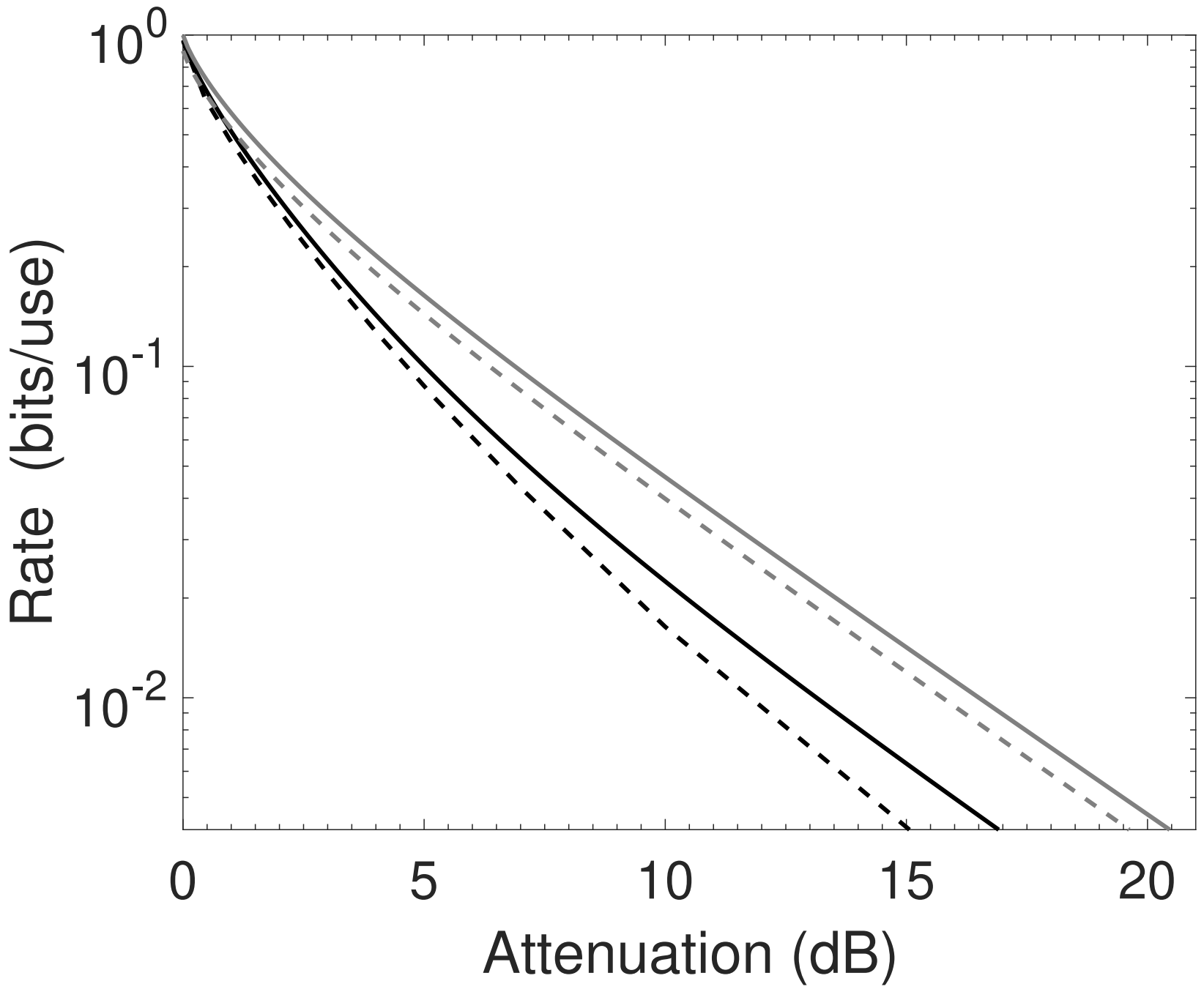

*Figure 11.5: Realistic secret key rate (bits/use) over the attenuation (decibels) in reverse reconciliation over the attenuation (decibels) for $N = 4$ and $z = 1$. We have plotted the rate for a pure-loss channel (lower solid line) and a thermal-loss channel with excess noise $\epsilon = 0.01$ (lower dashed line). The corresponding secret key rate for the protocol with Gaussian modulation ($V_M = 2$) has also been plotted for the case of pure-loss channel (upper solid line) and thermal-loss channel with excess noise $\epsilon = 0.01$ (upper dashed line). We see that, for this regime of energies, the rate of the four-state protocol does not coincide with the rate of the Gaussian protocol.*

## 11.3 Conclusion

In this chapter, we presented an extension of the security of the protocol using $N$ equidistant coherent states from the origin of the phase space to channel that has excess noise. More specifically, we used the assumption of the entangling cloner attack, which although is not the optimal attack, since the distribution of the signal coherent states is not a Gaussian, it still describes the realistic situation of a communication channel based on optical fibres. In this study, we did no make any assumption with respect to a Gaussian approximation and we found that the reverse reconciliation version of the protocol manifests the appropriate robustness regarding loss and channel noise for communication at metropolitan mid-range distances.

# Chapter 12

# Conclusion

In this thesis, we contributed in the research about CV-QKD. Our main goal was to obtain security analysis of the Gaussian protocols using thermal states and of the measurement device independent protocol incorporating finite size effects. This analysis allows us to assess the security and performance of the protocol in terms of achievable distance or level of the secret key rate in the realistic situation of finite-number exchanged signals as the core aspect of a practical application. Based on this analysis, we provided with results for both protocols showing their performance with respect to the relevant parameter of the block size. We noticed that for moderate block sizes their performance is quite close to the performance considered in the asymptotic regime, that made these protocols particularly promising for metropolitan area applications. More specifically, for the thermal state protocol, we noticed that for achieving reasonable key rates using signals in the lower frequencies, we should constrain the distance of the nodes to very small values and use larger block sizes.

In addition, such analysis can be considered as an intermediate step of achieving a stronger notion of security, that in a composable framework. We showed from which parameters such an analysis is dependent on for measurement-device-independent protocol and proposed a channel parameter estimation that is not based on the central limit theorem. As a secondary goal, we expanded the asymptotic secret key analysis of protocols such as the coherent states protocol with Gaussian modulation, the CV-MDI-QKD protocol and a three-user star network configuration to an asymmetric scenario for the users against the eavesdropper associated with the fast fading channels. We investigated the performance of such protocols and observed that it was close compared to the symmetric scenario of slow fading confirming that they are quite robust even to a such a pessimistic assumption. Finally, we assumed that the parties use a thermal loss channel for exchanging an arbitrary number of phase-encoded coherent states. To the best of our knowledge this has not been done without a Gaussian approximation, although such an assumed entangling cloner attack simulates the most realistic situation of a typical link in telecommunications. Here, we noticed the expected decrease in the performance of the protocol due to the additional excess channel noise but the reverse reconciliation version turns out to be able to achieve performances compatible with mid-range metropolitan distances.

## 12.1 Outlook

We have seen that protocols with discrete modulation can have almost the same performance with protocols with Gaussian modulation when the energy used for the encoding is quite small. It would be very interesting to see the implication of this for protocols using discrete encoding of thermal states or with discrete-modulation used in the CV-MDI-QKD protocol. Furthermore, a finite-size analysis for such protocols with discrete encoding would be a challenging step since in that case we cannot use the assumption of Gaussian variables for the channel parameter estimation.

### 12.1.1 Fast-fading channels

Our calculations in this chapter took into consideration a uniform distribution as the simplest case of fading but still incorporating the aspect of asymmetry that it was our main research goal. However, it would be very interesting in a future work to see a comparison of fast fading with other studies considering more complex distributions according to atmospheric turbulence phenomena. A starting point will be by assuming values for the variance of the uniform distribution so that it can describe such phenomena as above.

Furthermore, our calculations have been made only for symmetric configurations. It would be interesting to at some point an investigation to be made for asymmetric configurations in terms of distance from the relay. However, in any case, the parties can estimate the smallest of the transmissivities and build their secret key rate on this assumption.

# Appendix A

# Useful elements of estimation theory

## A.1 Bi-variate Normal distribution

According to the method of maximum likelihood estimation, for a bivariate normal distribution $X = (X_1, X_2)$, the estimators for the mean $\mu = (\mu_1, \mu_2)$ and the covariance matrix $\mathbf{V}$ are given by

$$\hat{\boldsymbol{\mu}} = \frac{1}{m}\sum_{i=1}^{m}\mathbf{X}_i \tag{A.1.1}$$

$$\widehat{\mathbf{V}} = \frac{1}{m}\sum_{i=1}^{m}(\mathbf{X}_i - \hat{\boldsymbol{\mu}})(\mathbf{X}_i - \hat{\boldsymbol{\mu}})^T \tag{A.1.2}$$

where $\mathbf{X}_i$ is the $i$-th statistical realization out of $m$ realizations of $\mathbf{X}$.

## A.2 The Central Limit Theorem

The central limit theorem states that, assuming $m \gg 1$ realizations $X_1, X_2, \ldots, X_m$ of a random variable $X$ with unknown density function $f$, mean $\mu$ and variance $\sigma^2 < \infty$, the sample mean

$$\bar{X} = \frac{1}{m}\sum_{i=1}^{m}X_i \tag{A.2.3}$$

is approximately normal with mean $\mu$ and variance $\sigma^2/m$.

## A.3 Chi-squared distribution

In order to estimate the mean value of a variable $Y$, that depends on the square of a variable $X$ for which we have $m$ realizations, we can use the following result: For $m$ realizations $X_i$, for $i = 1, 2, \ldots, m$, of a normally distributed variable $X$, having mean $\mu$ and unit variance, the variable

$$Y = \sum_{i=1}^{m} X_i^2 \sim \chi^2(k, \lambda) \tag{A.3.4}$$

is distributed according to the $\chi^2$ distribution with $k = m$ degrees of freedom and $\lambda = m\mu^2$. The mean value and variance of the chi-squared distribution is given by

$$\mathbb{E}(Y) = k + \lambda, \tag{A.3.5}$$

and

$$\mathrm{Var}(Y) = 2(k + 2\lambda). \tag{A.3.6}$$

**Remark**:A chi-squared distributed variable $X \sim \chi^2(k, \lambda)$ tends to a normal distribution for $k \to \infty$ or $\lambda \to \infty$ (see Ref. [65]).

## A.4 Confidence intervals

An error of $\epsilon = 10^{-10}$ defines a significance level $\delta = 2\epsilon$ for the symmetric confidence interval of a Gaussian distributed variable $X \sim G(\mu, \sigma^2)$ with mean $\mu$ and variance $\sigma^2$. This means that the probability of $X$ to be in the interval $\{\mu - x, \mu + x\}$ is given by

$$P(\mu - x < X < \mu + x) = 1 - \delta. \tag{A.4.7}$$

Subsequently, we have that

$$P(X > \mu + x) = \delta/2 \tag{A.4.8}$$

$$P(Y > x/\sigma) = \delta/2 \tag{A.4.9}$$

$$\Phi(x/\sigma) = 1 - \delta/2, \tag{A.4.10}$$

where $Y = \frac{X-\mu}{\sigma}$ is following a standard normal distribution and $\Phi(x/\sigma)$ is the corresponding cumulative distribution function. This results in

$$x = \Phi^{-1}(1 - \delta/2)\sigma$$

where $\Phi^{-1}(1 - \delta/2)$ is approximated to be $\sim 6.5$ for the given error $\epsilon$.

## A.5 Optimal combination of estimators

Let us assume to have two estimators $\hat{s}_1$ and $\hat{s}_2$, with variances $\sigma_1^2$ and $\sigma_2^2$, for the same quantity $s$ acquired by different processes. We then compute the optimal linear combination of the variances by the following formula

$$\sigma_{\text{opt}}^2 = \frac{\sigma_1^2 \sigma_2^2}{\sigma_1^2 + \sigma_2^2}. \tag{A.5.11}$$

# Appendix B

# Calculations for phase encoded states

## B.1 Orthonormal basis for $N$ coherent states

Suppose that we have $N$ coherent states described by amplitudes $a_k$ for $k = 0, 1 \dots N-1$. Since these states are non-orthogonal we can have a matrix $\mathbf{V}$ that describes their overlaps, which are given by

$$V_{ij} = \langle a_i | a_j \rangle = \exp\left[-\frac{1}{2}\left(|a_i|^2 + |a_j|^2 - 2a_i^* a_j\right)\right]. \tag{B.1.1}$$

For a constellation of states as described before and after the attenuation due to the propagation through a pure-loss channel, the overlaps for Bob are given by

$$V_{ij}^B = \langle \sqrt{\tau} a_i | \sqrt{\tau} a_j \rangle = \exp\left[\tau z^2 \left(e^{\mathrm{i}\frac{2\pi}{N}(j-i)} - 1\right)\right], \tag{B.1.2}$$

while for Eve we may write

$$\begin{aligned} V_{ij}^E =& \langle \sqrt{1-\tau} a_i | \sqrt{1-\tau} a_j \rangle = \\ &= \exp\left[(1-\tau) z^2 \left(e^{\mathrm{i}\frac{2\pi}{N}(j-i)} - 1\right)\right]. \end{aligned} \tag{B.1.3}$$

Then, according to the Gram-Schmidt procedure, we can derive an orthonormal basis $\{|i\rangle\} = \{|0\rangle, |1\rangle, \dots |N-1\rangle\}$ for the subspace spanned by these $N$ coherent states. As a result, each state will be expressed as a superposition of this basis vectors as

$$|a_k\rangle = \sum_{i=0}^{k} M_{ki} |i\rangle \tag{B.1.4}$$

where the $M_{ki}$ can be computed by the algorithm

$M_{k0} = V_{0k}$,

$M_{ki} = \frac{1}{M_{ii}}\left(V_{ik} - \sum_{j=0}^{i-1} M_{ij}^* M_{kj}\right)$ if $1 \leq i < k$,

$M_{ki} = 0$ otherwise,

$M_{kk} = \sqrt{1 - \sum_{i=0}^{k-1} |M_{ki}|^2}$ for $k > 0$.

Then the density matrix $\rho(a_k) = |a_k\rangle\langle a_k|$ is given by

$$\rho(a_k) = \sum_{i,j=0}^{k} M_{k,i} M_{k,j}^* |i\rangle\langle j|, \tag{B.1.5}$$

and the average state takes the form

$$\rho = \frac{1}{N}\sum_{k=0}^{N-1} \rho(a_k) = \frac{1}{N}\sum_{k=0}^{N-1}\sum_{i,j=0}^{k} M_{k,i} M_{k,j}^* |i\rangle\langle j|. \tag{B.1.6}$$

Diagonalizing the previous state, we then compute its von Neumann entropy.

## B.2 Asymptotic state for a continuous alphabet

Let us express a coherent state in the Fock basis, i.e,

$$\Pi(a) := |a\rangle\langle a| = e^{-|a|^2} \sum_{n,m=0}^{\infty} \frac{a^n (a^\dagger)^m}{\sqrt{n!}\sqrt{m!}} |n\rangle\langle m| \tag{B.2.7}$$

In order to be able to do numerical calculations, we have to truncate the Fock space and a very good approximation is given by $n \sim 2|\alpha|^2$. As a result, in this truncated Fock basis, the state will be

$$\Pi^{\text{tranc}}(a) \simeq e^{-|a|^2} \sum_{n,m=0}^{2\lfloor |a|^2 \rfloor} \frac{a^n (a^\dagger)^m}{\sqrt{n!}\sqrt{m!}} |n\rangle\langle m|. \tag{B.2.8}$$

For $N$ coherent states in a constellation with radius $z$, the average state can be written as

$$\rho = \frac{e^{-z^2}}{N} \sum_{n,m=0}^{2\lfloor z^2 \rfloor} \frac{z^{(n+m)} \sum_{j=0}^{N-1} e^{\mathrm{i}\frac{2\pi}{N}(n-m)j}}{\sqrt{n!}\sqrt{m!}} |n\rangle\langle m|, \tag{B.2.9}$$

where the non zero terms are the terms with $m - n = N$ and $n = m$. For a continuous distribution $p(a_\phi) = \frac{1}{2\pi}$ of phase-encoded coherent states $|a_\phi\rangle$ with fixed radius $z = |a|$ and $\phi = \arg(a_\phi)$, Eq. (B.2.9) becomes

$$\begin{aligned} \rho =& \frac{e^{-z^2}}{2\pi} \sum_{n,m=0}^{2\lfloor z^2 \rfloor} \frac{z^{(n+m)} \int_0^{2\pi} e^{\mathrm{i}\phi(n-m)} d\phi}{\sqrt{n!}\sqrt{m!}} |n\rangle\langle m| = \\ =& e^{-z^2} \sum_{n=0}^{2\lfloor z^2 \rfloor} \frac{z^{2n}}{n!} |n\rangle\langle n|. \end{aligned} \tag{B.2.10}$$

## B.3 Displaced thermal state

A thermal state with mean number of photons $\bar{n}$ may be expressed as a convex sum of coherent states $|a\rangle$ according to the P-Glauber representation as

$$\rho(\bar{n}) = \int p(a,\bar{n})|a\rangle\langle a|d^2a, \quad p(a,\bar{n}) = \frac{1}{\bar{n}\pi}e^{-|a|^2/\bar{n}}. \tag{B.3.11}$$

Applying the displacement operator $D(d)$, which displaces a coherent state $|a\rangle$ with amplitude $a$ into a coherent state $|a+d\rangle$ with amplitude $a+d$, we obtain a displaced thermal state

$$\begin{aligned}\rho(d,\bar{n}) &= D(d)\rho(\bar{n})D^\dagger(d) = \\ &= \int p(a,\bar{n})D(d)|a\rangle\langle a|D^\dagger(d)\ d^2a = \\ &= \int p(a,\bar{n})|a+d\rangle\langle a+d|d^2a = \\ &= \int p(c-d,\bar{n})|c\rangle\langle c|d^2c\end{aligned} \tag{B.3.12}$$

with $p(c-d,\bar{n}) = \frac{1}{\bar{n}\pi}e^{-|c-d|^2/\bar{n}}$. According to equation Eq. (B.2.7), we can have a representation of this state in Fock basis, so that

$$\rho(d,\bar{n}) = \int \sum_{n,m=0}^{\infty} p(a-d,\bar{n})e^{-|a|^2}\frac{a^n(a^*)^m}{\sqrt{n!}\sqrt{m!}}|n\rangle\langle m|\ d^2a \tag{B.3.13}$$

The state after projecting to a coherent state $|b\rangle$ (heterodyne measurement), i.e., $\Pi(b)\rho(d,\bar{n})\Pi^\dagger(b)$, will be calculated as

$$\begin{aligned}&\int d^2a \sum_{n,m,k,l,i,j=0}^{\infty} p(a-d,\bar{n})e^{-|\alpha|^2}\frac{\alpha^n(\alpha^*)^m}{\sqrt{n!}\sqrt{m!}}e^{-|b|^2}\frac{b^k(b^*)^l}{\sqrt{k!}\sqrt{l!}}\times \\ &\times e^{-|b|^2}\frac{b^i(b^*)^j}{\sqrt{i!}\sqrt{j!}}|k\rangle\langle l||n\rangle\langle m||i\rangle\langle j| =\end{aligned} \tag{B.3.14}$$

$$\begin{aligned}&\int d^2a\ p(a-d,\bar{n})e^{-|a|^2}e^{-2|b|^2}\sum_{n,m=0}^{\infty}\frac{(ab^*)^n(ba^*)^m}{\sqrt{n!}\sqrt{m!}}\times \\ &\times \sum_{k,j=0}^{\infty}\frac{b^k(b^*)^j}{\sqrt{k!}\sqrt{j!}}|k\rangle\langle j|,\end{aligned} \tag{B.3.15}$$

and, applying the trace operation, we obtain the probability distribution

$$\begin{aligned}p(b|d)(\bar{n}) &= \int d^2a\ \frac{1}{\bar{n}\pi}e^{-|a-d|^2/\bar{n}}e^{-(|a|^2+|b|^2-b^*a-ba^*)} = \\ &= \frac{1}{\bar{n}\pi}\int e^{-|a-d|^2/\bar{n}}e^{-|a-b|^2}d^2a = \\ &= \frac{1}{(\bar{n}+1)\pi}\exp\left(-|b-d|^2/(\bar{n}+1)\right).\end{aligned} \tag{B.3.16}$$

Let us write this probability distribution for the thermal output state of a thermal-loss channel with transmissivity $\tau$ and mean thermal photon number $\bar{n}$ when applied to an input coherent state $|a_k\rangle$ ($d := \sqrt{\tau} a_k$). We find Eq. (11.9).

# Abbreviations

**TMSV** two-mode squeezed vacuum

**CV** Continuous-variables

**QKD** quantum key distribution

**MDI** measurement-device-independent

**CM** covariance matrix